\documentclass[a4paper,11pt]{article}
\pdfoutput=1

\usepackage{jheppub}
\usepackage[T1]{fontenc} 
\usepackage{dsfont}

\usepackage{comment}
\usepackage{stackengine,xcolor}
\usepackage{Settings}
\usepackage{blkarray}
\usepackage{todonotes}
\usepackage{subcaption}
\usepackage{tabularx}
\usepackage{mathbbol}
\usepackage{cancel}

\usepackage{array, cellspace, makecell, caption, booktabs}
\setcellgapes{5pt}
\usepackage{tikz,pgf}   
\usepackage{tikz-feynman,contour}   

\newcommand{\HalfLen}{2.2}   
\newcommand{\BlobR}{7mm}     

\tikzset{
  kernelline/.style={line width=1.0pt, line cap=round, line join=round},
  blobstyle/.style={circle, draw=black, line width=0.9pt, fill=gray!15, minimum size=\BlobR},
  gluon/.style={decorate, decoration={snake, aspect=0.8, segment length=2.3mm, amplitude=0.6mm}, line width=0.7pt},
  impactstyle/.style={ellipse, draw=black, line width=0.9pt, fill=gray!15, minimum width=20mm, minimum height=8mm},
  jstyle/.style={circle, draw=black, line width=0.7pt, fill=white, minimum size=\BlobR, inner sep=0pt},
  smalljstyle/.style={circle, draw=black, line width=0.9pt, fill=white, minimum size=9pt, inner sep=0.05pt},
  qarrow/.style={line width=0.8pt, -{Latex[length=5pt, width=6pt]}},
  darrow/.style={line width=0.4pt, {Latex[length=5pt, width=4pt]}-{Latex[length=5pt, width=4pt]}}
}

\usepackage{tcolorbox}
\usepackage{xcolor}

\def\beq{\begin{equation}}
\def\eeq{\end{equation}}
\def\beqa{\begin{eqnarray}}
\def\eeqa{\end{eqnarray}}
\newcommand{\nn}{\nonumber}

\newcommand{\badat}{\begin{alignedat}}
\newcommand{\eadat}{\end{alignedat}}

\newcommand{\hd}{\hat{\mathrm{d}}}

\def\eqn#1{eq.~(\ref{#1})}

\graphicspath{ {Pictures/} }

\usepackage[colorlinks = true,
            linkcolor = blue,
            urlcolor  = blue,
            citecolor = blue,
            anchorcolor = blue]{hyperref}

\title{Gravitational amplitudes in the Regge limit:\\ waveforms, shock waves and unitarity cuts}
\date{\today}

\author[1]{Francesco Alessio,}
\author[1]{Vittorio Del Duca,}
\author[2,3]{Riccardo Gonzo,}
\author[1,4]{Emanuele Rosi}

\affiliation[1]{INFN, Laboratori Nazionali di Frascati, 00044 Frascati (RM), Italy }%
\affiliation[2]{Centre for Theoretical Physics, Department of Physics and Astronomy,
Queen Mary University of London, London E1 4NS, United Kingdom}
\affiliation[3]{Higgs Centre for Theoretical Physics, School of Physics and Astronomy, The University of Edinburgh, EH9 3FD, Scotland}
\affiliation[4]{Dipartimento di Fisica, Sapienza Università di Roma, Piazzale Aldo Moro 5, 00185, Roma, Italy}

\emailAdd{Francesco.Alessio@lnf.infn.it}\emailAdd{emanuele.rosi@uniroma1.it}\emailAdd{Vittorio.Del.Duca@cern.ch}\emailAdd{r.gonzo@qmul.ac.uk}

\abstract{Motivated by recent progress in the high-energy description of gravitational scattering, we develop a systematic Regge-theory framework for $2\to2+n$ amplitudes describing the scattering of two massive particles with $n$ graviton emissions, including spin effects. Working in the ultra-relativistic limit at leading logarithmic accuracy, the massive result smoothly reduces to its massless counterpart. We describe both quantum (Regge trajectory and BFKL $t$-channel evolution) and classical ($s$-channel multi-$H$ evolution) contributions using both an exponential representation of the S-matrix and a shock-wave formalism in light-cone quantisation. In the latter approach, gravitational Wilson lines evolve in rapidity space under a boost-invariant Hamiltonian, providing a space-time realisation of the high-energy dynamics and making contact with recent effective field theory descriptions in the forward limit. As an application, we compute the leading-logarithmic contribution to the massive spinless $2\to2$ amplitude at 5PM–2SF order, recovering the previously determined  massless result, and derive the tree-level $2\to3$ amplitude and its associated scattering waveform for Kerr black holes in the ultra-relativistic limit. }
 
\begin{document}

\preprint{QMUL-PH-26-03}

\clearpage
\maketitle

\section{Introduction and motivation}
\label{sec:intro}

The observations and precision measurements of gravitational-wave emissions by the LIGO-Virgo-KAGRA (LVK) network~\cite{LIGOScientific:2018mvr,LIGOScientific:2020ibl,KAGRA:2021vkt} and the prospect of a yet improved precision coupled to a much larger kinematic reach, as expected by the next generation of ground-\cite{Punturo:2010zz,Reitze:2019iox} and space-based~\cite{LISA:2017pwj} gravitational-wave observatories, have prompted the quest for a more accurate theoretical understanding of the gravitational waveform models of compact binary systems. At a fundamental level, this effort amounts to solving the relativistic gravitational two-body problem across a wide parameter space, encompassing different mass ratios, velocities and interaction strengths.

A particularly powerful weak-coupling description of relativistic two-body dynamics is provided by the post-Minkowskian (PM) expansion, which organises gravitational interactions as an expansion in Newton’s constant $G$ at arbitrary velocities. From a field theory perspective, the PM expansion is naturally realised through relativistic scattering amplitudes: the classical contribution of an $\ell$-loop four-point amplitude scales as ${\cal O}(G^{\ell+1})$ and contributes to the $(\ell+1)$PM order. Over the past years, a combination of on-shell amplitude methods~\cite{Neill:2013wsa,Bjerrum-Bohr:2018xdl,Cheung:2018wkq,Bern:2019crd}, worldline approaches~\cite{Kalin:2020mvi,Mogull:2020sak}, and effective field theory formulations of the two-body problem~\cite{Goldberger:2004jt,Porto:2016pyg} has enabled a rapid advance of higher PM results for the gravitational scattering of two massive scalar particles, with complete expressions for the $2 \to 2$ amplitude now available through 2PM~\cite{Westpfahl:1985tsl}, 3PM~\cite{Bern:2019nnu,Kalin:2020fhe,Damour:2020tta,DiVecchia:2021bdo,Brandhuber:2021eyq,Damgaard:2021ipf} and 4PM~\cite{Bern:2021dqo,Dlapa:2021npj,Bern:2021yeh,Dlapa:2021vgp,Dlapa:2022lmu,Damgaard:2023ttc,Dlapa:2024cje} order.

An independent and complementary organisation of relativistic two-body dynamics is provided by self-force (SF) theory~\cite{Poisson:2011nh,Barack:2018yvs,Pound:2021qin}, which treats the motion of a small compact object as a perturbation of geodesic motion in a black hole background and encodes corrections as an expansion in the symmetric mass ratio $\nu$, with the $n$SF order corresponding to the $\mathcal{O}(\nu^{\,n})$ deviation from geodesic motion. In this language, a fixed PM order of the $2 \to 2$ amplitude decomposes into SF sectors: the $(2n+1)$PM and $(2n+2)$PM orders contain contributions up to the $n$SF level~\cite{Damour:2017zjx}. Consequently, 5PM amplitudes probe dynamics through the 2SF sector. At present, the complete 1SF contribution at 5PM is known~\cite{Driesse:2024xad,Driesse:2024feo,Dlapa:2025biy}, and partial 2SF information has been obtained in the potential region~\cite{Bern:2025zno,Bern:2025wyd} including tail effects~\cite{Driesse:2026qiz}.

A closely related and physically central extension of this programme concerns radiative processes encoded in the $2\to3$ amplitude, which serves as the fundamental building block for scattering waveforms. In recent years, novel approaches based on both the KMOC formalism~\cite{Cristofoli:2021vyo} and worldline methods~\cite{Mogull:2020sak} have enabled the analytic computation of tree-level scattering waveforms~\cite{Mougiakakos:2021ckm,Jakobsen:2021smu,DeAngelis:2023lvf,DiVecchia:2021bdo,Brunello:2025cot}, reproducing the classic result of Kovács and Thorne obtained with traditional techniques~\cite{Kovacs:1977uw,Kovacs:1978eu}. This progress has since been pushed beyond tree-level, with explicit calculations at one loop~\cite{Brandhuber:2023hhy,Herderschee:2023fxh,Georgoudis:2023lgf,Elkhidir:2023dco,Caron-Huot:2023vxl,Georgoudis:2023ozp,Bini:2023fiz,Bini:2024rsy,Brunello:2025eso} and partial results at two and three loops~\cite{Alessio:2024onn,Bini:2024ijq,Georgoudis:2025vkk}, showing agreement with the probe limit~\cite{Georgoudis:2023eke,Fucito:2024wlg,Heissenberg:2025fcr}.  A further conceptual milestone has been the incorporation of spin effects through the identification of the Kerr geometry with a spinning point particle~\cite{Vines:2017hyw,Arkani-Hamed:2017jhn,Guevara:2018wpp,Chung:2018kqs,Guevara:2019fsj,Aoude:2020onz}, allowing the computation of scattering waveforms for spinning particles both at tree-level~\cite{Jakobsen:2021lvp,DeAngelis:2023lvf,Aoude:2023dui,Brandhuber:2023hhl} and at one loop~\cite{Bohnenblust:2023qmy,Bohnenblust:2025gir}. 

While the PM and SF frameworks provide complementary perturbative descriptions of the two-body motion, their fixed-order truncations do not uniformly capture the dynamics across the full parameter space. In particular, there exist regimes -- such as highly relativistic encounters at comparable masses -- where neither a finite PM or SF expansion are sufficient on their own. In practice, these limitations are exposed by comparisons with numerical relativity (NR) simulations~\cite{Pretorius:2005gq,Campanelli:2005dd,Baker:2005vv},
motivating the development of resummation strategies and alternative theoretical methods that extend beyond fixed-order perturbation theory. 

A sharp manifestation of this breakdown arises in the high-energy regime, where observables computed in the PM expansion cease to be reliable at large center-of-mass energies, beyond the D’Eath bound~\cite{DEath:1976bbo,Kovacs:1977uw,Kovacs:1978eu,Gruzinov:2014moa,Ciafaloni:2015xsr,Ciafaloni:2018uwe,DiVecchia:2022nna,Gonzo:2023cnv}, signalling a breakdown of the fixed-order expansion outside its domain of convergence. While physically motivated resummations -- such as those employed in self-force and EOB-based approaches -- can partially improve agreement with numerical simulations~\cite{Damour:2022ybd,Long:2024ltn,Clark:2025kvu,Swain:2024ngs,Long:2025nmj}, the appearance of high-energy logarithms points to the need for a theoretical framework that directly targets the underlying high-energy structure. 

This naturally motivates the study of the Regge limit~\cite{Regge:1959mz}, defined by $s \gg |t|, m_1^2, m_2^2$ for $2 \to 2$ amplitudes. A first example of these high-energy logarithms is provided by the 1SF sector at 3PM, where the imaginary part of the two-loop four-point amplitude develops a logarithmic enhancement $\sim \log(s/|t|)$ in the Regge limit~\cite{DiVecchia:2020ymx}. This logarithm is universal -- independent of the masses and spins of the scattering particles -- and was first identified in the context of graviton–graviton scattering by Amati, Ciafaloni and Veneziano~\cite{Amati:1990xe}. Diagrammatically, it arises from the three-particle unitarity cut of the so-called $H$ diagram~\cite{Amati:1990xe}, where a soft graviton is exchanged across an eikonal ladder. At higher orders, this mechanism generalises to an $s$-channel multi-$H$ structure with $N$ soft graviton exchanges, giving rise to an infinite tower of classical diagrams scaling as ${\cal O}\left[Gs^2(G^2 s \log(s/|t|))^N\right]$~\cite{Amati:2007ak,Ciafaloni:2015xsr,Rothstein:2024nlq}.

This emerging tower of logarithmic corrections was recently studied in Ref.~\cite{Rothstein:2024nlq}, using an adaptation of soft-collinear effective theory (SCET) to gravity in the forward limit introduced by Rothstein and Saavedra~\cite{Rothstein:2016bsq} and rapidity renormalisation group equations (RRGEs) for Glauber modes~\cite{Chiu:2011qc,Chiu:2012ir,Gao:2024qsg}. 
Remarkably, the same $s$-channel structure admits a complementary and purely on-shell interpretation. It can be reproduced by iteratively sewing tree-level amplitudes through three-particle unitarity cuts in the $s$-channel, making the connection to unitarity and factorisation manifest. Both the RRGE approach and the three-particle cut construction have been exploited to determine the leading-logarithmic behaviour of massless gravitational scattering amplitudes through 5PM order~\cite{Alessio:2025isu}.

In this paper, we develop a unified framework to organise the high-energy expansion of 
$2\to2$ and $2\to3$ gravitational amplitudes in the Regge limit. Our approach combines Regge theory and shock-wave methods in light-cone quantisation, providing a space-time realisation of high-energy gravitational dynamics and making contact  with recent effective field theory descriptions in the forward limit. We extend the QCD shock-wave formalism~\cite{Balitsky:1998ya,Balitsky:2001mr,Caron-Huot:2013fea} to gravity~\cite{Dray:1984ha,Lodone:2009qe}, formulating the dynamics in terms of a boost-invariant Hamiltonian governing the rapidity evolution of gravitational Wilson lines, and develop a Regge-theory description of multi-$H$ diagrams in massive scattering that clarifies the interplay  between $t$-channel (BFKL-type) evolution and classical $s$-channel unitarity cuts. As an application, we show that at 5PM order in the 2SF sector the leading-logarithmic contribution  to the massive $2\to2$ amplitude reduces to the corresponding massless result~\cite{Alessio:2025isu}, and extend the formalism to radiative processes by computing the ultra-relativistic limit of tree-level scattering waveforms from $2\to3$ amplitudes, including all-order spin effects.

The paper is organised as follows. In sec.~\ref{sec:MRK}, we review gravitational scattering amplitudes in the Regge limit and introduce multi-Regge kinematics (MRK), with particular emphasis on tree-level $2\to2$ and $2\to2+n$ amplitudes. In sec.~\ref{sec:regge}, we reformulate the Regge-theory description in terms of the $\hat{N}$ operator, which provides a generating functional for the iterative evolution along the $t$-channel and $s$-channel. In sec.~\ref{sec:shock-wave}, we then develop the shock-wave formalism for gravitational scattering, introducing the Reggeised graviton field and the associated boost Hamiltonian, and compute both $2\to2$ and $2\to3$ amplitudes within this framework. Finally, in sec.~\ref{sec:waveform} we apply these results to the computation of tree-level scattering waveforms in the ultra-relativistic limit, incorporating also spin effects for the first time . We conclude in sec.~\ref{sec:conclusions} with a summary and a discussion of future directions.

\paragraph{Conventions:} We adopt the metric convention $\eta=\mathrm{diag}(+,-,-,-)$ and set $\kappa^2=8\pi G$. Calculations are performed using dimensional regularisation in $D=4-2\epsilon$ dimensions, with renormalisation scale $\mu$. We also write $\hat{\mathrm{d}}^{D}k\equiv \mathrm{d}^{D}k/(2\pi)^D$, $\hat{\delta}^{(D)}(\cdot)\equiv (2\pi)^D\delta^{(D)}(\cdot)$ and $\delta^+(k^2)\equiv \Theta(k^0)\delta(k^2)$. Light-cone components are $p^\pm = p^0\pm p^{D-1}$, so that the momentum naturally decomposes as $p^\mu=\tfrac12(p^+ n_-^\mu+p^- n_+^\mu)+p_{\perp}^\mu$ with $n_\pm^\mu=(1,0,\ldots,0,\pm1)$ and $n_\pm\!\cdot p_{\perp}=0$. Thus $p^\mu_{\perp} = (0,0,\vec{p})$, with $\vec{p} \in \mathbb{R}^d$ and $d=D-2$, and we use $\int_{q_{\!\perp}}\equiv e^{\epsilon\gamma_E}\!\int \hat{\mathrm{d}}^{\,d}q_{\!\perp}$ as a convention for the transverse integrals. When specialising to $D=4$ ($\epsilon = 0$) we occasionally use the complex transverse variable $p_{\mathrm{T}} \equiv p^x+i p^y$; unless stated otherwise, expressions are $SO(d)$-covariant. We adopt the conventions of~\cite{Kosower:2018adc} for the $\hbar$ scaling, {\it i.e.} $\kappa\mapsto\kappa/\sqrt{\hbar}$ and $q\mapsto\hbar \bar{q}$ for graviton momenta. 
Finally, throughout the paper we use the shorthand notation $\simeq$ to denote equalities that hold at leading order in the Regge expansion.

\section{Gravity amplitudes in multi-Regge kinematics}
\label{sec:MRK}

We study the $2 \to 2 + n$ gravitational scattering process in Einstein gravity,
\beqa\label{eq:process-2to2n}
i(-p_1,-\lambda_1) + j(-p_2,-\lambda_2)
\;\longrightarrow\;
 i(p_4,\lambda_4) + j(p_3,\lambda_3) + \sum_{r=1}^{n} g(k_r,h_r) \,,
\eeqa
where $i,j\in\{\phi,g,\phi_{\rm Kerr}\}$ label the external species (a minimally coupled scalar $\phi$, a graviton $g$ or a massive scalar $\phi_{\rm Kerr}$  whose non-minimal couplings encode the spin-induced multipole structure of a Kerr black hole in the classical limit), and we adopt the all-outgoing convention for the scattering amplitude. The symbols $\lambda_{1,\dots,4}$ denote the helicities of the external legs (absent for scalars), or more generally the spin degrees of freedom for $\phi_{\rm Kerr}$ states, while $h_r$ label the helicities of the emitted gravitons. With our conventions, the on-shell conditions and momentum conservation read
\begin{align}
&p_1^2=p_4^2=m_1^2,\qquad p_2^2=p_3^2=m_2^2,\qquad 
k_r^2=0 \,\, \text{for} \,\, r=1,\ldots,n\,, \nonumber \\
&\qquad \qquad \qquad  p_1+p_2+p_3+p_4+\sum_{r=1}^{n} k_r = 0\,.
\label{eq:on-shell_mom}
\end{align}
To describe the high-energy behaviour, it is convenient to introduce light-cone coordinates,
\beq
p^\mu = (p^+,p^-,\vec{p}), \qquad 
p^\pm = p^0 \pm p^{D-1}, \qquad 
m_{\perp}^2 = m^2+\vec{p}^{\,2},
\eeq
so that an on-shell momentum takes the form
\beq
p^\mu = (m_{\perp} e^y,\, m_{\perp} e^{-y},\,\vec{p}\,),
\label{eq:rapidity}
\eeq
with rapidity $y=\tfrac12\log(p^+/p^-)$, as detailed in app.~\ref{sec:appa}. We define the $\hat{S}$-matrix as $\hat{S} = 1 + i \hat{T}$, so that the scattering amplitude $\mathcal{M}$ is, as usual, the $\hat{T}$-matrix element,
\beq
\langle p_3\,p_4\, k_1, \dots, k_n|\, \hat{T} \,| p_1\,p_2\rangle 
= \sum_{\ell=0}^{\infty} \mathcal{M}^{(\ell)}_{2 \to 2+n}\,\hat{\delta}^{(D)}\!
(p_1 + p_2 + p_3 + p_4 + \sum_{r=1}^n k_r)\,,
\label{eq:defM}
\eeq
where $\ell$ denotes the loop order and $\hat{\delta}^{(D)}(P)$ enforces the overall momentum conservation and where
\begin{align}
\label{eq:defMloops}
\mathcal{M}_{2\rightarrow2+n}^{(\ell)}=\mathcal{O}( G^{\ell+1+n/2})\,.
\end{align}

\subsection{Elastic $2\rightarrow 2$ amplitudes in the Regge limit}
\label{sec:4amps}
\begin{figure}[t]
\centering
\hspace{-15pt}
\centering
\begin{tikzpicture}[x=1cm,y=1cm]
  \node[anchor=east] at (-2.5,0) {\large $\mathcal M^{(0)}_{2\to2}\;=$};

  \def\TopY{ 1.5}
  \def\BotY{-1.5}

  \draw[kernelline] (-\HalfLen-0.5, \TopY) -- (\HalfLen+0.5, \TopY);
  \draw[kernelline] (-\HalfLen-0.5, \BotY) -- (\HalfLen+0.5, \BotY);

  \draw[kernelline,qarrow] (-1,\TopY+0.4) -- (-2.5,\TopY+0.4)
    node[midway, yshift=9pt] {\large $p_1$};
  \draw[kernelline,qarrow] ( 1,\TopY+0.4) -- ( 2.5,\TopY+0.4)
    node[midway, yshift=9pt] {\large $p_4$};

  \draw[kernelline,qarrow] (-1,\BotY-0.4) -- (-2.5,\BotY-0.4)
    node[midway, yshift=-9pt] {\large $p_2$};
  \draw[kernelline,qarrow] ( 1,\BotY-0.4) -- ( 2.5,\BotY-0.4)
    node[midway, yshift=-9pt] {\large $p_3$};

  \draw[kernelline,qarrow] (-1,\TopY-0.6) -- (-1,\BotY+0.6)
      node[midway, xshift=-10pt] {\large $q$};

  \node[impactstyle] (Ctop) at (0,\TopY) {\large $C_{i}$};
  \node[impactstyle] (Cbot) at (0,\BotY) {\large $C_j$};

  \draw[gluon] (0,\TopY-0.45) -- (0,\BotY+0.45);

\end{tikzpicture}

\caption{Structure of the elastic amplitude $\mathcal M^{(0)}_{2\to2}$ in MRK, represented as two impact factors connected by a graviton line.}
\label{fig:RK_amplitude}
\end{figure}
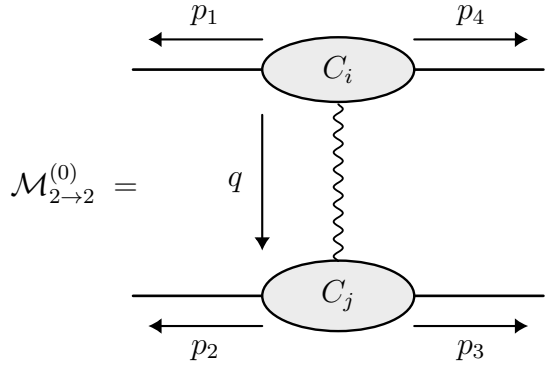

We start by considering the elastic $2\to2$ scattering with no additional graviton emissions, 
\beq
i(-p_1,-\lambda_1) + j(-p_2,-\lambda_2) \;\longrightarrow\; i(p_4,\lambda_4) + j(p_3,\lambda_3) \,,
\eeq
obtained by setting $n=0$ in eq. \eqref{eq:process-2to2n} and represented in Fig. \ref{fig:RK_amplitude}. The process is described in terms of the Mandelstam invariants,
\beq
s=(p_1+p_2)^2,\qquad t=(p_2+p_3)^2,\qquad u=(p_1+p_3)^2,
\eeq
together with the masses $m_1^2$ and $m_2^2$, which satisfy $s+t+u=2(m_1^2+m_2^2)$ as a consequence of momentum conservation and the on-shell relations \eqref{eq:on-shell_mom}.

The Regge limit of the $2\to 2$ amplitude is defined by
\beq
s \gg |t|,\, m_1^2,\, m_2^2 \,,
\label{eq:reggelim}
\eeq
that is, the centre-of-mass energy dominates over the momentum transfer and the particle masses. In this limit, the momenta exhibit a strong ordering in their light-cone components,
\beq
p_4^+ \gg p_3^+ ,\qquad p_4^- \ll p_3^- \,,
\label{eq:strordpm}
\eeq
which is equivalent, by the on-shell relations and transverse-momentum conservation,\footnote{
The second inequality in \eqref{eq:strordpm} follows from the first, since on shell one has 
$p_3^+p_3^- = \vec{p}_3{^2}+m_2^2$ and $p_4^+p_4^- = \vec{p}_{4}^{\,2}+m_1^2$, together with $\vec{p}_3+\vec{p}_4=0$.  Using $p^\pm=m_{\perp} e^{\pm y}$ with $m_{\perp}^2=m^2+\vec{p}^{\,2}$, the condition \eqref{eq:strordpm} is equivalent to a strong ordering of the rapidities of the final-state particles. } to a strong ordering of the rapidities of the two initial and final-state particles,
\begin{align}
y_{p_1} \gg y_{p_2}\,, \qquad y_{p_4} \gg y_{p_3}.
\end{align}
For massless states the transverse mass reduces to $m_{\perp}^2=\vec{p}^{\,2}$, and the ordering condition is unchanged.

Just as QCD amplitudes factorise in the Regge limit into impact factors coupled to a $t$-channel gluon~\cite{Kuraev:1976ge,Kuraev:1977fs,Balitsky:1978ic,DelDuca:1995zy,DelDuca:1996km}, gravity amplitudes factorise into impact factors coupled to a $t$-channel graviton~\cite{Lipatov:1982vv,Lipatov:1982it,Barcaro:2025ifi}. At tree-level one finds
\beq
{\cal M}_{2 \to 2}^{(0)}=-\frac{1}{\hbar}\bigl[\kappa\, C_i(p_1,p_4)\bigr]\,\frac{s^2}{t}\,\bigl[\kappa\, C_j(p_2,p_3)\bigr] + \mathcal{O}\left(\frac{t}{s},\frac{m_1^2}{s},\frac{m_2^2}{s}\right) \,,
\label{eq:4reggea}
\eeq
with $C_f(p_a,p_b)$ the impact factor for the transition $a\!\to\!b$ through the emission of an off-shell graviton from a line with field $f = \phi,\phi_{\mathrm{Kerr}},g$. As in the massless case of QCD amplitudes, the factorisation of eq.~(\ref{eq:4reggea}) holds up to power corrections of ${\cal O}(t/s)$. In addition, since the incoming and outgoing particles are massive, eq.~(\ref{eq:4regge}) holds up to further power corrections of ${\cal O}(m_1^2/s)$ and ${\cal O}(m_2^2/s)$ as well~\cite{Barcaro:2025ifi}. In other words, corrections depending on $t/m_1^2$, $t/m_2^2$ or $m_1^2/m_2^2$ are power suppressed in the Regge expansion of eq.~\eqref{eq:4reggea}. For brevity, here and throughout the paper we use the symbol $\simeq$ to denote equality at leading power in the (multi-)Regge expansion, so that \eqref{eq:4reggea} can be equivalently written as
\beq
{\cal M}_{2 \to 2}^{(0)}\simeq-\frac{1}{\hbar}\bigl[\kappa\, C_i(p_1,p_4)\bigr]\,\frac{s^2}{t}\,\bigl[\kappa\, C_j(p_2,p_3)\bigr] \,.
\label{eq:4regge}
\eeq
For scalar external states the impact factors in eq.~(\ref{eq:4regge}) are
\beq
C_{\phi}(p_1,p_4)=1,\qquad C_{\phi}(p_2,p_3)=1\,.
\label{eq:gravifscal}
\eeq
The factorisation in \eqref{eq:4regge} also holds for the $2\rightarrow2$ scattering of Kerr black holes \cite{Vines:2017hyw,Guevara:2018wpp,Guevara:2019fsj}. From the high-energy limit of the tree-level amplitude one can extract the impact factors (see \textit{e.g}. \cite{Guevara:2019fsj}) and they read
\begin{align}
\label{eq:impactfactorsKerr}
C_{\phi_{\mathrm{Kerr}}}(p_1,p_4;a_1)=e^{a_1\cdot q},\qquad C_{\phi_{\mathrm{Kerr}}}(p_2,p_3;a_2)=e^{a_2\cdot q},
\end{align}
where $q=-p_1-p_4=p_2+p_3$ and $a_i$ are the ring radii of the two spinning black holes satisfying the supplementary spin condition (SSC) $p_i\cdot a_i=0$. They reproduce the correct ultra-relativistic limit of the $2\rightarrow2$ Kerr amplitude \cite{Adamo:2022rob}. For gravitons, the impact factors depend on the direction of the transverse momentum. Let $\varepsilon^{(a)}_{ij}$ be symmetric, transverse and traceless polarisation tensors of the little group $SO(d)$, and define $\hat n^{i} \equiv p_3^i/|\vec{p}_3|$. Here $a,b=1,\ldots,d(d+1)/2-1$ label graviton polarisation states, i.e. a basis of such $SO(d)$ tensors. Then, with a convenient little-group choice for the $(1,4)$ leg, we adopt the convention,\footnote{Equivalently, a manifestly $SO(d)$–covariant form for $C_g(p_1,p_4)$ identical to eq.\eqref{eq:gravifD} with $\hat n'^{\,i}\equiv p_{4
}^{i}/|\vec{p}_{4}|=-\hat n^{i}$ may be used instead, yielding the same factorised amplitude.}
\beq
C_g(p_1^{(a)},p_4^{(b)})=1,\qquad
C_g\big(p_2^{(a)},p_3^{(b)}\big)
=\Big(\varepsilon^{(a)}_{ij}\,\hat n^i \hat n^j\Big)
 \Big(\varepsilon^{(b)}_{kl}\,\hat n^k \hat n^l\Big).
\label{eq:gravifD}
\eeq
At leading power in the Regge limit there is no polarisation mixing along the $s$-channel -- as ``flip'' components are power suppressed in $t/s$ -- so $C_g\big(p_2^{(a)},p_3^{(b)}\big) \propto v^{a}v^{b}$ with $v^{a}=\varepsilon^{(a)}_{ij}\hat n^i\hat n^j$. 
In $D=4$, with little group $U(1)$, one may use complex transverse coordinates $p_{\mathrm{T}}= p^x+i p^y$ and replace the real $SO(d)$ polarisation basis $\varepsilon^{(a)}_{ij}$ by a helicity basis $\varepsilon^{(h)}_{ij}$ with $h=\pm$. We then obtain the familiar expressions,
\beq
C_g(p_1^{(h)},p_4^{(h')})=1,\qquad 
C_g(p_2^{(-)},p_3^{(+)})=\left(\frac{p_{3\mathrm{T}}^*}{p_{3\mathrm{T}}}\right)^2,
\label{eq:gravif}
\eeq
which are pure phases related by parity and exhibit helicity conservation at leading power~\cite{Barcaro:2025ifi},
\begin{align}
    \big[C_g(p_i^{(h)},p_j^{(h')})\big]^\ast=C_g(p_i^{(-h)},p_j^{(-h')})\,,\qquad C_g(p_i^{(h)}, p_j^{(h')})\ \propto\ \delta_{h,-h'}\,.
\end{align}
Finally, it is interesting to note that the graviton impact factor in eq.~\eqref{eq:gravifD} is the double copy of the gluon impact factors in QCD~\cite{Barcaro:2025ifi,DelDuca:1995zy}.

\subsection{Higher-point $2 \to 2 + n$ tree amplitudes in multi-Regge kinematics}
\label{sec:reggemrk}

We now consider the generic $2 \to 2 + n$ process in eq.~\eqref{eq:process-2to2n} with
$n$ emitted gravitons between the two massive legs, with the on-shell constraint \eqref{eq:on-shell_mom}. For notational convenience, we set
\beq
k_0 \equiv p_4,\qquad k_{n+1}\equiv p_3,
\eeq
so that the emitted gravitons are $k_1,\dots,k_n$. Multi–Regge kinematics (MRK) is defined by\footnote{On-shell, one has $k_r^+k_r^-=m_{r\perp}^2$ with $m_{r\perp}^2=m_r^2+|\vec{k}_r|^2$. Therefore, given that in MRK the transverse scales are common to leading power -- differences are $\mathcal{O}(|t|/s)$ -- the $k^+$ ordering directly implies the reverse $k^-$ ordering  $k_0^- \ll k_1^- \ll \cdots \ll k_{n+1}^-$. See appendix~\ref{sec:appa} for more details on the multi-particle kinematics.}
\beq
k_0^+ \gg k_1^+ \gg \cdots \gg k_n^+ \gg k_{n+1}^+\,,
\qquad
|\vec{k}_{0}|\simeq |\vec{k}_{1}|\simeq\cdots\simeq |\vec{k}_{n}|\simeq |\vec{k}_{n+1}|\,,
\label{eq:mrk}
\eeq
or, equivalently, by the strong rapidity ordering,
\begin{align}\label{eq:y-ordering}
    y_0 \gg y_1 \gg \cdots \gg y_n \gg y_{n+1}.
\end{align}
At leading power in $t/s$, it is immaterial to use $|\vec{p}\,|$  or $m_{\perp}$ for the common transverse scale of the outgoing massive particles of momenta $p_3$ and $p_4$.
    
In ref.~\cite{Kuraev:1976ge}, Fadin, Kuraev and Lipatov (FKL) conjectured that in QCD, $2\to 2+n$ amplitudes in MRK iterate the emission of a final-state gluon along the $t$-channel gluon ladder, first occurring in a $2 \to 3$ amplitude. The FKL conjecture has been verified for six~\cite{DelDuca:1995ki}, seven~\cite{DelDuca:1999iql}
and, more generally, multi-leg~\cite{Byrne:2025phh} tree amplitudes with exchange of a gluon ladder in the $t$-channel as well as for $n$-point MHV amplitudes~\cite{DelDuca:1995zy}. The emission of a gluon along the ladder is described by the {\it Lipatov vertex}, also known as {\it central-emission vertex} (CEV)~\cite{Lipatov:1976zz}. The resulting gluon ladder is {\it universal}, in the sense that its structure depends solely on the gauge coupling and kinematics, and is independent of the flavour or mass of the external scattering particles 1 and 2. 

\begin{figure}[t]
\centering
\hspace{-5pt}
\begin{minipage}{0.48\linewidth}
\centering
\begin{tikzpicture}[x=1cm,y=1cm]
  \node[anchor=east] at (-2.5,0) {\large $\mathcal M^{(0)}_{2\to2+n}\;=$};

  \def\TopY{ 2.5}
  \def\BotY{-2.5}

  \draw[kernelline] (-\HalfLen-0.5, \TopY) -- (\HalfLen+0.5, \TopY);
  \draw[kernelline] (-\HalfLen-0.5, \BotY) -- (\HalfLen+0.5, \BotY);

  \draw[kernelline,qarrow] (-1,\TopY+0.4) -- (-2.5,\TopY+0.4)
    node[midway, yshift=9pt] {\large $p_1$};
  \draw[kernelline,qarrow] ( 1,\TopY+0.4) -- ( 2.5,\TopY+0.4)
    node[midway, yshift=9pt] {\large $p_4$};

  \draw[kernelline,qarrow] (-1,\BotY-0.4) -- (-2.5,\BotY-0.4)
    node[midway, yshift=-9pt] {\large $p_2$};
  \draw[kernelline,qarrow] ( 1,\BotY-0.4) -- ( 2.5,\BotY-0.4)
    node[midway, yshift=-9pt] {\large $p_3$};

  \draw[kernelline,qarrow] (-1,\TopY-1) -- (-1,\BotY+1)
      node[midway, xshift=-10pt] {\large $q$};

  \node[impactstyle] (Ctop) at (0,\TopY) {\large $C_i$};
  \node[impactstyle] (Cbot) at (0,\BotY) {\large $C_j$};

  \foreach \i/\yy in {1/1.05, 2/0.00, 3/-1.05} {%
    \node[jstyle] (J\i) at (0,\yy) {\large $J$};
    \draw[gluon, draw=blue] ($(J\i.east)$) -- ++(1.25,0);
  }

  \draw[gluon] (0,\TopY-0.45) -- (J1.north);
  \draw[gluon] (J1.south) -- (J2.north);
  \draw[gluon] (J2.south) -- (J3.north);
  \draw[gluon] (J3.south) -- (0,\BotY+0.45);

  \node[anchor=west] at (1,0.6) {\large $\vdots$};
  \node[anchor=west] at (1,-0.35) {\large $\vdots$};

  \draw[kernelline,qarrow] (1,\BotY+3.8) -- ( 2.5,\BotY+3.8)
    node[midway, yshift=9pt] {\large $k_1$};
    
  \draw[kernelline,qarrow] (1,\BotY+1.7) -- ( 2.5,\BotY+1.7)
    node[midway, yshift=9pt] {\large $k_n$};
  
\end{tikzpicture}
\end{minipage}\hfill
\begin{minipage}{0.48\linewidth}\centering

\vspace{0.6em}

\begin{tikzpicture}[x=1cm,y=1cm]
  \node[anchor=east] at (0,0) {\large $J^{\mu\nu}(q_r,k_r,q_{r+1}) = $};

  \begin{scope}[xshift=0.7cm]
    \draw[gluon] (0,0) -- (0, 1.35);    
    \draw[gluon] (0,0) -- (0,-1.35);    
    \draw[gluon, draw=blue] (0,0) -- (1.70,0);   

    \draw[kernelline,qarrow] (0.35,1.50) -- (0.35,0.65)
      node[midway, xshift=12pt] {\large $q_r$};
    \draw[kernelline,qarrow] (0.35,-0.65) -- (0.35,-1.50)
      node[midway, xshift=17pt] {\large $q_{r+1}$};
    \draw[kernelline,qarrow] (0.75,0.25) -- (1.60,0.25)
      node[midway, yshift=9pt] {\large $k_r$};

    \node[jstyle] at (0,0) {\large $J$};
    \node[] at (2.1,0) {$\mu \nu $};
  \end{scope}
\end{tikzpicture}
\end{minipage}
\caption{Structure of the amplitude $\mathcal M^{(0)}_{2\to2+n}$ in MRK, represented as a ladder of impact factors  $C_{i,j}$ connected by Lipatov currents $J^{\mu\nu}(q_r,k_r,q_{r+1})$ each emitting one graviton.}
\label{fig:MRK_amplitude}
\end{figure}
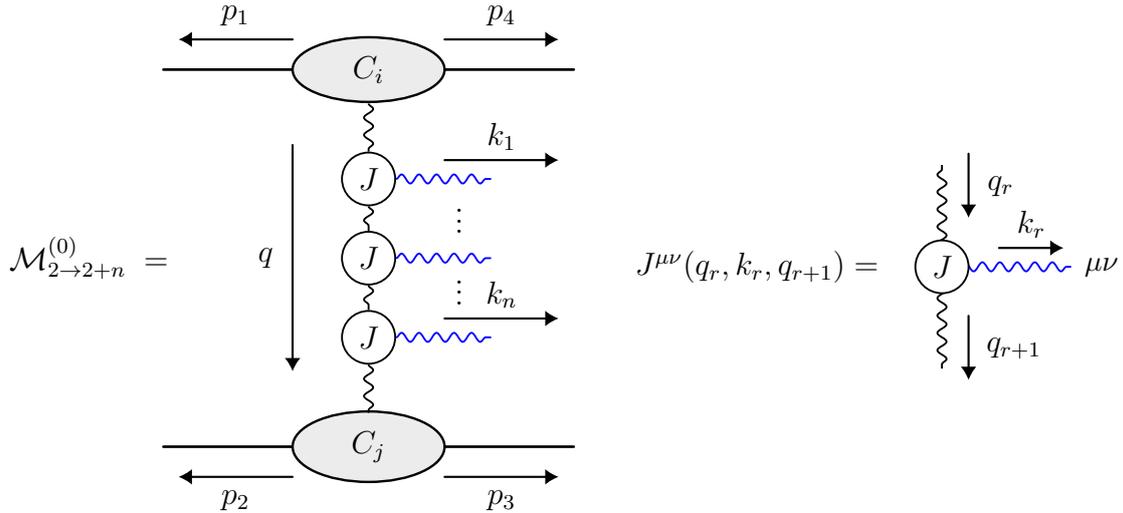

In a gravity theory, Lipatov~\cite{Lipatov:1982vv,Lipatov:1982it,Lipatov:1991nf} conjectured that $2 \to 2 + n$ amplitudes in MRK display the same structure as in QCD, iterating the emission of a final-state graviton -- called the graviton CEV -- along the $t$-channel graviton ladder, first occurring in a $2 \to 3$ amplitude. Just like the gluon ladder, also the graviton ladder is universal, i.e. independent of the mass and spin of the scattering particles 1 and 2. Lipatov's conjecture has been verified for five-point tree amplitudes, with four massive scalars and a graviton,
and for six-graviton tree amplitudes~\cite{Barcaro:2025ifi}, and we will assume that it holds for $n$-point tree amplitudes with exchange of a graviton ladder in the $t$-channel. Then the tree-level amplitudes in MRK with emission of $n$ gravitons along the $t$-channel ladder are expected to take the factorised ladder form~\cite{Raj:2025hse} (see Fig.~\ref{fig:MRK_amplitude}), 

\begin{align}
\nonumber
\mathcal{M}_{2 \to 2 +n}^{(0)}
\simeq&-\,\frac{s^2}{\hbar}\;
\big[\kappa\,C_i(p_1,k_0)\big]\;\frac{1}{t_1}
\\&\times \;
\prod_{r=1}^{n}\!\left\{\frac{\kappa}{\sqrt{\hbar}}\,\varepsilon^{(h_r)}_{\mu\nu}(k_r)\,J^{\mu\nu}(q_r,k_r,q_{r+1})\;\frac{1}{t_{r+1}}\right\}
\big[\kappa\,C_j(p_2,k_{n+1})\big],
\label{eq:npt}
\end{align}
in the kinematic region specified by \eqref{eq:y-ordering}, where the graviton polarisation tensor is expressed in factorised form,
$\varepsilon^{(h_r)}_{\mu\nu}(k_r)=\varepsilon^{(h_r)}_\mu(k_r)\varepsilon^{(h_r)}_\nu(k_r)$. Here $q_r$ is the exchanged momentum along the ladder,
\begin{align}
   &q_1 = -(p_1+k_0),\qquad
q_{r+1}=q_r-k_r \,\, \text{for} \,\, r=1,\ldots,n\,,\qquad
q_{n+1}=p_2+k_{n+1}\,, 
\label{eq:qchain}
\end{align}
and their virtualities yield the $1/t_r$ propagators with $t_r \equiv q_r^2 \simeq -\,\vec{q}^{\,2}_r$ in MRK (see appendix~\ref{sec:appa}). In arbitrary dimension $D$, the graviton CEV admits the double–copy form, 
\beq   
J^{\mu\nu}(q_1,k,q_2)
= \frac{1}{2} J^\mu(q_1,k,q_2)\,J^\nu(q_1,k,q_2)
- \frac{1}{2} j^\mu(q_1,k,q_2)\,j^\nu(q_1,k,q_2)\,,
\label{eq:gravcurr}
\eeq
with $k=q_1-q_2$ and where the gluon CEV $J^\mu$~\cite{Lipatov:1976zz} and $j^\mu$ are given by
\begin{subequations}
\label{eq:CEV_sub}
\begin{align}
\label{eq:CEV_suba}J^\mu(q_1,k,q_2)
&=-\big(q_1^\mu+q_2^\mu\big)_{\!\perp}
+\Big(\frac{s_1}{s}+2\,\frac{t_2}{s_2}\Big)p_2^\mu
-\Big(\frac{s_2}{s}+2\,\frac{t_1}{s_1}\Big)p_1^\mu,\\
\label{eq:CEV_subb}j^\mu(q_1,k,q_2)
&=2\,\big(q_{1\perp}^2 q_{2\perp}^2\big)^{1/2}
\!\left(\frac{p_1^\mu}{s_1}-\frac{p_2^\mu}{s_2}\right),
\end{align}
\end{subequations}
where $s\simeq 2\,p_1\!\cdot p_2$, $s_1=(p_4+k)^2\simeq -2p_1\cdot k$ and $s_2=(p_3+k)^2\simeq -2p_2\cdot k$ are positive definite and in agreement with \eqref{eq:svariables}. The auxiliary current $j^{\mu}$ was introduced in~\cite{Lipatov:1982vv,Lipatov:1982it} to remove the overlapping-channel singularities in the $s_1$ and $s_2$ channels. Both currents are gauge invariant, $J\!\cdot k=j\!\cdot k=0$. Although the amplitude in \eqref{eq:npt} corresponds to the emission of $n$ gravitons with strongly ordered rapidities, in the remainder of the paper we will also be interested in the more general case $2+\alpha\rightarrow2+\beta$ with $\alpha$ incoming and $\beta$ outgoing gravitons, as they will appear as building blocks of the higher loop $2\to 2$ case, see e.g. the $H^2$ diagram represented below in Fig. \ref{fig:H2diagram}. These amplitudes are simply obtained from \eqref{eq:npt} using crossing symmetry, i.e. by taking the complex conjugate of those polarisations corresponding to the incoming gravitons (recall that our external momenta are outgoing). Explicit examples at six points are carried out in the next section, see e.g. eqs. \eqref{eq:N6-LLMRK}. 

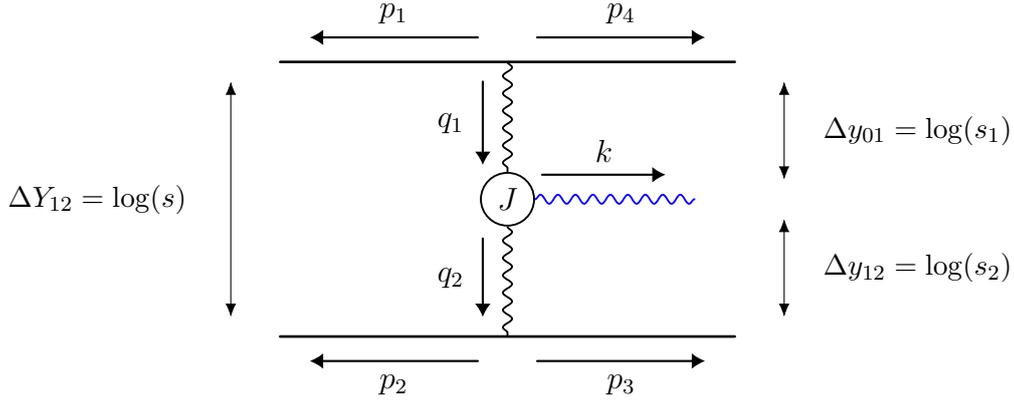
\begin{figure}[t!]
\centering
  \begin{tikzpicture}[scale=1.3,baseline=(mid.center),x=1cm,y=1cm]
    \coordinate (in1)  at (-2.3,  1.4);
    \coordinate (v1)   at ( 0,  1.4);
    \coordinate (out1) at ( 2.3,  1.4);
    \coordinate (in2)  at (-2.3, -1.4);
    \coordinate (w1)   at ( 0, -1.4);
    \coordinate (out2) at ( 2.3, -1.4);

    \node[jstyle] (mid) at (0,0)  {\large $J$};
    \coordinate (outk)  at (1.9,0);
    \coordinate (outkk) at (2.3,0);

    \draw[kernelline] (in1) -- (v1) -- (out1);
    \draw[kernelline] (in2) -- (w1) -- (out2);

    \draw[gluon] (v1) -- (mid); 
    \draw[gluon] (mid) -- (w1);
    \draw[gluon, draw=blue] (mid) -- (outk);

    \draw[kernelline,qarrow] ($(v1)+(-0.3,0.25)$) -- ($(in1)+(0.3,0.25)$) node[midway, yshift=9pt] {\large $p_1$};
    \draw[kernelline,qarrow] ($(v1)+(0.3,0.25)$) -- ($(out1)+(-0.3,0.25)$) node[midway, yshift=9pt] {\large $p_4$};
    \draw[kernelline,qarrow] ($(w1)+(-0.3,-0.25)$) -- ($(in2)+(0.3,-0.25)$) node[midway, yshift=-9pt] {\large $p_2$};
    \draw[kernelline,qarrow] ($(w1)+(0.3,-0.25)$) -- ($(out2)+(-0.3,-0.25)$) node[midway, yshift=-9pt] {\large $p_3$};

    \draw[kernelline,qarrow] ($(v1)+ (-0.25,-0.20)$) -- ($(mid)+ (-0.25, 0.35)$) node[midway, xshift= -12pt] {\large $q_1$};
    \draw[kernelline,qarrow] ($(mid)+(-0.25,-0.40)$) -- ($(w1) + (-0.25, 0.20)$) node[midway, xshift= -12pt] {\large $q_2$};
    \draw[kernelline,qarrow] ($(mid)+( 0.35, 0.25)$) -- ($(outk)+(-0.3,  0.25)$) node[midway, yshift= 9pt] {\large $k$};
    
    \draw[kernelline,darrow] ($(in1)+ (-0.5 ,-0.20)$) -- ($(in2)+ (-0.5, 0.20)$) node[midway, xshift= -50pt] {$\Delta Y_{12} = \log (s)$};
    \draw[kernelline,darrow] ($(out1)+ (0.5 ,-0.20)$) -- ($(outkk)+ (0.5, 0.20)$) node[midway, xshift= 50pt] {$\Delta y_{01} = \log (s_1)$};
    \draw[kernelline,darrow] ($(outkk)+(0.5 ,-0.20)$) -- ($(out2) + (0.5, 0.20)$) node[midway, xshift= 50pt] {$\Delta y_{12} = \log (s_2)$};
  \end{tikzpicture}
  \caption{Conventions for the tree-level five-point amplitude in MRK corresponding to the inelastic $2\rightarrow 3$ process involving two scalars and a graviton. All the rapidity gaps are large in this regime due to eq. \eqref{eq:MRK5pt2}.}
  \label{fig:1}
\end{figure}

We end this section by explicitly considering the inelastic $2\rightarrow 3$ scattering between scalars, depicted in Fig.\ref{fig:1}, with $q_1=-(p_1+p_4)$, $q_2=(p_2+p_3)$ and $k=q_1-q_2$. The multi-Regge regime of the tree-level $2\rightarrow3$ amplitude describing the emission of a real graviton corresponds to the kinematic region where 
\begin{align}
\label{eq:MRK5pt}
y_0\gg y_1\gg y_2,
\end{align}
which is equivalent to \cite{DiVecchia:2020ymx} 
\begin{align}
\label{eq:MRK5pt2}
  s,s_1,s_2\rightarrow\infty, \qquad \frac{s_1 s_2}{s}\simeq\vec{k}^{\,2},
  \end{align}
with $\vec{k}^{\,2}$ fixed and of order $|t|$. eq.\eqref{eq:npt} yields
\begin{align}
\mathcal{M}_{2\rightarrow3}^{(0)\,h}\simeq-\frac{\kappa^3}{\hbar^{3/2}}\frac{s^2}{t_1t_2}\varepsilon_{\mu\nu}^{(h)}(k)J^{\mu\nu}(q_1,k,q_2).
\end{align}
Working in light-cone ``plus'' gauge\footnote{Equivalently, one may work in a basis of light-cone minus polarisations~\cite{DelDuca:1995zy}.} $n_+\!\cdot\varepsilon^{(h)}(k)=0$~\cite{Lipatov:1991nf,DelDuca:1995zy,Barcaro:2025ifi}, we introduce the circular polarization basis $h=(\oplus,\ominus)$ defined by (see app. \ref{appDvectors} for more details)
\begin{align}
\label{eq:polLG}
\varepsilon^{\mu \,(\oplus)}(k)=\frac{1}{\sqrt{2}}\bigg(\frac{2k_{\mathrm{T}}}{k^-},0;1,i\bigg),\qquad \varepsilon^{\mu \,(\ominus)}(k)=[\varepsilon^{\mu \,(\oplus)}(k)]^*,
\end{align}
which satisfy transversality, $k_{\mu}\varepsilon^{\mu (h)}(k)=0$, orthogonality $\varepsilon_{\mu}^{(h)}(k) [ \varepsilon^{\mu (h')}(k) ]^\ast = - \delta^{h h'}$. 
In turn, that implies the orthogonality condition on the graviton polarisation tensor,
\begin{align}
\label{eq:Lbasis2}
 \varepsilon^{(h)}_{\mu\nu}(k) \big[\varepsilon^{(h')\mu\nu}(k)\big]^\ast = \delta^{h h'}.
\end{align}

Therefore one can expand the free-index amplitude $\mathcal{M}_{2\rightarrow3}^{\mu\nu}$ in the basis spanned by the polarisation tensors $\varepsilon^{(h)}_{\mu\nu}$ as
\begin{align}
\mathcal{M}^{\mu\nu}_{2\rightarrow3}=\mathcal{M}^{\ominus}_{2\rightarrow 3}\varepsilon^{\mu\nu}_{(\oplus)}(k)+\mathcal{M}^{\oplus}_{2\rightarrow 3}\varepsilon^{\mu\nu}_{(\ominus)}(k),
\end{align}
where the positive and negative helicity components can be retrieved by (see eq. \eqref{eq:relationsplusminus2}) 
\begin{align}
\mathcal{M}^{h}_{2\rightarrow 3}=\varepsilon_{\mu\nu}^{(h)}(k)\mathcal{M}_{2\rightarrow 3}^{\mu\nu}.
\end{align}
By explicit computation, we retrieve the very simple tree-level expressions~\cite{Ciafaloni:2015xsr},
\begin{align}
\label{eq:5pt+}\mathcal{M}^{(0)\,\oplus}_{2\rightarrow3}\simeq \frac{s^2\kappa^3}{\hbar^{3/2}}\frac{k_{\mathrm{T}}^* \, q_{1\mathrm{T}}-q^*_{1\mathrm{T}} \, k_{\mathrm{T}}}{(k_{\mathrm{T}}^*)^2 \, q^*_{1\mathrm{T}} \, q_{2\mathrm{T}}},\qquad\mathcal{M}_{2\rightarrow3}^{(0)\,\ominus}=[\mathcal{M}_{2\rightarrow3}^{(0)\,\oplus}]^*.
\end{align}
%

\section{Gravity amplitudes in the high-energy limit: Regge theory}
\label{sec:regge}

In this section, we show how to construct higher-loop amplitudes in the high-energy limit
employing, as building blocks, the tree-level $2\to2+n$ amplitudes introduced above in \eqref{eq:npt}. Although we could work with generic impact factors, below we focus on the scalar case. As clear from eqs. \eqref{eq:defM} and \eqref{eq:defMloops}, the amplitude describing a gravitational scattering process admits a perturbative PM expansion in powers of the gravitational constant $G$. To connect directly with the expected semi-classical behaviour at high-energies, we find it convenient to switch to the exponential representation of the S-matrix~\cite{Damgaard:2021ipf},
\begin{align}\label{eq:N-operator}
\hat{S} = \exp\left(i \frac{\hat{N}}{\hbar}\right)\,,
\end{align}
and compute the tree-level matrix elements of $\hat{N}$ in MRK.  
The advantage of this reformulation is that only two-massive-particle-irreducible (2MPI) $\hat{T}$-matrix elements contribute to $\hat{N}$, isolating the parts that exponentiate in impact-parameter space in the classical regime (see \cite{Cristofoli:2021jas,Damgaard:2021ipf,DiVecchia:2022nna,DiVecchia:2023frv,Brandhuber:2023hhy,Adamo:2024oxy} for details). Following \cite{Alessio:2025flu}, we introduce the momentum-space generating functionals,
\begin{align}
\label{eq:kernels}
\langle p_3 p_4| \hat{N} | p_1 p_2 \rangle &= \mathcal{K}_{2\rightarrow 2} (s,q_1) \, \hat{\delta}^{(D)}(q_1 - q_2) \,,  \nonumber  \\
\langle p_3 p_4 k| \hat{N} | p_1 p_2 \rangle &=  \mathcal{K}_{2\to 3}\left(s,q_1, q_2, k\right) \hat{\delta}^{(D)}(q_1 - q_2 - k) \,, \nonumber \\
\langle p_3 p_4 k_1 k_2| \hat{N} | p_1 p_2 \rangle &=  \mathcal{K}_{2\to 4}\left(s,q_1, q_2, k_1, k_2\right)  \hat{\delta}^{(D)}(q_1 - q_2 - k_1 - k_2) \,, \nonumber \\
\langle p_3 p_4 k_2| \hat{N} | p_1 p_2 k_1\rangle &=  \mathcal{K}_{3\to 3}\left(s, q_1, q_2, k_1, k_2\right) \hat{\delta}^{(D)}(q_1 -q_2 - k_1 -k_2)\,,
\end{align}
where the momentum transfers to each scalar line are $q_1 = -p_1 - p_4$ and $q_2 = p_2 + p_3$ and the overall momentum conservation gives $q=q_1=q_2$ in the elastic case. A similar construction applies to higher-points matrix elements.

At tree-level, the four-point generating functional $\mathcal{K}^{(0)}$ is identical to the single-graviton exchange amplitude~\eqref{eq:4regge}, i.e.
\begin{align}
\label{eq:1PMelastic}
    \mathcal{K}^{(0)}_{2\to 2}(s,q)\simeq -\kappa^2\frac{s^2}{t}\Theta(y_{p_1}-y_{p_2}).
\end{align}
The $\Theta$ function in rapidity appearing in \eqref{eq:1PMelastic} explicitly enforces the ordering of emissions\footnote{Note that we are working at leading logarithmic accuracy, where shifts of the external rapidities by $\mathcal{O}(t/s)$ from momentum transfer do not affect the result. Thus one may identify the incoming and outgoing rapidities $(y_{p_1},y_{p_2}) \simeq (y_{p_4},y_{p_3})$ for the purpose of implementing the MRK ordering at this order.}, and makes it clear that the support of the MRK kernels is the kinematical region specified by such ordering. The next non-trivial object is the five-point $\hat{N}$-operator matrix element, which in MRK becomes\footnote{We introduce the notation $\Theta(y_{0}>y_{1}>\dots>y_n>y_{n+1})\equiv\prod_{i=0}^{n}\Theta(y_i-y_{i+1})$ and the arguments $k^{h_i}_i$ of the various $N$-matrix elements denote that the external gravitons with momenta $k_i$ have polarisation $h_i$.}
\begin{align}
\label{eq:N5-LLMRK}
\mathcal{K}^{(0)}_{2\to 3}(s, q_1, q_2, k^{h})
\simeq - s^2 \kappa^2 \Theta(y_{p_1}\!>\!y_{k}\!>\!y_{p_2}) \;\mathcal{J}^{(h)}(q_1,k)\frac{1}{t_2}\,,
\end{align}
where we defined
\begin{align}
\mathcal{J}^{(h)}(q,k)\equiv\frac{1}{t}\;
\Big[\kappa \,\varepsilon^{(h)}_{\mu\nu}(k)\,J^{\mu\nu}(q,k,q-k)\Big]\,,\qquad t=q^2\,,
\end{align}
 while the six-point tree-level matrix elements are
\begin{subequations}
\label{eq:N6-LLMRK}
\begin{align}
& \mathcal{K}^{(0)}_{2\to 4}(s, q_1, q_2, k_1^{h_1}, k_2^{h_2}) & \\
&\quad \simeq - s^2 \kappa^2\Big[
\Theta(y_{p_1}\!>\!y_{k_1}\!>\!y_{k_2}\!>\!y_{p_2})\;
\,\mathcal{J}^{(h_1)}(q_1,k_1)\mathcal{J}^{(h_2)}(q_1-k_1,k_2) +(k_{1}^{h_1}\longleftrightarrow k_{2}^{h_2})\Big]\frac{1}{t_2}\,, \nonumber \\
& \mathcal{K}^{(0)}_{3\rightarrow3}(s, q_1, q_2, k_1^{h_1}, k_2^{h_2})  \\
&\quad \simeq  -\, s^2 \kappa^2 \Big[ \Theta(y_{p_1}\!>\!y_{k_1}\!>\!y_{k_2}\!>\!y_{p_2})\,  [\mathcal{J}^{(h_1)} (q_1,k_1)]^{*} \mathcal{J}^{(h_2)} (q_1-k_1,k_2) + (k_1^{h_1} \longleftrightarrow k_2^{h_2} )^* \Big]\frac{1}{t_2}\;. \nonumber 
\end{align}
\end{subequations}
This structure hints at major simplifications once the evolution is formulated in terms of the boost operator, since the phase space factorizes into ordered intervals. We will exploit this simplification in sec.~\ref{sec:shock-wave} using the shock–wave formalism.

\subsection{The generating functional of amplitudes in multi-Regge kinematics}

Having introduced the $n$-point tree-level generating functionals in MRK, we are now ready to understand how they enter the $2\to2$ amplitude at loop level in the Regge limit.
We first introduce the impact-parameter (Fourier-conjugate) representation,
\begin{subequations}
\label{eq:Fourier_transform}
\begin{align}
\label{eq:Fourier_transformel}
&\widetilde{\mathcal{K}}_{2\to 2}(b) \;=\; \int  \hat{\mathrm{d}}^D q \,\hat{\delta}\!\left(2 \bar{p}_1 \cdot q\right) \hat{\delta}\!\left(2 \bar{p}_2 \cdot q\right) \exp\!\Big[\frac{i}{\hbar}\,(q \cdot b)\Big] \,\mathcal{K}_{2\to 2}(q)\,, \\
\label{eq:Fourier_transforminel}
&\widetilde{\mathcal{K}}_{2+\alpha\to 2+\beta}\!\left(b_1, b_2; \{k_j\}\right)
 = \int  \hat{\mathrm{d}}^D q_1 \hat{\mathrm{d}}^D q_2 \,  \hat{\delta}\!\left(2 \bar{p}_1 \cdot q_1\right) \hat{\delta}\!\left(2 \bar{p}_2 \cdot q_2\right)  \hat{\delta}^{(D)}\!\Big(q_1{-}q_2{-}\sum_{j=1}^n k_j\Big) \nonumber \\
& \qquad \qquad \qquad \qquad \qquad \qquad  \times \exp\!\Big[\frac{i}{\hbar}\,(- q_1 \cdot b_1+q_2 \cdot b_2)\Big] \,\mathcal{K}_{2+\alpha\to 2+\beta}(q_1, q_2; \{k_j\})\,, 
\end{align}
\end{subequations}
with $\alpha+\beta=n$, where $b=b_1-b_2$ is the impact parameter and  $\bar p_i^\mu \simeq p_i$ denotes the classical momentum associated with the trajectory of particle $i$\footnote{At leading logarithmic accuracy in the multi-Regge expansion, $\bar p_i^\mu$ may be identified with the corresponding external momentum $p_i^\mu$, since their difference is a longitudinal shift of order $\mathcal{O}(t/s)$ that does not affect the transverse projection enforced by the delta functions. }. We then assume that the semiclassical S-matrix admits a coherent state expansion,
\begin{align}
\label{eq:HEFT_exp}
\hat{S}^{\mathrm{cl}}&=\exp\Bigg\{\frac{i}{\hbar} \Bigg[\widetilde{\mathcal{K}}_{2\to 2}(b) + \sum_{h} \frac{1}{\sqrt{\hbar}} \int_{k} \hat{\alpha}_{5}(k^{h})
\nonumber \\
&\qquad \qquad \qquad + \sum_{n=2}^{\infty} \frac{1}{n! \hbar^{n/2}} \sum_{h_1, \ldots, h_n}  \int_{k_1,\dots , k_n} \hat{\alpha}_{4+n}(k_1^{h_1}, \ldots, k_n^{h_n})\Bigg] \Bigg\}\,,
\end{align}
where we have introduced operators $\hat{\alpha}_{4+n}(k_1, \ldots, k_n)$, whose explicit form will be given below, which are in one-to-one correspondence with the
$n$-graviton matrix elements of $\hat N$. These operators are expressed in terms of the modes of the on-shell graviton field,
\begin{gather}
\label{eq:field_mode}
\hat{h}_{\mu \nu}(x) = \frac{1}{\sqrt{\hbar}}\sum_{h=\pm}\int_{k} \Big[\varepsilon^{(h)*}_{\mu \nu}(k) \hat{a}_{(h)}^{\dagger}(k) e^{-i \frac{k \cdot x}{\hbar}} + \mathrm{h.c.} \Big] \\
\int_k \equiv \int  \hat{\mathrm{d}}^D k \,\hat{\delta}_+\!\left(k^2\right) \, \hspace{20pt} \left[\hat{a}_{(h)}(k),\hat{a}^\dag_{(h')}(k')\right] = (2k^0) \delta_{h,h'} \hat{\delta}^{(D-1)}\!\big(k-k^\prime\big). \nonumber
\end{gather}
in which $\hat{\delta}^{(D-1)}$ as usual involves the spatial components in Minkowski coordinates. For our purposes, we will only consider explicitly loop-level amplitudes up to 5PM order, and therefore only the terms with $n=1$ and $n=2$ will be relevant. The $n=1$ operator corresponds to a single coherent state while the $n=2$ operator describes a two-mode coherent squeezed state~\cite{Alessio:2025flu,Fernandes:2024xqr},
\begin{align}
\hat{\alpha}_{5}(k^{h})
&= \widetilde{\mathcal{K}}_{2\to 3}(b_1, b_2;k^{h})\, \hat{a}_{(h)}^{\dagger}(k)
+  \,\widetilde{\mathcal{K}}^{*}_{2\to 3}(b_1, b_2; k^{h}) \hat{a}_{(h)}(k)\,,  \label{eq:alphaN}\\
\hat{\alpha}_{6}(k_1^{h_1}, k_2^{h_2}) &= \widetilde{\mathcal{K}}_{2\rightarrow 4}(b_1, b_2; k_1^{h_1}, k_2^{h_2}) \, \hat{a}_{(h_1)}^{\dagger}(k_1) \hat{a}_{(h_2)}^{\dagger}(k_2) + \mathrm{h.c.}
\nonumber \\
&\hspace{2cm} + \widetilde{\mathcal{K}}_{3\rightarrow3}(b_1, b_2; k_1^{h_1}, k_2^{h_2}) \, \hat{a}^{\dagger}_{(h_2)}(k_2) \hat{a}_{(h_1)}(k_1) + \mathrm{h.c.}\,,
\nonumber\end{align}
which are determined in general by the $\hat{N}$-matrix elements in \eqref{eq:kernels}. When working in MRK and in the leading logarithmic approximation, however, we only need to take the leading high-energy limit of the tree-level generating functionals in the $\hat{S}$-matrix expansion in eq. \eqref{eq:HEFT_exp} and \eqref{eq:alphaN}, as they are given in eqs. \eqref{eq:1PMelastic},\eqref{eq:N5-LLMRK} and \eqref{eq:N6-LLMRK}.  
Before proceeding further to define the loop-level $2\to2$ amplitude in MRK, there is an important caveat to be made. In general, the integration measures for the elastic and inelastic Fourier transforms defined in \eqref{eq:Fourier_transform} are different. In fact, the $\hat{\delta}$-functions in \eqref{eq:Fourier_transformel} force the integrated transfer momentum $q$ to lie on the $d$-dimensional plane perpendicular to $p_1$ and $p_2$, while those appearing in \eqref{eq:Fourier_transforminel} would give the integrated momentum $q_1$ (solving for instance the overall momentum conservation as $q_2=q_1-\sum_{j=1}^nk_j$) a non-trivial longitudinal component along $p_2$. In turn, the latter depends on the light-cone components of the emitted gravitons momenta. 
However, the generating kernels $\mathcal{K}_{2+\alpha\to2+\beta}$ in MRK depend on the exchanged momenta $q_i$ only through their transverse components. This follows directly from the structure of the Lipatov current in eq.~\eqref{eq:CEV_sub} and from the relations \eqref{eq:trMRK}, which suppress longitudinal contributions by powers of $t/s$. As a result, the longitudinal components induced by the inelastic $\hat\delta$-functions decouple, and the elastic and inelastic Fourier transforms effectively reduce to the same $(D-2)$-dimensional transverse integration measure.

\begin{figure}[t]
\centering

\begin{tikzpicture}[scale=1.3,baseline=(b.center),x=1cm,y=1cm]
  \coordinate (in1)  at (-2,  1.2);
  \coordinate (v1)   at (-1,  1.2);
  \coordinate (v2)   at ( 1,  1.2);
  \coordinate (out1) at ( 2,  1.2);
  \coordinate (in2)  at (-2, -1.2);
  \coordinate (w1)   at (-1, -1.2);
  \coordinate (w2)   at ( 1, -1.2);
  \coordinate (out2) at ( 2, -1.2);

  \node[jstyle] (b) at (-1,0) {\large $J$};
  \node[jstyle] (d) at ( 1,0) {\large $J$};

  \draw[kernelline] (in1) -- (v1) -- (v2) -- (out1);
  \draw[kernelline] (in2) -- (w1) -- (w2) -- (out2);

  \draw[gluon] (v1) -- (b);  \draw[gluon] (w1) -- (b);
  \draw[gluon] (v2) -- (d);  \draw[gluon] (w2) -- (d);

  \draw[gluon, draw=blue] (b) -- (d);

  \draw[kernelline,qarrow] ($(v1)+(-0.2,0.25)$) -- ($(v1)+(-0.9,0.25)$) node[midway, yshift=9pt] {\large $p_1$};
  \draw[kernelline,qarrow] ($(v1)+(0.30,0.25)$) -- ($(v2)+(-0.30,0.25)$) node[midway, yshift=9pt] {\large $k_0$};
  \draw[kernelline,qarrow] ($(v2)+(0.2,0.25)$) -- ($(v2)+(0.9,0.25)$) node[midway, yshift=9pt] {\large $p_4$};
  \draw[kernelline,qarrow] ($(w1)+(-0.2,-0.25)$) -- ($(w1)+(-0.9,-0.25)$) node[midway, yshift=-9pt] {\large $p_2$};
  \draw[kernelline,qarrow] ($(w1)+(0.30,-0.25)$) -- ($(w2)+(-0.30,-0.25)$) node[midway, yshift=-9pt] {\large $k_2$};
  \draw[kernelline,qarrow] ($(w2)+(0.2,-0.25)$) -- ($(w2)+(0.9,-0.25)$) node[midway, yshift=-9pt] {\large $p_3$};

  \draw[kernelline,qarrow] ($(v1)+(-0.22,-0.12)$) -- ($(b)+(-0.22,0.28)$) node[midway, xshift=-11pt] {\large $q_1$};
  \draw[kernelline,qarrow] ($(b)+(-0.22,-0.28)$) -- ($(w1)+(-0.22, 0.12)$) node[midway, xshift=-11pt] {\large $q_2$};
  \draw[kernelline,qarrow] ($(v2)+( 0.22,-0.12)$) -- ($(d)+( 0.22,0.28)$) node[midway, xshift= 11pt] {\large $q_3$};
  \draw[kernelline,qarrow] ($(d)+( 0.22,-0.28)$) -- ($(w2)+( 0.22, 0.12)$) node[midway, xshift= 11pt] {\large $q_4$};

  \draw[kernelline,qarrow] ($(b)+(0.32,0.25)$) -- ($(d)+(-0.32,0.25)$) node[midway, yshift=9pt] {\large $k$};

  \node[purple] at ($ (v1)!0.5!(v2) $) {\textbar};
  \node[purple] at ($ (w1)!0.5!(w2) $) {\textbar};
  \node[purple] at ($ (b)!0.5!(d) $) {\textbar};
\end{tikzpicture}
$\quad\longrightarrow\quad$
\hspace{-5pt}
\begin{tikzpicture}[scale=0.9,x=0.95cm,y=0.95cm,baseline=(C)]

  \coordinate (Tstub) at (0, 2.7);
  \coordinate (Top)   at (0, 1.70);
  \coordinate (Bot)   at (0,-1.70);
  \coordinate (Bstub) at (0,-2.7);

  \node[jstyle] (JL) at (-1.60, 0) {\large $J$};
  \node[jstyle] (JR) at ( 1.60, 0) {\large $J$};

  \draw[kernelline] (Tstub) -- (Top);
  \draw[kernelline] (Bot)   -- (Bstub);

  \draw[kernelline,qarrow] (-0.3, 2.7) -- (-0.3, 1.9) node[midway, left=4pt] {\large $q$};
  \draw[kernelline,qarrow] (-0.3,-1.9) -- (-0.3,-2.7) node[midway, left=4pt] {\large $q$};

  \draw[gluon] (Top) -- (JL);
  \draw[gluon] (Top) -- (JR);
  \draw[gluon] (Bot) -- (JL);
  \draw[gluon] (Bot) -- (JR);
    
  \def\dX{0.25}\def\dY{0.25}
   
  \draw[kernelline,qarrow]
  ($ (Top)!0.25!(JL) + (-\dX,\dY) $) -- ($ (Top)!0.75!(JL) + (-\dX,\dY) $)
  node[pos=0, left=6pt] {\large $q_1$};

  \draw[kernelline,qarrow]
  ($ (Top)!0.25!(JR) + (\dX,\dY) $) -- ($ (Top)!0.75!(JR) + (\dX,\dY) $)
  node[pos=0, right=6pt] {\large $q_3$};

  \draw[kernelline,qarrow]
  ($ (Bot)!0.75!(JL) + (-\dX,-\dY) $) -- ($ (Bot)!0.25!(JL) + (-\dX,-\dY) $)
  node[pos=0, left=-9pt, yshift=-15pt] {\large $q_2$};

  \draw[kernelline,qarrow]
  ($ (Bot)!0.75!(JR) + (\dX,-\dY) $) -- ($ (Bot)!0.25!(JR) + (\dX,-\dY) $)
  node[pos=0, right=-9pt, yshift=-15pt] {\large $q_4$};

  \draw[gluon, draw=blue] (JL) -- (JR);
  \draw[kernelline,qarrow] ($(JL)+(0.68,0.34)$) -- ($(JR)+(-0.68,0.34)$)
    node[midway, yshift=9pt] {\large $k$};

  \node[purple] at (0,0) {\textbar};
  \coordinate (C) at (0,0);
\end{tikzpicture}
\caption{
The $H$ diagram. Vertical wavy lines denote Reggeized gravitons, and the horizontal blue line represents the on-shell soft graviton exchanged between the Lipatov currents $J$. After integrating over rapidities, the external lines contract to an effective vertex carrying the transverse momentum $q_{\perp}$, yielding the representation $\mathcal{H}_{\mathrm{GR}}(q_{\perp}^{2})$ in eq.~(3.18).
}
\label{fig:Hdiagram}
\end{figure}
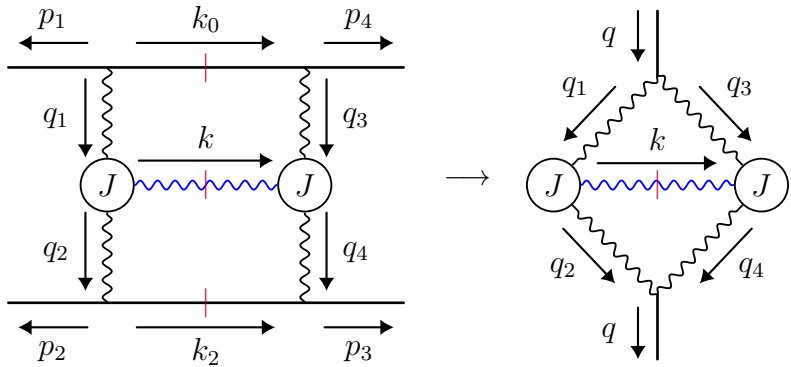

Using the representation \eqref{eq:HEFT_exp} of the S-matrix, we can now construct the contribution to the $2 \to 2$ loop-level amplitude at leading logarithmic accuracy in MRK,
\begin{align}
i\mathcal{M}_{2\rightarrow2}(s,q^2) \simeq 2 s \int\, \frac{\mathrm{d}^{d} b}{\hbar^d}\;e^{i\,\vec{q}\cdot \vec{b} /\hbar} \langle 0 | \hat{S}^{\mathrm{cl}} - 1 | 0 \rangle\,,
\label{eq:Smatrix_MRK}
\end{align}
and, more generally, for the $2\to 2+n$ amplitude,
\begin{align}
i\mathcal{M}_{2\rightarrow2+n}(s,q,q-\sum_{j=1}^n k_j;k^{h_1}_1,\dots,k^{h_n}_n) 
&\simeq 2 s \int\, \frac{\mathrm{d}^{d} b}{\hbar^d} \;e^{i\,\vec{q} \cdot \vec{b}/\hbar} \langle k_1^{h_1} \dots k_n^{h_n} | \hat{S}^{\mathrm{cl}} - 1 | 0 \rangle\,.
\label{eq:Smatrix_MRK2}
\end{align}
In \eqref{eq:Smatrix_MRK} and \eqref{eq:Smatrix_MRK2} it is understood that, without loss of generality, we can set $\vec{b}_2 = 0$ and $\vec{b}_1 = \vec{b}$ due to translation invariance. 

Focusing on the $2 \to 2$ scattering, the first term in the expansion of \eqref{eq:Smatrix_MRK} reproduces the tree-level amplitude, while the higher-order terms dependent solely on $\mathcal{K}^{(0)}$ encode its semiclassical iterations. The next non-trivial contribution arises at two-loop order and is obtained by reordering the oscillator operators appearing in the single-emission terms, giving a purely imaginary term which corresponds to the cut $H$ diagram in MRK,\footnote{The classical $2\to2$ amplitudes scale as $\hbar^{-1}$ in impact parameter space. Indeed, there is an hidden $\hbar^2$ dependence in eq.~\eqref{eq:Smatrix_MRK} that makes the classical scaling manifest.
}
\begin{align}
\widetilde{\mathcal M}^{(2)}_{H}(b)
&\simeq \frac{i}{2\hbar^3}\sum_{h}\int_{k}\;
\widetilde{\mathcal K}^{(0)}_{5}(b;k^h)\;
\widetilde{\mathcal K}^{(0)\,*}_{5}(b;k^h),
\end{align}
which depends on the contraction of two graviton emission currents. We can then use \eqref{eq:Smatrix_MRK} and \eqref{eq:Fourier_transform} to perform the inverse Fourier transform to momentum space, yielding
\begin{align}\label{eq:3pc}
&\mathcal M^{(2)}_{H}(s,q^2)
\\
&\quad \simeq\frac{i}{8 s \hbar^3}  \sum_{h}
\int \hat{\mathrm{d}} k_+ \int \hat{\mathrm{d}} k_- \, \hat{\delta}^{+}(k^2)  \int_{q_{1\!\perp},q_{2\!\perp}} \;
\mathcal{K}^{(0)}_{2\rightarrow3} \big(s;\,q_1,q_2;\,k^h\big)
\mathcal{K}^{(0)\,*}_{2\rightarrow3} \big(s;\,q_3,q_4;\,-k^h\big)\,, \nn 
\end{align}
with the conventions of Fig.~\ref{fig:Hdiagram},
\begin{equation}
    k=q_1-q_2=-q_3+q_4,\qquad q_3=q-q_1,\qquad q_4=q-q_2,
\end{equation}
and where the cut on–shell conditions are $p_1^2=p_4^2=m_1^2$, $p_2^2=p_3^2=m_2^2$ and $k^2=0$. The natural building block appearing in the integrand of the $H$ diagram \eqref{eq:3pc}, using the explicit expansion in multi-Regge kinematics \eqref{eq:N5-LLMRK}, is then
\begin{align}
\label{eq:GR-kernel-contraction}
\hspace{-10pt}\sum_{h}\mathcal{K}^{(0)}_{2\rightarrow3}\big(s;\,q_1,q_2;\,k^h\big)\;
\mathcal{K}^{(0)\,*}_{2\rightarrow3}\big(s;\,q_3,q_4;\,-k^h\big) \simeq \frac{s^4\kappa^6}{q^2_{1\perp}q^2_{2\perp}q^2_{3\perp}q^2_{4\perp}}\mathcal{H}_{\mathrm{GR}}(q_1,q_2;q_3,q_4).
\end{align}
where $\mathcal{H}_{\mathrm{GR}}(q_1,q_2;q_3,q_4)$ is the so-called gravitational kernel,\footnote{We anticipate that $\mathcal{H}_{\mathrm{GR}}$ will serve as a central structural element also in the shock-wave formulation in sec.\ref{sec:shock-wave}, reappearing at higher orders in perturbation theory.}
\begin{align}
\label{eq:GR-kernel-def}
\mathcal{H}_{\mathrm{GR}}(q_1,q_2;q_3,q_4)
&\;\equiv\;
J^{\mu\nu}(q_1,k,q_2)\,
\mathcal{P}_{\mu\nu\rho\sigma}\,
J^{\rho\sigma}(q_3,-k,q_4)\,, \\
\mathcal{P}_{\mu\nu\rho\sigma}
&=\frac{1}{2}(\eta_{\mu\rho}\eta_{\nu\sigma}+\eta_{\mu\sigma}\eta_{\nu\rho})
-\frac{1}{D-2}\eta_{\mu\nu}\eta_{\rho\sigma}\,. \nonumber
\end{align}
built from the graviton CEV $J^{\mu\nu}$ and the de Donder graviton projector $\mathcal{P}_{\mu\nu\rho\sigma}$.\footnote{Alternatively, instead of the de Donder projector, one can also sum over light-cone polarisations \eqref{eq:polLG}. } We note that the kernel $\mathcal{H}_{\mathrm{GR}}$ is symmetric under the exchange $(q_1,q_2)\leftrightarrow(q_3,q_4)$, as well as separately under $q_1\leftrightarrow q_2$ and $q_3\leftrightarrow q_4$. Imposing momentum conservation as $q_3 = q - q_1$ and $q_4 = q - q_2$, we recover the Regge form of the gravitational kernel first introduced in eq.~(47) of~\cite{Lipatov:1982it},
\begin{align}
\mathcal{H}_{\mathrm{GR}}(q_1,q_2;q)
\;\equiv\;
\mathcal{H}_{\mathrm{GR}}(q_1,q_2;q{-}q_1,q{-}q_2)\,,
\end{align}
with
\begin{align}
\label{eq:gravk}
\mathcal{H}_{\mathrm{GR}}(q_1,q_2;q)
&= \Big[\mathcal{H}_{\mathrm{YM}}(q_1,q_2;q)\Big]^2
+\frac{4}{(q_{1\perp}-q_{2\perp})^4}
\bigg[
q_{1\perp}^2 q_{2\perp}^2 (q_{\perp}{-}q_{1\perp})^2 (q_{\perp}{-}q_{2\perp})^2
\\[-4pt]
&\qquad\quad
- (q_{\perp}{-}q_{1\perp})^2 (q_{\perp}{-}q_{2\perp})^2 (q_{1\perp}\!\cdot\! q_{2\perp})^2
- q_{1\perp}^2 q_{2\perp}^2 \big[(q_{\perp}{-}q_{1\perp})\!\cdot\!(q_{\perp}{-}q_{2\perp})\big]^2
\bigg], \nonumber
\end{align}
written in terms of the Yang-Mills (or QCD) kernel at $t \neq 0$~\cite{Lipatov:1976zz,Kuraev:1976ge,Kuraev:1977fs,Balitsky:1978ic},
\begin{align}
\mathcal{H}_{\mathrm{YM}}(q_1,q_2;q)
= - q_{\perp}^2
+ \frac{
q_{2\perp}^2 (q_{\perp}{-}q_{1\perp})^2
+ q_{1\perp}^2 (q_{\perp}{-}q_{2\perp})^2
}{
(q_{1\perp}-q_{2\perp})^2
}\,.
\end{align}
Equation~\eqref{eq:gravk} can equivalently be obtained by summing over graviton helicity states in the current–emission vertex of eq.~\eqref{eq:gravcurr}~\cite{Barcaro:2025ifi}.

We can now evaluate~\eqref{eq:3pc}, recovering the result for the $H$ diagram \cite{Amati:1990xe,DiVecchia:2020ymx,Alessio:2025isu}, 
\begin{align}
\label{eq:3pca}
&\,\mathcal M^{(2)}_{H}(s,q^2) \simeq i\frac{(8 \pi G)^3 s^3}{8 \hbar^3} \mu^{4\epsilon} \int^{y_2}_{y_0} \hat{\mathrm{d}} y \int_{q_{1\!\perp},q_{2\!\perp}} \; \frac{\mathcal{H}_{\mathrm{GR}}(q_1,q_2;q_3,q_4)}{q^2_{1\perp}q^2_{2\perp}q^2_{3\perp}q^2_{4\perp}}\,, 
\end{align}
where $y_0 < y< y_2$ represents the total rapidity interval available to the exchanged graviton and $y_0$ and $y_2$ are the rapidities of the incoming massive particles, see \eqref{eq:rapidity}. As explained in appendix \ref{sec:appa}, in the high-energy limit this interval yields $y_2 - y_0 \simeq \log(s/|t|)$, giving\footnote{Note that the renormalization scheme in this work is slightly different than what has been done in \cite{Alessio:2025isu}, where $\mu^2 \exp(\gamma \epsilon)$ in place of $\mu^3 \exp(\gamma \epsilon)$ appeared together for each transverse integral.}
\begin{align}\label{eq:3pcc}
\mathcal M^{(2)}_{H}&(s,q^2) \simeq  i\frac{\left(8 \pi G\right)^3}{16\pi\hbar^3}s^3\log\Big(\frac{s}{|t|}\Big)  H^1(q^2)\,, \\
H^1(q^2) & = \mu^{4\epsilon} \int_{q_{1\!\perp},q_{2\!\perp}} \frac{\mathcal{H}_{\mathrm{GR}}(q_{1},q_{2};q_{3},q_{4})}{q^2_{1\perp}q^2_{2\perp}q^2_{3\perp}q^2_{4\perp}} = \label{eq:h1} \\
& = \frac{1}{(4\pi)^2} \left(\frac{4\pi\mu^2 e^{\gamma_E}}{\vec{q}^{\,2}}\right)^{2\epsilon} \left( \frac{\Gamma (-\epsilon )^4 \Gamma (\epsilon +1)^2}{\Gamma (-2 \epsilon )^2} + \frac{2 (2 \epsilon -3) \Gamma (-\epsilon )^3 \Gamma (2 \epsilon +1)}{3 \Gamma (-3 \epsilon )}\right),\nonumber
\end{align}
in dimensional regularization with $D=4-2\epsilon$, see appendix \ref{app:integrals} for details on the integrals. Expanding around $\epsilon=0$, we obtain
\begin{align}\label{eq:H1expanded}
\,\mathcal M^{(2)}_{H}(s,q^2)&\simeq 4i (G/\hbar)^3 s^3 \log\left(\frac{s}{|t|}\right) \left(\frac{4\pi\mu^2}{\vec{q}^{\,2}}\right)^{2\epsilon} \left(-\frac{1}{\epsilon ^2}+\frac{2}{\epsilon }+\zeta _2 +\mathcal{O}(\epsilon)\right)\,.
\end{align}
It is also worth noticing that the leading infrared divergent $1/\epsilon^2$ part of $\mathrm{Im}\,\mathcal M^{(2)}_{H}$ is entirely captured by the exchange of Weinberg soft gravitons \cite{DiVecchia:2021ndb,Alessio:2022kwv}.  Indeed, in the strict soft limit $\omega\to0$, the Lipatov current correctly reduces to the Weinberg soft-graviton current multiplying the tree-level elastic amplitude \cite{Amati:1990xe} (see also sec.~3.5 of \cite{Raj:2025hse}),
\begin{align}
\label{eq:softlimit}
-\kappa^3 \frac{s^2}{t_1 t_2}\,
J^{\mu\nu}(q_1,k,q_2)
\stackrel{\omega \to 0}{\longrightarrow}
J^{\mu\nu}_{1/\omega}(p_i,k)\,
\mathcal{M}^{(0)}_{2\to 2}(s,q^2)\, .
\end{align}
Here $\mathcal{M}^{(0)}_{2\to 2}$ is the tree-level elastic amplitude given in
\eqref{eq:4regge}, and
\begin{align}
J^{\mu\nu}_{1/\omega}(p_i,k)
=
\kappa\bigg[
q\!\cdot\! k
\bigg(
\frac{p_1^{\mu}p_1^{\nu}}{(p_1\!\cdot\! k)^2}
-
\frac{p_2^{\mu}p_2^{\nu}}{(p_2\!\cdot\! k)^2}
\bigg)
-
\frac{p_1^{\mu}q^{\nu}+p_1^{\nu}q^{\mu}}{p_1\!\cdot\! k}
+
\frac{p_2^{\mu}q^{\nu}+p_2^{\nu}q^{\mu}}{p_2\!\cdot\! k}
\bigg],
\end{align}
with $q=-p_1-p_4=p_2+p_3$ the momentum transfer. This expression coincides with the Weinberg soft-graviton factor expanded to leading order in $q$, i.e.
\begin{align}
\mathcal{S}^{\mu\nu}_{1/\omega}(k)
=
\kappa\sum_{i=1}^4
\frac{p_i^{\mu}p_i^{\nu}}{p_i\!\cdot\! k}
\quad\Longrightarrow\quad
J^{\mu\nu}_{1/\omega}(p_i,k)
=
\mathcal{S}^{\mu\nu}_{1/\omega}(k)
+\mathcal{O}(q^2)\, .
\end{align}
Therefore, in the leading soft approximation, the $\hat N$-matrix element entering \eqref{eq:3pc} factorises as $\mathcal{K}^{(0)}_{2\to3}(s;q_1,q_2,k^h) \stackrel{\omega\to0}{\longrightarrow} \varepsilon_{\mu\nu}^{(h)}(k)\, \mathcal{S}^{\mu\nu}_{1/\omega}(k)\, \mathcal{K}^{(0)}(s;q)$.

The $H$-diagram just discussed can be regarded as the starting point of two different iterations, in the $t$- and $s$-channels respectively, on which we focus next.

\subsection{Iteration along the $t$-channel: gravitational BFKL equation}

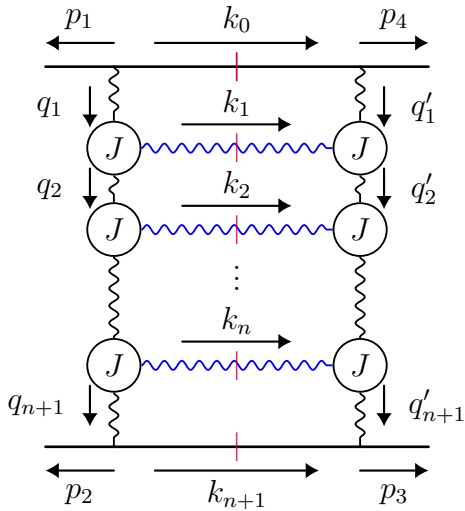
\begin{figure}[t]
\centering
\begin{tikzpicture}[scale=0.9,x=1cm,y=1cm,baseline=(current bounding box.center)]

  \coordinate (Ltop) at (-1.8,  2.8);
  \coordinate (Lbot) at (-1.8, -2.8);
  \coordinate (Rtop) at ( 1.8,  2.8);
  \coordinate (Rbot) at ( 1.8, -2.8);
  \coordinate (UL) at (-2.8,  2.8);
  \coordinate (LL) at (-2.8, -2.8);
  \coordinate (UR) at ( 2.8,  2.8);
  \coordinate (LR) at ( 2.8, -2.8);

  \node[jstyle] (J1L) at (-1.8,  1.6) {\large $J$};
  \node[jstyle] (J2L) at (-1.8,  0.4) {\large $J$};
  \node[jstyle] (JnL) at (-1.8, -1.6) {\large $J$};

  \node[jstyle] (J1R) at ( 1.8,  1.6) {\large $J$};
  \node[jstyle] (J2R) at ( 1.8,  0.4) {\large $J$};
  \node[jstyle] (JnR) at ( 1.8, -1.6) {\large $J$};

  \draw[kernelline] (UL) -- (Ltop);
  \draw[kernelline] (UR) -- (Rtop);
  \draw[kernelline] (LL) -- (Lbot);
  \draw[kernelline] (LR) -- (Rbot);
  \draw[kernelline] (Ltop) -- (Rtop);
  \draw[kernelline] (Lbot) -- (Rbot);

  \draw[gluon] (Ltop) -- (J1L);
  \draw[gluon] (J1L) -- (J2L);
  \draw[gluon] (J2L) -- (JnL);
  \draw[gluon] (JnL) -- (Lbot);
  \draw[gluon] (Rtop) -- (J1R);
  \draw[gluon] (J1R) -- (J2R);
  \draw[gluon] (J2R) -- (JnR);
  \draw[gluon] (JnR) -- (Rbot);

  \draw[kernelline,qarrow] ($(Ltop)+(-0.35,-0.3)$) -- ($(J1L)+(-0.35,0.3)$)
    node[midway, xshift=-15pt] {\large $q_1$};
  \draw[kernelline,qarrow] ($(J1L)+(-0.35,-0.3)$) -- ($(J2L)+(-0.35,0.3)$)
    node[midway, xshift=-15pt] {\large $q_2$};
  \draw[kernelline,qarrow] ($(JnL)+(-0.35,-0.3)$) -- ($(Lbot)+(-0.35,0.3)$)
    node[midway, xshift=-20pt] {\large $q_{n+1}$};
  \draw[kernelline,qarrow] ($(Rtop)+(0.35,-0.3)$) -- ($(J1R)+(0.35,0.3)$)
    node[midway, xshift=15pt] {\large $q'_1$};
  \draw[kernelline,qarrow] ($(J1R)+(0.35,-0.3)$) -- ($(J2R)+(0.35,0.3)$)
    node[midway, xshift=15pt] {\large $q'_2$};
  \draw[kernelline,qarrow] ($(JnR)+(0.35,-0.3)$) -- ($(Rbot)+(0.35,0.3)$)
    node[midway, xshift=20pt] {\large $q'_{n+1}$};

  \draw[gluon, draw=blue] (J1L) -- node[above=2pt, yshift=6pt] {\large $k_1$} (J1R);
  \draw[gluon, draw=blue] (J2L) -- node[above=2pt, yshift=6pt] {\large $k_2$} (J2R);
  \draw[gluon, draw=blue] (JnL) -- node[above=2pt, yshift=6pt] {\large $k_n$} (JnR);

  \draw[kernelline,qarrow] ($(J1L)+(1.0,0.35)$) -- ($(J1R)+(-1.0,0.35)$);
  \draw[kernelline,qarrow] ($(J2L)+(1.0,0.35)$) -- ($(J2R)+(-1.0,0.35)$);
  \draw[kernelline,qarrow] ($(JnL)+(1.0,0.35)$) -- ($(JnR)+(-1.0,0.35)$);

  \draw[kernelline,qarrow] ($(Ltop)+(0,0.35)$) -- ($(Ltop)+(-1.0,0.35)$)
    node[midway, yshift=9pt] {\large $p_1$};
  \draw[kernelline,qarrow] ($(Ltop)+(0.6,0.35)$) -- ($(Rtop)+(-0.6,0.35)$)
    node[midway, yshift=9pt] {\large $k_0$};
  \draw[kernelline,qarrow] ($(Rtop)+(0,0.35)$) -- ($(Rtop)+(1.0,0.35)$)
    node[midway, yshift=9pt] {\large $p_4$};
  \draw[kernelline,qarrow] ($(Lbot)+(0,-0.35)$) -- ($(Lbot)+(-1.0,-0.35)$)
    node[midway, yshift=-9pt] {\large $p_2$};
  \draw[kernelline,qarrow] ($(Lbot)+(0.6,-0.35)$) -- ($(Rbot)+(-0.6,-0.35)$)
    node[midway, yshift=-9pt] {\large $k_{n+1}$};
  \draw[kernelline,qarrow] ($(Rbot)+(0,-0.35)$) -- ($(Rbot)+(1.0,-0.35)$)
    node[midway, yshift=-9pt] {\large $p_3$};

  \node at ($ (Ltop)!0.5!(Rtop) +(0,-3)$) {\large $\vdots$};

  \node[purple] at ($ (J1L)!0.5!(J1R) $) {\textbar};
  \node[purple] at ($ (J2L)!0.5!(J2R) $) {\textbar};
  \node[purple] at ($ (JnL)!0.5!(JnR) $) {\textbar};

  \node[purple] at ($ (Ltop)!0.5!(Rtop) $) {\textbar};
  \node[purple] at ($ (Ltop)!0.5!(Rtop) $) {\textbar};
  \node[purple] at ($ (Lbot)!0.5!(Rbot) $) {\textbar};

\end{tikzpicture}

\caption{
The BFKL ladder obtained by sewing two amplitudes $\mathcal{M}_{2\to2+n}^{(0)}$ in MRK.
Vertical wavy lines denote Reggeized gravitons, while horizontal blue wavy lines represent soft graviton exchanges between Lipatov currents $J$.
}
\label{fig:bfkl-ladder}
\end{figure}

The first iteration in the $t$-channel arises from pairwise contractions of equal-multiplicity $\hat{N}$ matrix elements in the coherent state expansion \eqref{eq:Smatrix_MRK}, leading to the gravitational BFKL equation \cite{Lipatov:1982it}. This captures the leading logarithmic piece of the discontinuity of the $2\rightarrow2$ amplitude due to infinitely many gravitons emitted along the $t$-channel (see Fig. \ref{fig:bfkl-ladder}). However, as confirmed by power counting in $\hbar$, or by observing that the cut diagram in Fig. \ref{fig:bfkl-ladder} with $n>1$ gravitons contains graviton loops, this iteration yields purely quantum corrections of the form $\mathcal{O}\!\left(G^{2}s^{3}(G\log(s/|t|))^n\right)$ to the gravitational scattering \cite{Cristofoli:2021jas,Britto:2021pud}. We take this opportunity to revisit the original Lipatov construction of such iteration.  The starting point is the optical theorem,
\begin{align}
  &2\mathrm{Im}\,\mathcal{M}_{2\rightarrow2}(s,q^2)
  \nn \\
  & \, = \sum_{n \geq 1} \sum_{h_1,\dots,h_n}\int \mathrm{d} \mathcal{P}_{2+n} \,\hat{\delta}^{(D)}\Big(p_1+p_2+\sum_{j=0}^{n+1}k_j\Big)|\mathcal{M}_{2\rightarrow 2+ n}(p_1,p_2,\{k_1^{h_1},\dots, k_n^{h_n}\})|^2,
\end{align}
where the $(n+2)$-particle phase space becomes in MRK (see eq. \eqref{eq:phase-space})
\allowdisplaybreaks
\begin{align}
\nonumber\int \mathrm{d} \mathcal{P}_{2+n} &=
  \int  \hat{\mathrm{d}}^D k_0 \,\hat{\delta}_+\!\left(k_0^2-m_1^2\right) \int  \hat{\mathrm{d}}^D k_{n+1} \,\hat{\delta}_+\!\left(k_{n+1}^2-m_2^2\right) \left[\prod_{r=1}^{n}  \int  \hat{\mathrm{d}}^D k_r \,\hat{\delta}_+\!\left(k_r^2\right) \right]\,\\&\simeq\frac{1}{2^{n+1}s}\frac{1}{n!}\bigg(\frac{\log{(s/|t|)}}{2\pi}\bigg)^n
\left( \prod_{i=1}^{n+1}\int_{q_{i \perp}} \right),
\end{align}
and the $2\to 2+n$ amplitudes are taken in MRK with a strong rapidity ordering as in \eqref{eq:npt}. While the $n=1$ term corresponds to the two-loop cut $H$ diagram considered earlier (see e.g. eq. \eqref{eq:GR-kernel-contraction}), the terms with $n>1$ correspond to more gravitons emitted along the $t$-channel. In order to include all possible sources of high-energy logarithms, we take the graviton propagators to be dressed by their Regge trajectory $\alpha(q_{\perp})$,
\begin{align}
\label{eq:dressing}
\frac{1}{q^2_{i \perp}}\rightarrow\frac{1}{q^2_{i \perp}}e^{(y_{i-1}-y_i)\alpha(q_{i \perp})} \,.
\end{align}
where $y_i$ are the ordered rapidities of the produced states, so that each factor $y_{i-1}-y_i$ is the rapidity gap associated with the propagator of momentum $q_i$. At leading logarithmic order, it is sufficient to consider the one-loop Regge trajectory~\cite{Lipatov:1982it},
\begin{align}
\label{eq:amplitude_BFKL}
    &\mathcal{M}^{(0),\alpha\text{-dressed}}_{2 \to 2 +n}\big(1,2;\,k_0,k_1^{h_1},\ldots,k_n^{h_n},k_{n+1}\big) \simeq - \Theta(y_{0} >y_1 \dots > y_{n} > y_{n+1})\,\\
&\qquad \times \frac{s^2 \kappa^2}{\hbar}\; \frac{1}{t_1} e^{(y_{0}-y_1)\alpha^{(1)}(q_{1 \perp})}\; \prod_{r=1}^{n}\!\left\{\frac{\kappa}{\sqrt{\hbar}}\,\varepsilon^{(h_r)}_{\mu\nu}(k_r)\,J^{\mu\nu}(q_r,k_r,q_{r+1})\;\frac{1}{t_{r+1}} e^{(y_{r}-y_{r+1})\alpha^{(1)}(q_{r+1 \,\perp})}\right\}\,,\nonumber
\end{align}
and a direct calculation shows that \cite{Bartels:2012ra,Rothstein:2024nlq,Raj:2025hse}
\begin{align}
\alpha^{(1)}(q_{\perp})&=-\frac{\kappa^2 }{4\pi\hbar}(\vec{q}^{\,2})^2(d+1) \mathcal{B}_{1,1}(q) \nonumber \\
&= \frac{\kappa^2 \, \vec{q}^{\,2}}{16\pi^2\hbar} \left(\frac{4\pi\mu^2e^{\gamma_E}}{\vec{q}^{\,2}}\right)^\epsilon\left[(3-2\epsilon)\frac{\Gamma(1-\epsilon)\Gamma(-\epsilon)\Gamma(\epsilon)}{\Gamma(-2\epsilon)}\right] \nonumber \\
&=\frac{\kappa^2 \, \vec{q}^{\,2}}{8\pi^2\hbar}\left[\frac{3}{\epsilon}-3\log\left(\frac{\vec{q}^{\,2}}{4\pi\mu^2}\right)+2+{\cal O}(\epsilon)\right]
\label{eq:gravitonRegge}
\end{align}
in terms of the generalised bubble integrals $\mathcal{B}_{a,b}$ defined in Appendix \ref{app:integrals}. Unlike the gluon trajectory in QCD, which is purely infrared divergent at one loop, the graviton trajectory carries both UV and IR divergences coming from virtual graviton loops.  Notice that the radiative corrections obtained by using the graviton Regge trajectory in \eqref{eq:dressing} are purely quantum and therefore will never contribute to classical gravitational observables, i.e.
\begin{align}
    \hspace{-10pt}\mathcal{M}^{(0),\alpha\text{-dressed}}_{2 \to 2 +n}\big(1,2;\,k_0,k_1^{h_1},\ldots,k_n^{h_n},k_{n+1}\big) \stackrel{\hbar \to 0}{\to} \mathcal{M}^{(0)}_{2 \to 2 +n}\big(1,2;\,k_0,k_1^{h_1},\ldots,k_n^{h_n},k_{n+1}\big)\,.
\end{align}
Squaring the amplitude \eqref{eq:amplitude_BFKL} and summing over the graviton helicities, we obtain
\begin{align}
&\sum_{h_1,\dots,h_n} |\mathcal{M}^{(0),\alpha\text{-dressed}}_{2\rightarrow 2+ n}(p_1,p_2,\{k_1^{h_1},\dots, k_n^{h_n}\})|^2\\
& \qquad \qquad \simeq s^4 \frac{\kappa^{2(n+2)}}{\hbar^{n+2}} \prod_{i=1}^{n+1}\frac{1}{q^2_{i \perp}(q+q_i)^2_{\perp}}e^{[\alpha^{(1)}(q_{i \perp})+\alpha^{(1)}(q+q_{i \perp})](y_{i-1}-y_i)}\prod_{j=1}^n\mathcal{H}_{\mathrm{GR}}(q_j,q_{j+1};q)\,,\nonumber 
\end{align}
where we have used the definition of the kernel \eqref{eq:GR-kernel-def}. One can then derive the BFKL ladder contribution,
\begin{multline}
\label{eq:BFKL1}
\frac{2}{s^3}\mathrm{Im}\,\mathcal{M}_{2\rightarrow2}(s,q^2)\Big|_{\rm BFKL} \simeq \sum_{n=0}^{\infty}\frac{\kappa^{2(n+2)}}{2^{n+1}\hbar^{n+2}}\left(\prod_{i=1}^{n+1}\int_{q_{i\!\perp}}\right) \left(\prod_{i=1}^{n}\int \hat{\mathrm{d}}y_i \right) \Theta(y_{0} > \dots > y_{n} > y_{n+1}) \\
\times\prod_{i=1}^{n+1}\frac{1}{q^2_{i \perp}(q+q_i)^2_{\perp}}e^{[\alpha^{(1)}(q_{i\perp})+\alpha^{(1)}(q+q_{i\perp})](y_{i-1}-y_i)}\Bigg[\prod_{j=1}^n\mathcal{H}_{\mathrm{GR}}(q_j,q_{j+1};q) \Bigg].
\end{multline}
The strong ordering in the rapidities $y_0\gg y_1\gg\dots\gg y_{n+1}$ forces their integration domains to be entangled. To disentangle them, one performs a Laplace transform in the rapidity,\footnote{Equivalently, it is a Mellin transform in the Mandelstam variable $s$.}
\begin{align}
\mathcal{F}_l(q^2_{\perp})=\int_0^{\infty}\hat{\mathrm{d}}(\Delta y)\,\,e^{-l \Delta y}\frac{2}{s^3}\mathrm{Im\,\mathcal{M}}_{2\rightarrow2}(s,q^2_{\perp}),
\end{align}
where $\Delta y=y_0-y_{n+1}\simeq\log(s/|t|)$, yielding
\begin{align}
\mathcal{F}_l(q^2_{\perp})&=\sum_{n=0}^{\infty} \frac{\kappa^{2(n+2)}}{(4\pi)^{1+n}\hbar^{n+2}}\bigg(\prod_{i=1}^{n+1}\int_{q_{i\!\perp}}\bigg) \nonumber \\
&\times\prod_{j=1}^{n+1}\frac{1}{q^2_{j \perp}(q+q_j)^2_{\perp}}\frac{1}{l-\alpha^{(1)}(q_j)-\alpha^{(1)}(q+q_j)} \Bigg[\prod_{l=1}^n \mathcal{H}_{\mathrm{GR}}(q_l,q_{l+1};q)\Bigg].
\end{align}
Further introducing $f_l$ through
\begin{align}
\mathcal{F}_l(q^2_{\perp})=\frac{\kappa^4}{4\pi\hbar^2}\int_{k_{\perp}}\frac{f_l(k_{\perp},q_{\perp})}{k^2_{\perp}(q+k)^2_{\perp}},
\end{align}
then \eqref{eq:BFKL1} translates into an integral equation for $f_l$, namely
\begin{align}
[l-\alpha^{(1)}(k_{\perp})-\alpha^{(1)}(q_{\perp}+k_{\perp})]f_l(k_{\perp},q_{\perp})=1+\frac{\kappa^{4}}{4\pi\hbar^2}\int_{k_{\perp}'}\frac{\mathcal{H}_{\mathrm{GR}}(k,k';q)}{(k_{\perp}')^2(q+k')^2_{\perp}}f_l(k'_{\perp},q_{\perp}),
\end{align}
which is the gravitational BFKL equation \cite{Lipatov:1982vv,Lipatov:1982it,Lipatov:1991nf}. As discussed earlier, the Regge trajectory $\alpha(q_{\perp})$ is quantum in the classical limit $\hbar\!\to\!0$.
Consequently, apart from the $n=0$ term -- corresponding to the superclassical one-loop correction with no logarithmic enhancement -- and the $n=1$ term representing the $H$-diagram, the entire tower of logarithms accounted by the BFKL equation is quantum.

Despite these limitations, it remains of considerable interest -- from a purely amplitude perspective -- to investigate whether the full gravitational BFKL equation, incorporating both classical and quantum contributions, admits an analytic solution for generic $|t|\neq0$. Building on earlier developments in QCD~\cite{Kirschner:1982qf,Kirschner:1983di,Bartels:2008ce}, the first explicit solution of the gravitational BFKL equation at $|t|=0$ in terms of its eigenvalues was recently obtained in~\cite{Lipatov:1982vv,Raj:2025hse}.  Two important caveats, however, must be emphasized.  First, the presence of UV divergences may indicate the need for a string-theoretic completion, where the finite string scale provides a natural regulator of the high-energy region and potentially softens the gravitational BFKL spectrum.  Second, the BFKL solution cannot be unitary in isolation.  As the exponential representation of the S-matrix~\eqref{eq:Smatrix_MRK} makes explicit, even semiclassically there exist competing contributions associated with $s$-channel evolution which include contractions of three or more $\hat{N}$ matrix elements: neglecting these terms leads to an amplitude that violates perturbative unitarity at high energy.  

We now proceed to discuss the simplest of these $s$-channel iterations, leading to classical contributions to the gravitational scattering. 

\subsection{Iteration along the $s$-channel: multi-H diagrams}
\label{sec:multiH}

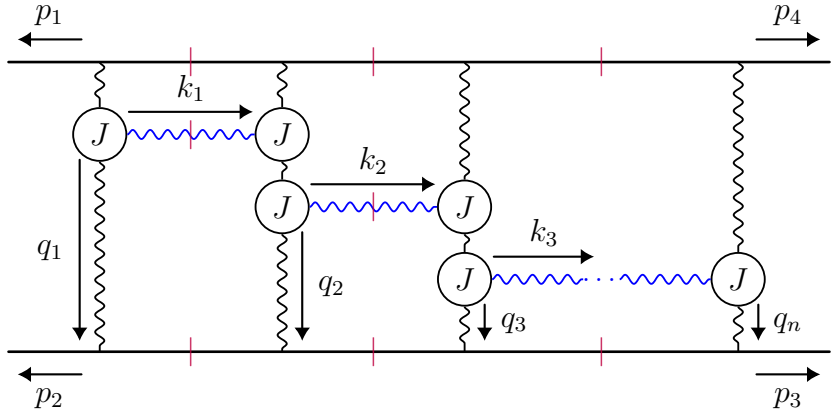
\begin{figure}[t]
\centering
\begin{tikzpicture}[scale=1.2,baseline=(bL.center),x=1cm,y=1cm]

  \def\h{0.80}  
  
  \coordinate (in1)  at (-2,  1.6);
  \coordinate (v1)   at (-1,  1.6);
  \coordinate (v2)   at ( 1,  1.6);
  \coordinate (v3)   at ( 3,  1.6);
  \coordinate (v4)   at ( 6,  1.6); 
  \coordinate (out1) at ( 7,  1.6);
  
  \coordinate (in2)  at (-2, -1.6);
  \coordinate (w1)   at (-1, -1.6);
  \coordinate (w2)   at ( 1, -1.6);
  \coordinate (w3)   at ( 3, -1.6);
  \coordinate (w4)   at ( 6, -1.6);
  \coordinate (out2) at ( 7, -1.6);

  \coordinate (first)   at ( 4.3, -\h);
  \coordinate (second) at ( 4.7, -\h);

  \node[jstyle] (bL)    at (-1, \h) {\large $J$};

  \node[jstyle] (dTop)  at ( 1, \h) {\large $J$};
  \node[jstyle] (dBot)  at ( 1,  0) {\large $J$};

  \node[jstyle] (eBot)  at ( 3, 0) {\large $J$};
  \node[jstyle] (fBot)  at ( 3, -\h) {\large $J$};

  \node[jstyle] (gBot)  at ( 6, -\h) {\large $J$};

  \draw[kernelline] (in1) -- (v1) -- (v2) -- (v3) -- (v4) -- (out1);
  \draw[kernelline] (in2) -- (w1) -- (w2) -- (w3) -- (w4) -- (out2);

  \draw[gluon] (v1) -- (bL);
  \draw[gluon] (bL) -- (w1);

  \draw[gluon] (v2) -- (dTop);
  \draw[gluon] (dTop) -- (dBot);
  \draw[gluon] (dBot) -- (w2);

  \draw[gluon] (v3) -- (eBot);
  \draw[gluon] (eBot) -- (fBot);
  \draw[gluon] (fBot) -- (w3);

  \draw[gluon] (v4) -- (gBot); 
  \draw[gluon] (gBot) -- (w4);  

  \draw[gluon, draw=blue] (bL)  -- (dTop); 
  \draw[gluon, draw=blue] (dBot)-- (eBot); 
  \draw[gluon, draw=blue] (fBot)-- (first);  
  \draw[gluon, draw=blue] (second)-- (gBot);  

  \draw[kernelline,qarrow] ($(v1)+(-0.2,0.25)$) -- ($(v1)+(-0.9,0.25)$) node[midway, yshift=9pt] {\large $p_1$};
  \draw[kernelline,qarrow] ($(w1)+(-0.2,-0.25)$) -- ($(w1)+(-0.9,-0.25)$) node[midway, yshift=-9pt] {\large $p_2$};
  \draw[kernelline,qarrow] ($(w4)+(0.2,-0.25)$) -- ($(w4)+(0.9,-0.25)$) node[midway, yshift=-9pt] {\large $p_3$};
  \draw[kernelline,qarrow] ($(v4)+(0.2,0.25)$) -- ($(v4)+(0.9,0.25)$) node[midway, yshift=9pt] {\large $p_4$};
  \draw[kernelline,qarrow] ($(bL)+(-0.22,-0.28)$) -- ($(w1)+(-0.22,0.12)$) node[midway, xshift=-11pt] {\large $q_1$};
  \draw[kernelline,qarrow] ($(dBot)+(0.22,-0.28)$) -- ($(w2)+(0.22,0.12)$) node[midway, xshift=11pt] {\large $q_2$};
  \draw[kernelline,qarrow] ($(fBot)+(0.22,-0.28)$) -- ($(w3)+(0.22,0.12)$) node[midway, xshift=11pt] {\large $q_3$};
  \draw[kernelline,qarrow] ($(gBot)+(0.22,-0.28)$) -- ($(w4)+(0.22,0.12)$) node[midway, xshift=11pt] {\large $q_n$};

  \draw[kernelline,qarrow] ($(bL)+(0.32,0.25)$)  -- ($(dTop)+(-0.32,0.25)$) node[midway, yshift=9pt] {\large $k_1$};
  \draw[kernelline,qarrow] ($(dBot)+(0.32,0.25)$) -- ($(eBot)+(-0.32,0.25)$) node[midway, yshift=9pt] {\large $k_2$};
  \draw[kernelline,qarrow] ($(fBot)+(0.32,0.25)$) -- ($(first)+(0.12,0.25)$)  node[midway, yshift=9pt] {\large $k_3$}; 

  \node[purple] at ($ (v1)!0.5!(v2) $) {\textbar};
  \node[purple] at ($ (w1)!0.5!(w2) $) {\textbar};
  \node[purple] at ($ (v2)!0.5!(v3) $) {\textbar};
  \node[purple] at ($ (w2)!0.5!(w3) $) {\textbar};
  \node[purple] at ($ (v3)!0.5!(v4) $) {\textbar};
  \node[purple] at ($ (w3)!0.5!(w4) $) {\textbar};
  \node[purple] at ($ (bL)!0.5!(dTop) $) {\textbar};
  \node[purple] at ($ (dBot)!0.5!(eBot) $) {\textbar};
  \node[blue] at ($ (first)!0.5!(second) $) {\dots};
  
\end{tikzpicture}
\caption{Multi-$H$ diagrams.}
\label{fig:multiHdiagram}
\end{figure}

The second iteration, recently analysed in~\cite{Rothstein:2024nlq,Alessio:2025isu}, encompasses all multi-$H$ diagrams\footnote{This class of diagrams was first identified in~\cite{Ciafaloni:2015xsr}, although they were not explicitly evaluated there.} containing the maximal number of three-particle cuts, as shown in Fig.~\ref{fig:multiHdiagram}. These diagrams arise from contractions of generating functionals with graviton emissions in the exponential representation~\eqref{eq:Smatrix_MRK} and are related to iterated $s$-channel discontinuities of the $2\!\to\!2$ amplitude. They occur at $(2N{+}1)$--PM order, where $N$ counts the soft graviton exchanges, and capture the classical contributions of the form $\mathcal{O}\left(Gs^2\big(G^{2}s\log(s/|t|)\big)^N\right)$. In what follows, we refer to these logarithmic enhancements as \emph{classical leading logarithms} because, at a certain PM order, they have the highest power of high-energy logarithms. All others logarithms, appearing with lower powers, are referred to as next-to-leading logarithms.

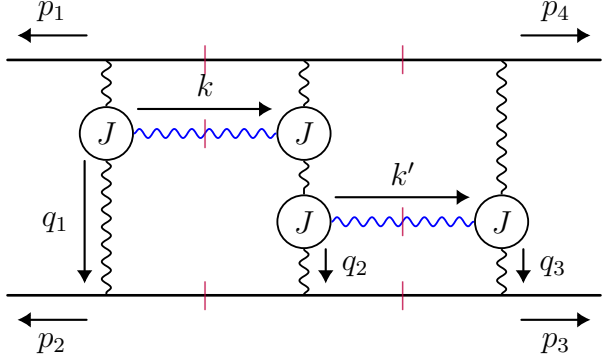
\begin{figure}[t]
\centering
\begin{tikzpicture}[scale=1.3,baseline=(bL.center),x=1cm,y=1cm]
  \def\h{0.45} 
  
  \coordinate (in1)  at (-2,  1.2);
  \coordinate (v1)   at (-1,  1.2);
  \coordinate (v2)   at ( 1,  1.2);
  \coordinate (v3)   at ( 3,  1.2);
  \coordinate (out1) at ( 4,  1.2);
  \coordinate (in2)  at (-2, -1.2);
  \coordinate (w1)   at (-1, -1.2);
  \coordinate (w2)   at ( 1, -1.2);
  \coordinate (w3)   at ( 3, -1.2);
  \coordinate (out2) at ( 4, -1.2);

  \node[jstyle] (bL)   at (-1,  \h) {\large $J$}; 
  \node[jstyle] (dTop) at ( 1,  \h) {\large $J$}; 
  \node[jstyle] (dBot) at ( 1, -\h) {\large $J$}; 
  \node[jstyle] (eR)   at ( 3, -\h) {\large $J$}; 

  \draw[kernelline] (in1) -- (v1) -- (v2) -- (v3) -- (out1);
  \draw[kernelline] (in2) -- (w1) -- (w2) -- (w3) -- (out2);

  \draw[gluon] (v1) -- (bL);
  \draw[gluon] (bL) -- (w1);

  \draw[gluon] (v2) -- (dTop);   
  \draw[gluon] (dTop) -- (dBot); 
  \draw[gluon] (dBot) -- (w2);   

  \draw[gluon] (v3) -- (eR);
  \draw[gluon] (eR) -- (w3);

  \draw[gluon, draw=blue] (bL) -- (dTop); 
  \draw[gluon, draw=blue] (dBot) -- (eR); 

  \draw[kernelline,qarrow] ($(v1)+(-0.2,0.25)$) -- ($(v1)+(-0.9,0.25)$) node[midway, yshift=9pt] {\large $p_1$};
  \draw[kernelline,qarrow] ($(v3)+(0.2,0.25)$) -- ($(v3)+(0.9,0.25)$) node[midway, yshift=9pt] {\large $p_4$};
  \draw[kernelline,qarrow] ($(w1)+(-0.2,-0.25)$) -- ($(w1)+(-0.9,-0.25)$) node[midway, yshift=-9pt] {\large $p_2$};
  \draw[kernelline,qarrow] ($(w3)+(0.2,-0.25)$) -- ($(w3)+(0.9,-0.25)$) node[midway, yshift=-9pt] {\large $p_3$};
  \draw[kernelline,qarrow] ($(bL)+(-0.22,-0.28)$) -- ($(w1)+(-0.22,0.12)$) node[midway, xshift=-11pt] {\large $q_1$};
  \draw[kernelline,qarrow] ($(dBot)+(0.22,-0.28)$) -- ($(w2)+(0.22,0.12)$) node[midway, xshift=11pt] {\large $q_2$};
  \draw[kernelline,qarrow] ($(eR)+(0.22,-0.28)$) -- ($(w3)+(0.22,0.12)$) node[midway, xshift=11pt] {\large $q_3$};

  \draw[kernelline,qarrow] ($(bL)+(0.32,0.25)$) -- ($(dTop)+(-0.32,0.25)$) node[midway, yshift=9pt] {\large $k$};
  \draw[kernelline,qarrow] ($(dBot)+(0.32,0.25)$) -- ($(eR)+(-0.32,0.25)$) node[midway, yshift=9pt] {\large $k'$};

  \node[purple] at ($ (v1)!0.5!(v2) $) {\textbar};
  \node[purple] at ($ (w1)!0.5!(w2) $) {\textbar};
  \node[purple] at ($ (v2)!0.5!(v3) $) {\textbar};
  \node[purple] at ($ (w2)!0.5!(w3) $) {\textbar};
  \node[purple] at ($ (bL)!0.5!(dTop) $) {\textbar};
  \node[purple] at ($ (dBot)!0.5!(eR) $) {\textbar};
\end{tikzpicture}
\caption{Double-$H$ diagram.}
\label{fig:H2diagram}
\end{figure}

Following the single $H$ diagram of Fig.~\ref{fig:Hdiagram} at 3PM, the next non-trivial term in this hierarchy appears at 5PM and corresponds to the double-$H$ diagram shown in Fig.~\ref{fig:H2diagram}. This arises from contracting two single-emission generating functionals with a double emission one, giving a purely real term which corresponds to the cut double $H$ diagram in MRK,
\begin{align}
\widetilde{\mathcal M}^{(4)}_{H^2}(s, b)
&\simeq\frac{i^3}{3!\,2!\,\hbar^5}\sum_{h_1,h_2}\int_{k_1,k_2}
  \Big[
    \widetilde{\mathcal{K}}^{(0)}_{2\to 3}(b;k_1^{h_1})\,
    \widetilde{\mathcal{K}}^{(0)}_{3\to 3}(b;k_1^{h_1};k_2^{h_2})\,
    \widetilde{\mathcal{K}}^{(0)\,*}_{2\to 3}(b;k_2^{h_2})
\nonumber \\
&\qquad \qquad
    +2\;\widetilde{\mathcal{K}}^{(0)\,*}_{2\to 3}(b;k_1^{h_1})\,
      \widetilde{\mathcal{K}}^{(0)\,*}_{2\to 3}(b;k_2^{h_2})\,
      \widetilde{\mathcal{K}}^{(0)}_{2\to 4}(b;k_1^{h_1}; k_2^{h_2})    \;+\;{\rm \mathrm{h.c.}}\Big]\,,
\end{align}
Upon Fourier transforming to transverse-momentum space, this expression reduces to the iterated three-particle discontinuity,
\begin{equation}\label{eq:mh2}
\mathcal M^{(4)}_{H^2}(s,q^2) \simeq - \frac{(8\pi G)^{5}s^4}{128 \pi^2 \hbar^5}\log^2\left(\frac{s}{|t|}\right)\;H^2(q^2)\,,
\end{equation}
with the scalar transverse integral $H^2(q^2)$ defined by
\begin{align}\label{eq:definition-H2-perp}
& H^2(q^2) \\
& \hspace{5pt} = \mu^{8\epsilon}\int_{q_{1\!\perp},q_{2\!\perp},k_{\perp},k'_{\perp}}\,
\frac{\mathcal{H}_{\mathrm{GR}}(q_1,q_1-k;k)\, \mathcal{H}_{\mathrm{GR}}(q_2+k,q_2+k-k';k')}{\vec{q}_1^{\,2} \vec{q}_2^{\,2} (\vec{q}_1-\vec{k})^2 (\vec{q}_2+\vec{k})^2 (\vec{q}_2+\vec{k}-\vec{k}')^2 (\vec{q}-\vec{q}_1-\vec{q}_2)^2 (\vec{q}-\vec{q}_1-\vec{q}_2+\vec{k}')^2}\,.\nonumber
\end{align}
As explained in~\cite{Alessio:2025isu}, to which we refer for additional technical details, the integral can be reduced to a linear combination of iterated bubble integrals and a (generalized) massless kite topology; see also Appendix~\ref{app:integrals}. Evaluating these contributions yields
\begin{align}
&H^2(q^2) = \frac{\vec{q}^{\,2}}{(4\pi)^4}\left(\frac{4\pi\mu^2 e^{\gamma_E}}{\vec{q}^{\,2}}\right)^{4\epsilon} \nonumber \\
&\qquad\times \Big\{ g(d)\left[f_1(d)+f_2(d)\left(h_a(d)+h_b(d)\right)\right]+\left[\bar{f}_1(d)+\bar{f}_2(d)\left(\bar{h}_a(d)+\bar{h}_b(d)\right)\right] \Big\}\,,
\label{eq:double-h}
\end{align}
where the functions $g(d)$, $f_i(d)$, $h_i(d)$, $\bar{f}_i(d)$ and $\bar{h}_i(d)$ -- defined in  Appendix~\ref{app:integrals} -- are expressed in terms of Gamma functions together with ${}_3 F_2$  hypergeometric functions at unit argument. The $\epsilon$-expanded result in $q$-space reads \cite{Alessio:2025isu}
\begin{align}
     \mathcal{M}_{H^2}^{(4)}(s,q^2) \simeq& -2 G^5 s^4 \log^2\left(\frac{s}{|t|}\right)\frac{\vec{q}^{\,2}}{\pi} \left(\frac{4\pi \mu^2}{\vec{q}^{\,2}}\right)^{4\epsilon} \nonumber \\
     &\qquad \times\left[-\frac{1}{\epsilon^3}-\frac{1}{\epsilon^2}-\frac{9}{\epsilon}+\frac{2}{\epsilon} \zeta_2+\frac{2}{\epsilon} \zeta_3+\mathcal{O}(1)\right] \label{eq:h2-result}
\end{align}

\section{Gravity amplitudes in the high-energy limit: shock-wave formalism}
\label{sec:shock-wave}
In this section, we adapt the shock-wave formalism, originally developed in the context of high-energy scattering in QCD~\cite{Balitsky:1995ub,Caron-Huot:2013fea,Caron-Huot:2017fxr}, to gravity amplitudes in the Regge limit. In particular, here show that the gravitational BFKL equation, as well as the tower of classical multi-$H$ diagrams appearing in sec.~\ref{sec:multiH}, can be reproduced by rapidity-evolution equations governed by a Hamiltonian composed of the graviton Regge trajectory and the gravitational kernel.

The shock-wave formalism describes the scattering of highly-energetic states as a projectile vs. target experiment. Such states are represented by Wilson lines (WLs) in the $+$ and $-$ light-cone directions. While in QCD, owing to the strong-coupling nature of the interactions among partons in the colliding hadrons, projectile and target states are given by a convolution of an infinite number of WLs~\cite{Korchemskaya:1994qp,Falcioni:2021buo}, in classical gravity it suffices to consider a single WL for each state. Therefore, for a general state we pick~\cite{Melville:2013qca} 
\begin{equation}
    |\tilde\Psi(b)\ra \simeq \tilde\Phi(b)\,|0\ra\,,
    \label{eq:singleWL}
\end{equation}
where $b$ is a fixed transverse position and $\tilde\Phi$ is a gravitational WL, defined as\footnote{With our choice $\kappa^2 = 8\pi G$, the normalization of the Wilson line exponent follows naturally from the the double--copy prescription $g \rightarrow \sqrt{16\pi G}/2 = \kappa/\sqrt{2}$~\cite{Monteiro:2014cda}.
Alternative normalizations, such as those adopted in~\cite{White:2011yy}, corresponds to a different rescaling of the Reggeised graviton field and its propagator. In this work we adhere to the QCD conventions of~\cite{Abreu:2024xoh} for the Reggeon two--point function $\langle W(q)\,|\,W(q')\rangle$.}
\begin{equation}\label{eq:grav-WL}
    \tilde\Phi(b) \equiv \exp\left(i\frac{\kappa}{\sqrt{2} \hbar}\int_{-\infty}^{\infty}ds\,p^\mu p^\nu h_{\mu\nu}\left(s\,p+b_{\perp}\right) \right) \,.
\end{equation}
At leading order in the Regge limit, eq.~\eqref{eq:singleWL} reduces to an exact equality, without extra modulation form factors,
\begin{equation}
  |\tilde\Psi(b)\ra = |\tilde\Psi^\text{\tiny LO}(b)\ra + \cdots = \big(1+\cdots\big)\,\tilde\Phi(b)\,|0\ra \,,
\end{equation}
where the $\dots$ refer to terms that might enter both as subleading terms in $t/s$, which are quantum corrections to the impact factors, and as mass terms $m_i^2/s$~\cite{White:2011yy}.
In what follows we will omit the subscript {\small LO}, that will always be implicit.

In contrast to semi-infinite null Wilson lines, the infinite Wilson lines in~\eqref{eq:grav-WL} exhibit rapidity divergences~\cite{Caron-Huot:2013fea}, reflecting the fact that lightlike (or nearly lightlike) Wilson lines require an additional rapidity regulator to separate modes by longitudinal momentum.
This is a purely geometrical effect and it does not depend on the number of transverse dimensions; in fact, rapidity divergences are not tamed in dimensional regularisation, but instead by introducing a rapidity cutoff $\eta$ having the physical meaning of tilting the WL trajectories~\cite{Caron-Huot:2013fea}.
Therefore, the states are renormalised through an evolution equation,
\begin{equation}
    -\frac{d}{d\eta} \tilde\Phi^\eta(b) = \hat{H}\,\tilde\Phi^\eta(b)\,,
    \label{eq:rrge}
\end{equation}
up to the value of $\eta=L$, with $L=\log|s/t|-i\pi/2$ the crossing symmetric large logarithm introduced in~\cite{Caron-Huot:2017fxr}, dictated by the signature symmetry of the gravitational amplitude in the high-energy limit \cite{Caron-Huot:2017fxr,Alessio:2025isu}. In the QCD case, the above evolution equation is named JIMWLK–Balitsky \cite{Balitsky:1995ub,Caron-Huot:2013fea,Caron-Huot:2017fxr}. Before moving on, we stress here that, while the latter exhibits an $SL(2,\mathbb{C})$ conformal invariance in the transverse plane, in gravity such symmetry is broken by the dimensionful nature of the gravitational coupling $\kappa$, as discussed in Appendix~\ref{app:conformal}. As a consequence, the rapidity evolution in gravity is not conformally invariant in the transverse plane and necessarily involves explicit dependence on the transverse momentum scale, leading to an Hamiltonian that differs qualitatively from its QCD counterpart.

We are now in a position to define an amplitude for the process involving the projectile and target Wilson lines, $\tilde\Phi_{i}$  and $\tilde\Phi_{j}$, respectively. First, we evolve e.g. the projectile state to the same rapidity of the target, getting  $\tilde\Phi^L_{i}=\mathrm{exp}(-\hat{H}L)\tilde\Phi_i.$ Then, the amplitude is given by their equal-rapidity inner product~\cite{Caron-Huot:2017fxr,Melville:2013qca}\footnote{All correlators in this section are vacuum expectation values of time-ordered products.},
\begin{equation}\label{eq:amplitude-from-inner-product}
    i\widetilde{\mathcal{M}}_{2\to2} (s,b) \simeq \big\langle \tilde\Phi_j(b_{2}) \, \tilde\Phi_i^L(b_{1}) \big\rangle\,,
\end{equation}
where $b = b_{1} - b_{2}$ and its momentum space counterpart is defined through 
\begin{equation}
    \label{eq:ampbspace}\widetilde{\mathcal{M}}_{2\to2} (s,b) = \frac{1}{2s} \int_{q_{\perp}} \, \mathcal{M}_{2\rightarrow 2}(s,q^2) \, e^{-i\vec{b}\cdot \vec{q}/\hbar}\,,
\end{equation}
as in \eqref{eq:Smatrix_MRK}. In sec.~\ref{sec:amplitudes-from-SWs} we will show that the perturbative expansion of such amplitude is identical to that obtained within the Regge theory formalism exploited in sec.~\ref{sec:regge}.
In momentum space, the amplitude takes the form,
\begin{equation}
    i\mathcal{M}_{2\to2} (s,q^2) \,\hat{\delta}^{(d)}\left(\vec{q}_1-\vec{q}_2\right) \simeq 2s \, \langle {\Psi}_j (q_2) | \,e^{-\hat{H} L} | {\Psi}_i (q_1) \rangle\,,
    \label{eq:amplitude-SWs-momentumSpace}
\end{equation}
in which $|{\Psi}(q)\rangle$ is the Fourier transform of \eqref{eq:singleWL}, i.e.
\begin{align}
|\Psi(q)\rangle=\Phi(q)|0\rangle,\qquad \Phi(q)\equiv \int \frac{d^db}{\hbar^d}\,\tilde\Phi(b)\,e^{i\vec{b}\cdot \vec{q}/\hbar} \,.
\label{eq:fourier_WL}
\end{align}

\subsection{The Reggeised graviton field and the boost Hamiltonian}\label{sec:boost}

\begin{figure}[t]
\centering
$\big|\Psi_i(q)\big\rangle  = $
\begin{tikzpicture}[scale=1.3,baseline=(base.center)]
  \coordinate (extL)  at (0, 0);
  \coordinate (extR)  at (1, 0);
  \coordinate (base)  at (0,-0.25);
  \draw[kernelline] (extL) -- (extR);
  \end{tikzpicture}
$\;+\;$
\begin{tikzpicture}[scale=1.3,baseline=(base.center)]
  \coordinate (extL)  at (-0.5, 0);
  \coordinate (i1)    at (0,  0);
  \coordinate (extR)  at (0.5,  0);
  \coordinate (b1)    at (0,  -0.5);
  \coordinate (base)  at (0,-0.25);
  \draw[kernelline] (extL) -- (i1) -- (extR);
  \draw[gluon] (i1) -- (b1);
\end{tikzpicture}
$\;+\;$
\begin{tikzpicture}[scale=1.3,baseline=(base.center)]
  \coordinate (extL)  at (-0.5, 0);
  \coordinate (i1)    at (0,  0);
  \coordinate (i2)    at (0.5,  0);
  \coordinate (extR)  at (1,  0);
  \coordinate (b1)    at (0,  -0.5);
  \coordinate (b2)    at (0.5,  -0.5);
  \coordinate (base)  at (0,-0.25);
  \draw[kernelline] (extL) -- (i1) -- (i2) -- (extR);
  \draw[gluon] (i1) -- (b1);
  \draw[gluon] (i2) -- (b2);
\end{tikzpicture}
$\;+\;$
\begin{tikzpicture}[scale=1.3,baseline=(base.center)]
  \coordinate (extL)  at (-0.5, 0);
  \coordinate (i1)    at (0,  0);
  \coordinate (i2)    at (0.5,  0);
  \coordinate (i3)    at (1,  0);
  \coordinate (extR)  at (1.5,  0);
  \coordinate (b1)    at (0,  -0.5);
  \coordinate (b2)    at (0.5,  -0.5);
  \coordinate (b3)    at (1,  -0.5);
  \coordinate (base)  at (0,-0.25);
  \draw[kernelline] (extL) -- (i1) -- (i2) -- (i3) -- (extR);
  \draw[gluon] (i1) -- (b1);
  \draw[gluon] (i2) -- (b2);
  \draw[gluon] (i3) -- (b3);
\end{tikzpicture}
$\;+\cdots$
\caption{Expansion of the projectile state in the number of Reggeon field insertions, as in \eqref{eq:princ}. At the leading order in the effective high-energy coupling $\kappa\sqrt{s}$, this expansion coincides with the expansion of the Wilson line $\Phi_i(q)$. In our conventions, we take the momentum $q$ to be outgoing.}
\label{fig:reggeon-states}
\end{figure}

We are interested in obtaining a perturbative expansion in powers of $G$ of the amplitude in \eqref{eq:amplitude-SWs-momentumSpace} in order to match the PM expansion of  gravitational amplitudes in the Regge limit discussed in sec. \ref{sec:regge}. 
The projectile and target states are expanded in Reggeon fields,
\begin{equation}
\left|\Psi(q)\right\rangle=\sum_{n=0}^{\infty}\left|\psi_{ n}(q)\right\rangle\,,
\label{eq:princ}
\end{equation}
 as represented in Fig.\ref{fig:reggeon-states}, where $\left|\psi_{ n}\right\rangle$ represents a state of $n$ Reggeised gravitons. Using the definition \eqref{eq:grav-WL} and parametrizing the incoming momenta in the centre-of-mass frame as
\begin{align}
\label{eq:mompar}
p_1^{\mu}=-\sqrt{s}(1,0,\vec{0}),\qquad p_2^{\mu}=-\sqrt{s}(0,1,\vec{0}),
\end{align} it follows that the WLs can be parametrised in terms of a field $\tilde{W}$ which plays the role of the source for Reggeised gravitons,
\begin{equation}\label{eq:W-field}
\tilde\Phi(b)=\exp\left[i \frac{\kappa}{\hbar}\,\sqrt{\frac{s}{2}}\,\tilde{W}(b)\right]
\,,
\end{equation}
where, considering for instance the projectile state,
\begin{equation}\label{eq:W-field-def}
\tilde{W}(b_1) = \int_{-\infty}^\infty dx^+ \, h_{++}(x^+,0,\vec{b}_1)\,.
\end{equation}
We expand the Fourier transform of the Wilson line as a convolution of Reggeon states,
\begin{subequations}
\label{eq:Reggeons}
\begin{align}
\label{eq:Reggeon1}
    & |\psi_{1}(q)\rangle = \frac{i\kappa}{\hbar}\sqrt{\frac{s}{2}} W(q) |0\rangle \,, \\
    & |\psi_{2}(q)\rangle = -\frac{\kappa^2}{2!\,\hbar^2} \frac{s}{2} \int_{q_{1\perp}} \, W(q_1)\, W(q-q_1) |0\rangle \,, \\
    & \cdots \nonumber \\
    & |\psi_{n}(q)\rangle = \frac{1}{n!}\left(\frac{i\kappa}{\hbar}\sqrt{\frac{s}{2}}\right)^n \prod_{j=1}^{n}\int_{q_{j \perp}} W(q_j)\hat{\delta}^{(d)}\Big(\vec{q}-\sum_{j=1}^n \vec{q}_j\Big)
    |0\rangle \,.
  \end{align}
\end{subequations}
Sometimes we will use $\left|W^n(q)\right\rangle$ as a shorthand for the convolution $W(q_1)\cdots W(q_{n})|0\rangle$ with total momentum $q$ and no prefactors.
For the purposes of this paper, it is sufficient to notice that the  leading order inner product in the Reggeised graviton basis is Gaussian,
\begin{equation}
    \big\langle W(q_1) \big| W(q_1') \big\rangle = -i\hbar \, \frac{\hat{\delta}^{(d)}\left(\vec{q}_1-\vec{q}_1^{\,\prime}\right)}{\vec{q}_1^{\,2}} + \mathcal{O}(G)\,,
\end{equation}
where we adopt the definition and normalisation of eq.~(6.37) of~\cite{Abreu:2024xoh}. Multi-Reggeon overlaps are obtained by Wick contraction of the elementary two-point function. For instance, the four-point overlap factorises as
\begin{equation}
\label{eq:W4pt}
\begin{aligned}
\big\langle W(q_1)\,W(q_2) \big| W(q_1')\,W(q_2') \big\rangle
&=
\big\langle W(q_1) \big| W(q_1') \big\rangle
\big\langle W(q_2) \big| W(q_2') \big\rangle
\\[4pt]
&\quad+
\big\langle W(q_1) \big| W(q_2') \big\rangle
\big\langle W(q_2) \big| W(q_1') \big\rangle
+\mathcal{O}(G)\,.
\end{aligned}
\end{equation}
By an appropriate choice of scheme, one may always define an orthogonal basis of multi-Reggeon states such that overlaps between states with different Reggeon number vanish. In particular, for generic momenta,
\begin{equation}
\label{eq:Worth}
\big\langle W(q_1)\cdots W(q_n) \,\big|\, W(q_1')\cdots W(q_m') \big\rangle
= \mathcal{O}(G)
\qquad \text{for } n\neq m \,,
\end{equation}
so that Reggeised gravitons behave as free fields at leading order.

We now give an explicit formulation for the evolution Hamiltonian $\hat{H}$ in eq.~\eqref{eq:rrge} acting on a state with a given number of Reggeons.
The amplitude \eqref{eq:amplitude-SWs-momentumSpace} must be $s\leftrightarrow u$ crossing symmetric and thus the Hamiltonian cannot mix even and odd numbers of Reggeons, i.e. $\hat{H}_{k,k\pm2n+1}=0$, for any integer $n$.
Furthermore we discard off-diagonal terms $\hat{H}_{k,k\pm2n}$ since they do not contribute to the amplitude at leading logarithmic accuracy. Moreover, the full Hamiltonian must be Hermitian,
\begin{equation}
\frac{d}{d\eta}\big\langle \mathcal{O}_1 \big| \mathcal{O}_2 \big\rangle = 0
\;\;\Longleftrightarrow\;\;
\big\langle \hat{H}\,\mathcal{O}_1 \big| \mathcal{O}_2 \big\rangle
= \big\langle \mathcal{O}_1 \big| \hat{H}\,\mathcal{O}_2 \big\rangle\,,
\end{equation}
for any operator $\mathcal{O}_i$. This implies projectile-target symmetry $i\leftrightarrow j$, namely that
\begin{equation}
  \hat{H}_{k \to k+2n} = \hat{H}_{k+2n \to k}\,.
\end{equation}
This symmetry ensures that the odd and even Reggeon sectors remain orthogonal and closed under the action of the Hamiltonian, so $\hat H$ is block-diagonal in Reggeon number parity. At leading logarithmic accuracy we further neglect number-changing transitions within each parity sector, keeping only $k\to k$ terms,
\begin{equation}\label{eq:Hamiltonian}
\hat{H}\!\left(\!\begin{array}{c}
W \\[2pt]
WW \\[2pt]
WWW \\[2pt]
\vdots
\end{array}\!\right)
=\left(\!\begin{array}{cccc}
\hat{H}_{1\to1} & 0 & 0 & \cdots \\
0 & \hat{H}_{2\to2} & 0 & \cdots \\
0 & 0 & \hat{H}_{3\to3} & \cdots \\
\vdots & \vdots & \vdots & \ddots
\end{array}\!\right)\!
\left(\!\begin{array}{c}
W \\[2pt]
WW \\[2pt]
WWW \\[2pt]
\vdots
\end{array}\!\right)
\;+\;\mathcal O(\kappa^4)\,.
\end{equation}
The diagonal terms of the Hamiltonian read
\begin{equation}\label{eq:Hkk}
    \hat{H}_{k\to k} = \hat{\mathcal{R}}_1 + \hat{\mathcal{R}}_2\,,
\end{equation}
where $\hat{\mathcal{R}}_1$ corrects the Reggeon propagator with a one-loop Regge trajectory, defined in \eqref{eq:gravitonRegge},
\begin{equation}\label{eq:hatR1}
    \hat{\mathcal{R}}_1 = \int_{q_\perp} \;\alpha^{(1)}(q)\,W(q)\,\frac{\delta}{\delta W(q)} \,,
\end{equation}
and $\hat{\mathcal{R}}_2$ represents the insertion of a soft graviton exchange between two Reggeons\footnote{We recall how the functional derivative acts in momentum space $\dfrac{\delta W(q')}{\delta W(q)}=\hat{\delta}^{(d)}(q-q')$.},
\begin{equation}
    \label{eq:hatr2}\hat{\mathcal{R}}_2 = -\frac{\kappa^2}{8\pi\hbar} \int_{\ell_\perp,q_{1\perp},q_{2\perp}} H_{22}(\ell;q_1,q_2) W(q_1\!+\!\ell) W(q_2\!-\!\ell) \frac{\delta}{\delta W(q_1)}\,\frac{\delta}{\delta W(q_2)} \,,
\end{equation}
up to orders $O(\kappa^4)$.
The function $H_{22}$, called \emph{boost Hamiltonian}, is proportional to the gravitational kernel of eq.~\eqref{eq:gravk},
\begin{equation}
H_{22}(\ell;\,q_1,q_2) = \frac{\mathcal{H}_{\mathrm{GR}}(q_1,q_1+\ell;q_1+q_2)}{\vec{q}_1^{\,2}\,\vec{q}_2^{\,2}} \,.
\end{equation}
Therefore the action of \eqref{eq:Hkk} on an $n$-Reggeon state $|\psi_{n}\rangle$ yields a one-loop correction which is the necessary ingredient for the leading logarithmic part of the amplitude \eqref{eq:amplitude-SWs-momentumSpace}, as represented for the three-Reggeon state in Figure~\ref{fig:combinatorics-H2}.

\subsection{$2 \to 2$ amplitudes in the shock-wave formalism}\label{sec:amplitudes-from-SWs}

We start by expanding eq.~\eqref{eq:amplitude-SWs-momentumSpace} in powers of the gravitational coupling $\kappa$ using the expansion of the evolution operator,
\begin{align}
e^{-\hat{H} L}
= 1 - L \hat{H} + \frac{L^2}{2} \hat{H}^2
- \frac{L^3}{6} \hat{H}^3
+ \frac{L^4}{24} \hat{H}^4 + \cdots,
\end{align}
and projecting onto external multi-Reggeon states finding, up to 5PM,
\begin{subequations}
\label{eq:loop-expanded-amplitude}
\begin{align}
\textbf{tree:}\;\;\;
&\frac{i}{2 s}\,\mathcal{M}^{(0)}_{2\to2} (s,q^2)
= \big\langle\psi_{j,1}\,\big|\,\psi_{i,1}\big\rangle,\label{eq:loop-expanded-amplitude-tree}
\\[6pt]
\textbf{1 loop:}\;\;
&\frac{i}{2 s}\,\mathcal{M}^{(1)}_{2\to2} (s,q^2)
=-L\,\big\langle\psi_{j, 1}\big| \hat{\mathcal{R}}_1 \big|\psi_{i, 1}\big\rangle
+  \big\langle\psi_{j, 2} \big| \psi_{i, 2}\big\rangle,\label{eq:loop-expanded-amplitude-one} \\[6pt]
\textbf{2 loops:}\;\;
&\frac{i}{2 s}\,\mathcal{M}^{(2)}_{2\to2} (s,q^2)
=\frac{L^2}{2}\,\big\langle\psi_{j, 1}\big|\hat{\mathcal{R}}_1^2\big|\psi_{i, 1}\big\rangle
 -\,L\,\big\langle\psi_{j, 2}\big| \big(\hat{\mathcal{R}}_1+\hat{\mathcal{R}}_2\big)\big|\psi_{i, 2}\big\rangle
+ \nonumber \\[3pt]
&\qquad\quad + \big\langle\psi_{j, 3} \big| \psi_{i, 3}\big\rangle, \label{eq:loop-expanded-amplitude-two}\\[6pt]
\textbf{3 loops:}\;\;
&\frac{i}{2 s}\,\mathcal{M}^{(3)}_{2\to2} (s,q^2)
=-\frac{L^3}{6}\,\big\langle\psi_{j,1}\big|\hat{\mathcal{R}}_1^{3}\big|\psi_{i,1}\big\rangle + \frac{L^2}{2}\,\big\langle\psi_{j,2}\big|\big(\hat{\mathcal{R}}_1+\hat{\mathcal{R}}_2\big)^{2}\big|\psi_{i,2}\big\rangle+
 \nonumber \\
&\qquad\quad -L\,\big\langle\psi_{j,3}\big|\big(\hat{\mathcal{R}}_1+\hat{\mathcal{R}}_2\big)\big|\psi_{i,3}\big\rangle
+ \big\langle\psi_{j,4} \big| \psi_{i,4}\big\rangle,
\label{eq:loop-expanded-amplitude-three}\\[6pt]
\textbf{4 loops:}\;\;
&\frac{i}{2 s}\,\mathcal{M}^{(4)}_{2\to2} (s,q^2)
=\frac{L^4}{24}\,\big\langle\psi_{j,1}\big|\hat{\mathcal{R}}_1^4\big|\psi_{i,1}\big\rangle + \nonumber \\
&\qquad\quad-\frac{L^3}{6}\,\big\langle\psi_{j,2}\big|\,\big(\hat{\mathcal{R}}_1+\hat{\mathcal{R}}_2\big)^{3}\,\big|\psi_{i,2}\big\rangle +\frac{L^2}{2}\,\big\langle\psi_{j,3}\big|\,\big(\hat{\mathcal{R}}_1+\hat{\mathcal{R}}_2\big)^{2}\,\big|\psi_{i,3}\big\rangle + \nonumber \\[3pt]
&\qquad\quad -\,L\,\big\langle\psi_{j,4}\big|\big(\hat{\mathcal{R}}_1+\hat{\mathcal{R}}_2\big)\big|\psi_{i,4}\big\rangle
+ \big\langle\psi_{j,5} \big| \psi_{i,5}\big\rangle.\label{eq:loop-expanded-amplitude-four} 
\end{align}
\end{subequations}
This explains clearly how all the leading logarithmic contributions - superclassical, classical and quantum - arise in the gravitational $2\to2$ amplitude, and it can be extended systematically to all loop orders. Notice that, since $\hat{\mathcal{R}}_1$ is a purely quantum graviton loop, if one is only interested in classical physics, it has to be set to zero in the above expressions\footnote{
All the terms involving $\hat{\mathcal{R}}_1$ correspond to genuine quantum effects. 
However, they are not the only quantum contributions in \eqref{eq:loop-expanded-amplitude}: $t$-channel Reggeon iterations generated by $\hat{\mathcal{R}}_2$ are also quantum in origin, as explicitly seen in the BFKL evolution discussed in sec.~\ref{sec:regge}. }.

The terms appearing in \eqref{eq:loop-expanded-amplitude} perfectly fit in a more general structure of the massless high-energy gravitational amplitude found in~\cite{Alessio:2025isu}\footnote{At leading logarithmic accuracy in the Regge expansion, it also coincides with the ultra relativistic limit of the massive scattering amplitude.},
\begin{equation}\label{eq:ampl-high-energies}
  \mathcal{M}_{2\rightarrow 2}^{(\ell)}(s,q^2) \Big|_{m_1=m_2=0} = \frac{G^{\ell+1}(tz_t)^{\ell + 2}}{t} \left(\frac{4\pi\mu^2e^{\gamma_E}}{-t}\right)^{\ell\epsilon}\sum_{j=0}^\infty\sum_{k = 0}^{\text{min}(\ell,j)}i^{\ell + j}f^{(\ell)}_{(j,k)}(\epsilon) \, z_t^{-j} \, L^k \,,
\end{equation}
where $z_t = (2s+t)/t$ is a dimensionless, crossing symmetric variable and $\ell$ is the loop order.
The shock-wave formalism only predicts the highest powers of $L$ contributions in \eqref{eq:ampl-high-energies}, namely
\begin{equation}\label{eq:ampl-high-energies-LL}
  \mathcal{M}_{2\rightarrow 2}^{(\ell)}(s,q^2) \simeq \frac{G^{\ell+1}(tz_t)^{\ell + 2}}{t} \left(\frac{4\pi\mu^2e^{\gamma_E}}{-t}\right)^{\ell\epsilon} \sum_{j=0}^\infty i^{\ell + j}f^{(\ell)}_{(j,k_\text{max})}(\epsilon) \, z_t^{-j} \, L^{k_\text{max}} \,,
\end{equation}
with $k_\text{max} = \text{min}(\ell,j)$. In fact, in the boost Hamiltonian $\hat{H}$ we have discarded off-diagonal terms $\hat{H}_{k\to k\pm2n}$ that do contribute at next-to-leading logarithmic (NLL) accuracy, as well as $O(\kappa^4)$ terms in the diagonal elements. Thus from the shock-wave perspective, LL accuracy has to be understood as the leading logarithmic part in front of a given power of $s/t$. 

Let us now analyse in detail the connection between \eqref{eq:loop-expanded-amplitude} and \eqref{eq:ampl-high-energies-LL}.
The leading terms in the Regge limit are captured, in impact-parameter space, by the exponentiation of leading eikonal phase $2\delta_0(s,b)$. In \eqref{eq:ampl-high-energies-LL}, they correspond to the terms  $f^{(\ell)}_{(0,0)}$, i.e. $j=0$ and no powers of $L$. Within the shock-wave formalism such terms are proportional to the scalar product of projectile and target $n$ Reggeon states,
\begin{equation}
\begin{aligned}
\label{eq:superclassical}
    \mathcal{M}_{2\rightarrow 2}^{(n-1)}(s,q^2)\Big|_{\text{S}^{n-1}\text{C}} & \simeq \frac{2s}{i}\big\langle\psi_{j,n} \big| \psi_{i,n}\big\rangle  \\
    & = \frac{1}{n!}\frac{8\pi s}{i\,\vec{q}^{\,2}} \left(\frac{i\kappa^2 s}{8\pi\hbar}\right)^n
    \frac{\Gamma(1+n\epsilon-\epsilon)\Gamma(-\epsilon)^n}{\Gamma(-n\epsilon)} \left(\frac{4\pi\mu^2e^{\gamma_E}}{\vec{q}^{\,2}} \right)^{(n-1)\epsilon}\,.
\end{aligned}
\end{equation}
where we used the bubble integrals family $\mathcal{B}_{1,1}^{(n)}$ in eq.~\eqref{eq:iterated-bubbles} and where, by $\mathrm{S}^{n}\mathrm{C}$ we mean a term that, after the replacement $\vec{q}\rightarrow \hbar\, \vec{q}$, is singular in the $\hbar\rightarrow 0$ limit as $\sim \hbar^{-3-n}$. This has to be compared with the FT of $(2i\delta_0(s,b))^n/n!$. 

At one loop \eqref{eq:loop-expanded-amplitude-one} also the Regge trajectory enters through $\hat{\mathcal{R}}_1$ and this contribution, expressed using wave vectors, is suppressed by two additional powers of $\hbar$ relative to the superclassical $n=2$ eikonal term in \eqref{eq:superclassical}. It is indeed the quantum (Q) contribution corresponding to the ``soft-eye graph'' in eq.~(5.1) of \cite{Rothstein:2024nlq},
\begin{equation}
    \mathcal{M}_{2\rightarrow 2}^{(1)}(s,q^2)\Big|_{\text{Q}} \simeq 2\,i\,s\,L\,\big\langle\psi_{j, 1}\big| \hat{\mathcal{R}}_1 \big|\psi_{i, 1}\big\rangle = L\,\frac{s^2\kappa^2}{\vec{q}^{\,2}} \alpha^{(1)}(q_\perp)\,.
\end{equation}
Notice that at one loop, there are no classical terms. This will be true for any odd number of loops, as predicted by eq.~\eqref{eq:ampl-high-energies}. At two-loop order, equation \eqref{eq:loop-expanded-amplitude-two} has three terms. The last one has already been discussed in~\eqref{eq:superclassical}, while the term proportional to $L^2$ is a double insertion of the one loop Regge trajectory, yielding another quantum term. The other contribution, which contains a single logarithm, has a classical (CL) scaling
that we show below to coincide with the $H$ diagram in Fig.~\ref{fig:Hdiagram}. The two-Reggeon projectile $|\psi_{i,2}\rangle$ is evolved in rapidity by adding a gravitational Kernel and then contracted with the target state,
\begin{align}
    \nonumber\mathcal{M}_{2\rightarrow 2}^{(2)}(s,q^2)\Big|_{\text{CL}} 
     &\simeq - \frac{2sL}{i} \frac{\kappa^4 s^2}{16\hbar^4}\int \hat{\mathrm{d}}^d\vec{q}_{1} \, \hat{\mathrm{d}}^d\vec{q}_{1}^{\,}{'} \, \big\langle W(q_1') W(q'-q_1') \big| \hat{\mathcal{R}}_2 \big| W(q_1) W(q-q_1) \big\rangle  \nn \\
    &  = i L \frac{\kappa^6 s^3}{16\pi\hbar^3} \, H^1(q^2)\,.
\end{align}
This expression is identical to equation \eqref{eq:3pcc}, up to the replacement of the logarithm with its crossing symmetric version $\log(s/|t|) \mapsto L$ which gives information on the real part of this amplitude, as Amati, Ciafaloni and Veneziano proved in 1990 using the optical theorem~\cite{Amati:1990xe}. 
Such substitution relates real and imaginary parts of the amplitude in an highly non-trivial way. It allows to extract part of the NLL amplitude with opposite reality with respect to the leading logarithmic one~\cite{Alessio:2025isu}. 

At three loops there are no classical pieces of the amplitude at leading logarithmic accuracy, while at four loops we have a double logarithmic contribution from $\big\langle\psi_{j,3}\big|\,\hat{\mathcal{R}}_2^{2}\,\big|\psi_{i,3}\big\rangle$ which has an overall classical scaling. However, it contains both true classical as well as graviton-loop terms which we discard, see Fig.\ref{fig:combinatorics-H2}. The final amplitude is
\begin{equation}
    \mathcal{M}_{2\rightarrow 2}^{(4)}(s,q^2)\Big|_{\text{CL}} \simeq - \frac{\kappa^{10} s^4 L^2}{512\cdot 3! \, \pi^2 \hbar^5}\, (4\cdot3!) \,H^2(q^2) = - \frac{\kappa^{10} s^4 L^2}{128 \,\pi^2 \hbar^5}\,H^2(q^2)\,,
    \label{eq:H2-from-shock-waves}
\end{equation}
where $H^2(q^2)$ is defined in \eqref{eq:definition-H2-perp}.
The non-trivial combinatorial factor $4\cdot3!$ takes into account how many ways one can generate $H^2$  by applying the functional derivatives in $\hat{\mathcal{R}}_2^2$ and it is pictured in Fig.~\ref{fig:combinatorics-H2}. From the whole set of diagrams we discard the two particle reducible ones which factorise in impact parameter space, as explained below equation \eqref{eq:N-operator}.
\begin{figure}[t]
\centering
$(\hat{\mathcal{R}}_2)^2\,\big| \psi_{i,3} \big\rangle = 2\hat{\mathcal{R}}_2 \Bigg\{ $
\begin{tikzpicture}[scale=1.3,baseline=(jL.center)]
  \def\h{0.5}
  \def\hs{0.4}
   
  \coordinate (in1)  at (-3*\hs,  1);
  \coordinate (v1)   at (-2*\hs,  1);
  \coordinate (v2)   at ( 0,      1);
  \coordinate (v3)   at ( 2*\hs,  1);
  \coordinate (out1) at ( 3*\hs,  1);

  \coordinate (b1)   at (-2*\hs,  0);
  \coordinate (b2)   at ( 0,      0);
  \coordinate (b3)   at ( 2*\hs,  0);

  \node[smalljstyle] (jL)   at (-2*\hs, \h) {\tiny $J$}; 
  \node[smalljstyle] (jR)   at ( 0,     \h) {\tiny $J$}; 

  \draw[kernelline] (in1) -- (v1) -- (v2) -- (v3) -- (out1);

  \draw[gluon] (v1) -- (jL);
  \draw[gluon] (jL) -- (b1);
  \draw[gluon] (v2) -- (jR);
  \draw[gluon] (jR) -- (b2);
  \draw[gluon] (v3) -- (b3);

  \draw[gluon, draw=blue] (jL) -- (jR); 

  \node[purple] at ($ (v1)!0.5!(v2) $) {\tiny\textbar};
  \node[purple] at ($ (v2)!0.5!(v3) $) {\tiny\textbar};
  \node[purple] at ($ (jL)!0.5!(jR) $) {\small\textbar};
\end{tikzpicture}
$+$
\begin{tikzpicture}[scale=1.3,baseline=(jL.center)]
  \def\h{0.5}
  \def\hs{0.4}
   
  \coordinate (in1)  at (-3*\hs,  1);
  \coordinate (v1)   at (-2*\hs,  1);
  \coordinate (v2)   at ( 0,      1);
  \coordinate (v3)   at ( 2*\hs,  1);
  \coordinate (out1) at ( 3*\hs,  1);

  \coordinate (b1)   at (-2*\hs,  0);
  \coordinate (b2)   at ( 0,      0);
  \coordinate (b3)   at ( 2*\hs,  0);

  \node[smalljstyle] (jL)   at (0,    \h) {\tiny $J$}; 
  \node[smalljstyle] (jR)   at (2*\hs,\h) {\tiny $J$}; 

  \draw[kernelline] (in1) -- (v1) -- (v2) -- (v3) -- (out1);

  \draw[gluon] (v1) -- (b1);
  \draw[gluon] (v2) -- (jL);
  \draw[gluon] (jL) -- (b2);
  \draw[gluon] (v3) -- (jR);
  \draw[gluon] (jR) -- (b3);

  \draw[gluon, draw=blue] (jL) -- (jR); 

  \node[purple] at ($ (v1)!0.5!(v2) $) {\tiny\textbar};
  \node[purple] at ($ (v2)!0.5!(v3) $) {\tiny\textbar};
  \node[purple] at ($ (jL)!0.5!(jR) $) {\small\textbar};
\end{tikzpicture}
$+$
\begin{tikzpicture}[scale=1.3,baseline=(jL.center)]
  \def\h{0.5}
  \def\hs{0.4}
   
  \coordinate (in1)  at (-3*\hs,  1);
  \coordinate (v1)   at (-2*\hs,  1);
  \coordinate (v2)   at ( 0,      1);
  \coordinate (v3)   at ( 2*\hs,  1);
  \coordinate (out1) at ( 3*\hs,  1);

  \coordinate (b1)   at (-2*\hs,  0);
  \coordinate (b2)   at ( 0,      0);
  \coordinate (b3)   at ( 2*\hs,  0);

  \node[smalljstyle] (jL)   at (-2*\hs, \h) {\tiny $J$}; 
  \coordinate (jR1)   at (-0.2*\hs, \h);
  \coordinate (jR2)   at (+0.2*\hs, \h); 
  \node[smalljstyle] (j2)   at (+2*\hs, \h) {\tiny $J$} ; 

  \draw[kernelline] (in1) -- (v1) -- (v2) -- (v3) -- (out1);

  \draw[gluon] (v1) -- (jL);
  \draw[gluon] (jL) -- (b1);
  \draw[gluon] (v2) -- (b2);
  \draw[gluon] (v3) -- (jR);
  \draw[gluon] (jR) -- (b3);

  \draw[gluon, draw=blue] (jL) -- (jR1); 
  \draw[gluon, draw=blue] (jR2) -- (jR); 

  \node[purple] at ($ (v1)!0.5!(v2) $) {\tiny\textbar};
  \node[purple] at ($ (v2)!0.5!(v3) $) {\tiny\textbar};
  \node[purple] at ($ (jR2)!0.45!(jR) $) {\small\textbar};
\end{tikzpicture}
$\Bigg\} = $ \\ 
$\hspace{5pt}\,4\,\times\;$
\begin{tikzpicture}[scale=1.3,baseline=(base.center)]
  \def\h{0.5}
  \def\hs{0.4}

  \coordinate (base) at (0, 0.5*\h);

  \coordinate (in1)  at (-3*\hs,  2*\h);
  \coordinate (v1)   at (-2*\hs,  2*\h);
  \coordinate (v2)   at ( 0,      2*\h);
  \coordinate (v3)   at ( 2*\hs,  2*\h);
  \coordinate (out1) at ( 3*\hs,  2*\h);

  \coordinate (b1)   at (-2*\hs,  -\h);
  \coordinate (b2)   at ( 0,      -\h);
  \coordinate (b3)   at ( 2*\hs,  -\h);

  \node[smalljstyle]  (jLU)  at (-2*\hs, \h)  {\tiny $J$};
  \node[smalljstyle]  (jRU)  at ( 0,     \h)  {\tiny $J$};
  \node[smalljstyle]  (jLL)  at (-2*\hs, 0)   {\tiny $J$};
  \node[smalljstyle]  (jRL)  at ( 0,     0)   {\tiny $J$};

  \draw[kernelline] (in1) -- (v1) -- (v2) -- (v3) -- (out1);

  \draw[gluon] (v1) -- (jLU);
  \draw[gluon] (jLU) -- (jLL);
  \draw[gluon] (jLL) -- (b1);
  \draw[gluon] (v2) -- (jRU);
  \draw[gluon] (jRU) -- (jRL);
  \draw[gluon] (jRL) -- (b2);
  \draw[gluon] (v3) -- (b3);

  \draw[gluon, draw=blue] (jLU) -- (jRU); 
  \draw[gluon, draw=blue] (jLL) -- (jRL); 

  \node[purple] at ($ (v1)!0.5!(v2) $) {\tiny\textbar};
  \node[purple] at ($ (v2)!0.5!(v3) $) {\tiny\textbar};
  \node[purple] at ($ (jLU)!0.5!(jRU) $) {\small\textbar};
  \node[purple] at ($ (jLL)!0.5!(jRL) $) {\small\textbar};
\end{tikzpicture}
$+\,4\,\times\;$
\begin{tikzpicture}[scale=1.3,baseline=(base.center)]
  \def\h{0.5}
  \def\hs{0.4}

  \coordinate (base) at (0, 0.5*\h);
   
  \coordinate (in1)  at (-3*\hs,  2*\h);
  \coordinate (v1)   at (-2*\hs,  2*\h);
  \coordinate (v2)   at ( 0,      2*\h);
  \coordinate (v3)   at ( 2*\hs,  2*\h);
  \coordinate (out1) at ( 3*\hs,  2*\h);

  \coordinate (b1)   at (-2*\hs,  -\h);
  \coordinate (b2)   at ( 0,      -\h);
  \coordinate (b3)   at ( 2*\hs,  -\h);

  \node[smalljstyle] (jL)   at (0,    \h) {\tiny $J$}; 
  \node[smalljstyle] (jR)   at (2*\hs,\h) {\tiny $J$}; 
  \node[smalljstyle]  (jL2)  at (-2*\hs, 0)   {\tiny $J$};
  \node[smalljstyle]  (jR2)  at ( 0,     0)   {\tiny $J$};

  \draw[kernelline] (in1) -- (v1) -- (v2) -- (v3) -- (out1);

  \draw[gluon] (v1) -- (jL2);
  \draw[gluon] (jL2) -- (b1);
  \draw[gluon] (v2) -- (jL);
  \draw[gluon] (jL) -- (jR2);
  \draw[gluon] (jR2) -- (b2);
  \draw[gluon] (v3) -- (jR);
  \draw[gluon] (jR) -- (b3);

  \draw[gluon, draw=blue] (jL) -- (jR);
  \draw[gluon, draw=blue] (jL2) -- (jR2); 

  \node[purple] at ($ (v1)!0.5!(v2) $) {\tiny\textbar};
  \node[purple] at ($ (v2)!0.5!(v3) $) {\tiny\textbar};
  \node[purple] at ($ (jL)!0.5!(jR) $) {\small\textbar};
  \node[purple] at ($ (jL2)!0.5!(jR2) $) {\small\textbar};
\end{tikzpicture}
$+\,4\,\times\;$
\begin{tikzpicture}[scale=1.3,baseline=(base.center)]
  \def\h{0.5}
  \def\hs{0.4}

  \coordinate (base) at (0, 0.5*\h);
   
  \coordinate (in1)  at (-3*\hs,  1);
  \coordinate (v1)   at (-2*\hs,  1);
  \coordinate (v2)   at ( 0,      1);
  \coordinate (v3)   at ( 2*\hs,  1);
  \coordinate (out1) at ( 3*\hs,  1);

  \coordinate (b1)   at (-2*\hs,  -\h);
  \coordinate (b2)   at ( 0,      -\h);
  \coordinate (b3)   at ( 2*\hs,  -\h);

  \node[smalljstyle] (jL)   at (-2*\hs, \h) {\tiny $J$}; 
  \coordinate (jR1)   at (-0.2*\hs, \h);
  \coordinate (jR2)   at (+0.2*\hs, \h); 
  \node[smalljstyle] (j2)   at (+2*\hs, \h) {\tiny $J$};
  \node[smalljstyle]  (jLlow)  at (-2*\hs, 0)   {\tiny $J$};
  \node[smalljstyle]  (jRlow)  at ( 0,     0)   {\tiny $J$}; 

  \draw[kernelline] (in1) -- (v1) -- (v2) -- (v3) -- (out1);

  \draw[gluon] (v1) -- (jL);
  \draw[gluon] (jL) -- (jLlow);
  \draw[gluon] (jLlow) -- (b1);
  \draw[gluon] (v2) -- (jRlow);
  \draw[gluon] (jRlow) -- (b2);
  \draw[gluon] (v3) -- (jR);
  \draw[gluon] (jR) -- (b3);

  \draw[gluon, draw=blue] (jL) -- (jR1); 
  \draw[gluon, draw=blue] (jR2) -- (jR); 
  \draw[gluon, draw=blue] (jLlow) -- (jRlow); 

  \node[purple] at ($ (v1)!0.5!(v2) $) {\tiny\textbar};
  \node[purple] at ($ (v2)!0.5!(v3) $) {\tiny\textbar};
  \node[purple] at ($ (jR2)!0.45!(jR) $) {\small\textbar};
  \node[purple] at ($ (jLlow)!0.5!(jRlow) $) {\small\textbar};
\end{tikzpicture}
$+$\\
$+\,4\,\times\;$
\begin{tikzpicture}[scale=1.3,baseline=(base.center)]
  \def\h{0.5}
  \def\hs{0.4}
  
  \coordinate (base) at (0, 0.5*\h);
   
  \coordinate (in1)  at (-3*\hs,  1);
  \coordinate (v1)   at (-2*\hs,  1);
  \coordinate (v2)   at ( 0,      1);
  \coordinate (v3)   at ( 2*\hs,  1);
  \coordinate (out1) at ( 3*\hs,  1);

  \coordinate (b1)   at (-2*\hs,  -\h);
  \coordinate (b2)   at ( 0,      -\h);
  \coordinate (b3)   at ( 2*\hs,  -\h);

  \node[smalljstyle] (jL)   at (-2*\hs, \h) {\tiny $J$}; 
  \node[smalljstyle] (jR)   at ( 0,     \h) {\tiny $J$}; 
  \node[smalljstyle] (jjL)  at (0,      0) {\tiny $J$}; 
  \node[smalljstyle] (jjR)  at (2*\hs,  0) {\tiny $J$}; 

  \draw[kernelline] (in1) -- (v1) -- (v2) -- (v3) -- (out1);

  \draw[gluon] (v1) -- (jL);
  \draw[gluon] (jL) -- (b1);
  \draw[gluon] (v2) -- (jR);
  \draw[gluon] (jR) -- (jjL);
  \draw[gluon] (jjL) -- (b2);
  \draw[gluon] (v3) -- (jjR);
  \draw[gluon] (jjR) -- (b3);

  \draw[gluon, draw=blue] (jL) -- (jR); 
  \draw[gluon, draw=blue] (jjL) -- (jjR); 

  \node[purple] at ($ (v1)!0.5!(v2) $) {\tiny\textbar};
  \node[purple] at ($ (v2)!0.5!(v3) $) {\tiny\textbar};
  \node[purple] at ($ (jL)!0.5!(jR) $) {\small\textbar};
  \node[purple] at ($ (jjL)!0.5!(jjR) $) {\small\textbar};
\end{tikzpicture}
$+\,4\,\times\;$
\begin{tikzpicture}[scale=1.3,baseline=(base.center)]
  \def\h{0.5}
  \def\hs{0.4}

  \coordinate (base) at (0, 0.5*\h);
   
  \coordinate (in1)  at (-3*\hs,  1);
  \coordinate (v1)   at (-2*\hs,  1);
  \coordinate (v2)   at ( 0,      1);
  \coordinate (v3)   at ( 2*\hs,  1);
  \coordinate (out1) at ( 3*\hs,  1);

  \coordinate (b1)   at (-2*\hs,  -\h);
  \coordinate (b2)   at ( 0,      -\h);
  \coordinate (b3)   at ( 2*\hs,  -\h);

  \node[smalljstyle] (jL)   at (0,    \h) {\tiny $J$}; 
  \node[smalljstyle] (jR)   at (2*\hs,\h) {\tiny $J$};
  \node[smalljstyle] (jjL)  at (0,     0) {\tiny $J$}; 
  \node[smalljstyle] (jjR)  at (2*\hs, 0) {\tiny $J$}; 

  \draw[kernelline] (in1) -- (v1) -- (v2) -- (v3) -- (out1);

  \draw[gluon] (v1) -- (b1);
  \draw[gluon] (v2) -- (jL);
  \draw[gluon] (jL) -- (jjL);
  \draw[gluon] (jjL) -- (b2);
  \draw[gluon] (v3) -- (jR);
  \draw[gluon] (jR) -- (jjR);
  \draw[gluon] (jjR) -- (b3);

  \draw[gluon, draw=blue] (jL) -- (jR); 
  \draw[gluon, draw=blue] (jjL) -- (jjR); 

  \node[purple] at ($ (v1)!0.5!(v2) $) {\tiny\textbar};
  \node[purple] at ($ (v2)!0.5!(v3) $) {\tiny\textbar};
  \node[purple] at ($ (jL)!0.5!(jR) $) {\small\textbar};
  \node[purple] at ($ (jjL)!0.5!(jjR) $) {\small\textbar};
\end{tikzpicture}
$+\,4\,\times\;$
\begin{tikzpicture}[scale=1.3,baseline=(base.center)]
  \def\h{0.5}
  \def\hs{0.4}

  \coordinate (base) at (0, 0.5*\h);
   
  \coordinate (in1)  at (-3*\hs,  1);
  \coordinate (v1)   at (-2*\hs,  1);
  \coordinate (v2)   at ( 0,      1);
  \coordinate (v3)   at ( 2*\hs,  1);
  \coordinate (out1) at ( 3*\hs,  1);

  \coordinate (b1)   at (-2*\hs,  -\h);
  \coordinate (b2)   at ( 0,      -\h);
  \coordinate (b3)   at ( 2*\hs,  -\h);

  \node[smalljstyle] (jL)   at (-2*\hs, \h) {\tiny $J$}; 
  \coordinate (jR1)   at (-0.2*\hs, \h);
  \coordinate (jR2)   at (+0.2*\hs, \h); 
  \node[smalljstyle] (j2)   at (+2*\hs, \h) {\tiny $J$};
  \node[smalljstyle] (jjL)  at (0,     0) {\tiny $J$}; 
  \node[smalljstyle] (jjR)  at (2*\hs, 0) {\tiny $J$}; 

  \draw[kernelline] (in1) -- (v1) -- (v2) -- (v3) -- (out1);

  \draw[gluon] (v1) -- (jL);
  \draw[gluon] (jL) -- (b1);

  \draw[gluon] (v2) -- (jjL);
  \draw[gluon] (jjL) -- (b2);
  \draw[gluon] (v3) -- (jR);
  \draw[gluon] (jR) -- (jjR);
  \draw[gluon] (jjR) -- (b3);

  \draw[gluon, draw=blue] (jL) -- (jR1); 
  \draw[gluon, draw=blue] (jR2) -- (jR);
  \draw[gluon, draw=blue] (jjL) -- (jjR); 

  \node[purple] at ($ (v1)!0.5!(v2) $) {\tiny\textbar};
  \node[purple] at ($ (v2)!0.5!(v3) $) {\tiny\textbar};
  \node[purple] at ($ (jR2)!0.45!(jR) $) {\small\textbar};
  \node[purple] at ($ (jjL)!0.5!(jjR) $) {\small\textbar};
\end{tikzpicture}
$+$\\
$+\,4\,\times\;$
\begin{tikzpicture}[scale=1.3,baseline=(base.center)]
  \def\h{0.5}
  \def\hs{0.4}

  \coordinate (base) at (0, 0.5*\h);
   
  \coordinate (in1)  at (-3*\hs,  1);
  \coordinate (v1)   at (-2*\hs,  1);
  \coordinate (v2)   at ( 0,      1);
  \coordinate (v3)   at ( 2*\hs,  1);
  \coordinate (out1) at ( 3*\hs,  1);

  \coordinate (b1)   at (-2*\hs,  -\h);
  \coordinate (b2)   at ( 0,      -\h);
  \coordinate (b3)   at ( 2*\hs,  -\h);

  \node[smalljstyle] (jL)   at (-2*\hs, \h) {\tiny $J$}; 
  \node[smalljstyle] (jR)   at ( 0,     \h) {\tiny $J$}; 
  \node[smalljstyle] (jjL)   at (-2*\hs,0) {\tiny $J$}; 
  \coordinate (jjR1)   at (-0.2*\hs, 0);
  \coordinate (jjR2)   at (+0.2*\hs, 0); 
  \node[smalljstyle] (jj2)   at (+2*\hs, 0) {\tiny $J$}; 

  \draw[kernelline] (in1) -- (v1) -- (v2) -- (v3) -- (out1);

  \draw[gluon] (v1) -- (jL);
  \draw[gluon] (jL) -- (jjL);
  \draw[gluon] (jjL) -- (b1);
  \draw[gluon] (v2) -- (jR);
  \draw[gluon] (jR) -- (b2);
  \draw[gluon] (v3) -- (jjR);
  \draw[gluon] (jjR) -- (b3);

  \draw[gluon, draw=blue] (jL) -- (jR);
  \draw[gluon, draw=blue] (jjL) -- (jjR1);
  \draw[gluon, draw=blue] (jjR2) -- (jjR);

  \node[purple] at ($ (v1)!0.5!(v2) $) {\tiny\textbar};
  \node[purple] at ($ (v2)!0.5!(v3) $) {\tiny\textbar};
  \node[purple] at ($ (jL)!0.5!(jR) $) {\small\textbar};
  \node[purple] at ($ (jjR2)!0.45!(jjR) $) {\small\textbar};
\end{tikzpicture}
$+\,4\,\times\;$
\begin{tikzpicture}[scale=1.3,baseline=(base.center)]
  \def\h{0.5}
  \def\hs{0.4}

  \coordinate (base) at (0, 0.5*\h);
   
  \coordinate (in1)  at (-3*\hs,  1);
  \coordinate (v1)   at (-2*\hs,  1);
  \coordinate (v2)   at ( 0,      1);
  \coordinate (v3)   at ( 2*\hs,  1);
  \coordinate (out1) at ( 3*\hs,  1);

  \coordinate (b1)   at (-2*\hs,  -\h);
  \coordinate (b2)   at ( 0,      -\h);
  \coordinate (b3)   at ( 2*\hs,  -\h);

  \node[smalljstyle] (jL)   at (0,    \h) {\tiny $J$}; 
  \node[smalljstyle] (jR)   at (2*\hs,\h) {\tiny $J$}; 
  \node[smalljstyle] (jjL)   at (-2*\hs,0) {\tiny $J$}; 
  \coordinate (jjR1)   at (-0.2*\hs, 0);
  \coordinate (jjR2)   at (+0.2*\hs, 0); 
  \node[smalljstyle] (jj2)   at (+2*\hs, 0) {\tiny $J$};  

  \draw[kernelline] (in1) -- (v1) -- (v2) -- (v3) -- (out1);

  \draw[gluon] (v1) -- (jjL);
  \draw[gluon] (jjL) -- (b1);
  \draw[gluon] (v2) -- (jL);
  \draw[gluon] (jL) -- (b2);
  \draw[gluon] (v3) -- (jR);
  \draw[gluon] (jR) -- (jjR);
  \draw[gluon] (jjR) -- (b3);

  \draw[gluon, draw=blue] (jL) -- (jR);
  \draw[gluon, draw=blue] (jjL) -- (jjR1); 
  \draw[gluon, draw=blue] (jjR2) -- (jjR); 

  \node[purple] at ($ (v1)!0.5!(v2) $) {\tiny\textbar};
  \node[purple] at ($ (v2)!0.5!(v3) $) {\tiny\textbar};
  \node[purple] at ($ (jL)!0.5!(jR) $) {\small\textbar};
  \node[purple] at ($ (jjR2)!0.45!(jjR) $) {\small\textbar};
\end{tikzpicture}
$+\,4\,\times\;$
\begin{tikzpicture}[scale=1.3,baseline=(base.center)]
  \def\h{0.5}
  \def\hs{0.4}

  \coordinate (base) at (0, 0.5*\h);
   
  \coordinate (in1)  at (-3*\hs,  1);
  \coordinate (v1)   at (-2*\hs,  1);
  \coordinate (v2)   at ( 0,      1);
  \coordinate (v3)   at ( 2*\hs,  1);
  \coordinate (out1) at ( 3*\hs,  1);

  \coordinate (b1)   at (-2*\hs,  -\h);
  \coordinate (b2)   at ( 0,      -\h);
  \coordinate (b3)   at ( 2*\hs,  -\h);

  \node[smalljstyle] (jL)   at (-2*\hs, \h) {\tiny $J$}; 
  \coordinate (jR1)   at (-0.2*\hs, \h);
  \coordinate (jR2)   at (+0.2*\hs, \h); 
  \node[smalljstyle] (jR)   at (+2*\hs, \h) {\tiny $J$};  
  \node[smalljstyle] (jjL)   at (-2*\hs,0) {\tiny $J$}; 
  \coordinate (jjR1)   at (-0.2*\hs, 0);
  \coordinate (jjR2)   at (+0.2*\hs, 0); 
  \node[smalljstyle] (jj2)   at (+2*\hs, 0) {\tiny $J$};  

  \draw[kernelline] (in1) -- (v1) -- (v2) -- (v3) -- (out1);

  \draw[gluon] (v1) -- (jL);
  \draw[gluon] (jL) -- (jjL);
  \draw[gluon] (jjL) -- (b1);
  \draw[gluon] (v2) -- (b2);
  \draw[gluon] (v3) -- (jR);
  \draw[gluon] (jR) -- (jjR);
  \draw[gluon] (jjR) -- (b3);

  \draw[gluon, draw=blue] (jL) -- (jR1); 
  \draw[gluon, draw=blue] (jR2) -- (jR);
  \draw[gluon, draw=blue] (jjL) -- (jjR1); 
  \draw[gluon, draw=blue] (jjR2) -- (jjR); 

  \node[purple] at ($ (v1)!0.5!(v2) $) {\tiny\textbar};
  \node[purple] at ($ (v2)!0.5!(v3) $) {\tiny\textbar};
  \node[purple] at ($ (jR2)!0.45!(jR) $) {\small\textbar};
  \node[purple] at ($ (jjR2)!0.45!(jjR) $) {\small\textbar};
\end{tikzpicture}
$\,.\hspace{5pt}$\\
\caption{Combinatorial factor in eq.~\eqref{eq:H2-from-shock-waves}. Applying two times the evolution operator $\hat{H}$ to $|\psi_{i,3}\rangle$ and discarding the Regge trajectory contribution, we get 36 diagrams. Among these, $4\cdot3!$ will contribute to the classical $2\to2$ amplitude, while the remaining $4\cdot3$ show a classical scaling as quantum times superclassical pieces, that we discard. Despite the graphical representation, high-energy factorisation allows the Reggeon lines to be permuted freely, implying that all contributing diagrams can be planarised.
}
\label{fig:combinatorics-H2}
\end{figure}

It is easy to generalise the multi-H construction to higher-loop orders. As discussed in \cite{Amati:1993tb,Rothstein:2024nlq,Alessio:2025isu}, it exhibits a recursive structure. The classical leading logarithmic terms of order $\mathcal{O}(G^{2n+1}s^{n+2}L^n)$ at $2n$ loops are built by applying $n$ times the boost Hamiltonian to an $(n+1)$-Reggeon state,
\begin{equation}
    \label{eq:Hn-on-psi-n+1}
    \frac{(-L\hat{H})^n}{n!}|\psi_{i,n+1}\big\rangle = \frac{(-i)^n}{n!(n+1)!} \frac{i\kappa}{\hbar} \left(\frac{\kappa^3}{8\pi\hbar^2}\right)^n \left(\frac{s}{2}\right)^\frac{n+1}{2}\,L^n \,\left(\hat{r}_2\right)^n \big| W^{n+1}(q) \big\rangle \,,
\end{equation}
in which we defined for convenience $\hat{\mathcal{R}}_2 = \left(\kappa^2/(8\pi\hbar)\right)\hat{r}_2$ and we discarded $\hat{\mathcal{R}}_1$ contributions.
Then \eqref{eq:Hn-on-psi-n+1} has to be contracted with the target state and multiplied by $2s/i$, giving
\begin{equation}
  \mathcal{M}_{2\rightarrow 2}^{(2n)}(s,q^2)\Big|_{\text{CL}} \simeq \frac{i}{n!\left((n+1)!\right)^2} \frac{\kappa^{4n+2}s^{n+2}L^n}{\left(16\pi\hbar^3\right)^n \hbar^2}\, \big\langle W^{n+1}(q') \big| \left(\hat{r}_2\right)^n \big| W^{n+1}(q) \big\rangle\,.
\end{equation}
The non-trivial transverse integrals that will appear at arbitrary number of loops, as well as the combinatorics of the diagrams, are entirely encoded in the angle brackets.
By Wick contractions, a factor of $(-i\hbar)^{n+1}(n+1)!$ is generated and, by dimensional analysis, the transverse integral scales as $(\vec{q}^{\,2})^{n-1} \propto \hbar^{2(n-1)} $ for $\epsilon=0$. This proves the classical scaling $\hbar^{-3}$ and the reality of the amplitude at leading logarithmic accuracy, which is
\begin{equation}
  \begin{aligned}
    \mathcal{M}^{(2n)}_{H^{n}}(s,q^2)
      & \simeq \mathrm{Im}\,\mathcal{M}^{(2n)}_{H^{n}}(s,q^2)\,, \qquad n \ \text{odd}, \\[0.3em]
    \mathcal{M}^{(2n)}_{H^{n}}(s,q^2)
      & \simeq \mathrm{Re}\,\mathcal{M}^{(2n)}_{H^{n}}(s,q^2)\,, \qquad n \ \text{even}.
  \end{aligned}
\end{equation}
It can be shown that at $2n$ loops there are in principle $[n(n+1)]^n$ diagrams, differing by the order in which the gravitational kernel is inserted, or equivalently by the action of the functional derivatives in $\hat{r}_2$.
The first non-trivial example is indeed at $n=2$, where there are 36 diagrams as shown in Fig.~\ref{fig:combinatorics-H2}.
Most of them turn out to be equivalent, while the others can be discarded \textit{ab initio} since they are two-particle reducible and correspond to iterations. In the case of $H^2$, the computation reduces to just one diagram, while for $H^3$ -- six loops, $\mathcal{O}(G^{7}s^{5}L^3)$ -- it can be shown  that there are only three inequivalent topologies, one of which is not planar. It would be interesting to explore this further by computing the corresponding analytic expressions and investigating their all-orders resummation.


\subsection{$2\rightarrow 3$ amplitudes in the shock-wave formalism}
\label{sec:2to3}

So far, we have reformulated the high-energy, leading logarithmic  approximation of the $2\rightarrow 2$ elastic scattering in terms of shock-waves. Here we further extend the formalism to include the inelastic $2\rightarrow 3$ case as well, taking again inspiration from the QCD case \cite{Caron-Huot:2013fea}. 
Compared to the $2\rightarrow 2$ process, where the high-energy limit is governed by only one large logarithm $L=\log|s/t|-i\pi/2$, corresponding to the large rapidity gap $\Delta Y_{12}$, the real emission of a graviton along the ladder introduces a new rapidity scale. With reference to Fig.\,\ref{fig:1}, we denote $L_{1,2}\equiv\log|s_{1,2}/t|-i\pi/4$\footnote{We are adopting the choice made e.g. in \cite{Buccioni:2024gzo,Abreu:2014cla} for the branch cuts.} the large logarithms associated with the rapidity gaps between the shock-waves and the emitted graviton. The leading-logarithm five-point amplitude is defined through the correlator,
\begin{align}
\frac{i}{2 s}\mathcal{M}^{h}_{2\to 3} (q_1,k,q_2) \,\hat{\delta}^{(d)}(\vec{k}-\vec{q}_1+\vec{q}_2) \simeq \langle \Psi_j(q_2) | \,e^{-\hat{H}L_2} \hat{a}^{(h)}(k)\,e^{-\hat{H}L_1}| \Psi_i (q_{1}) \rangle,
\label{eq:5ptshock}
\end{align}
where the Hamiltonian generating the evolution in the rapidities is again the one in
\eqref{eq:Hkk} and where $\hat{a}^{(h)}(k)$ is the annihilation operator for the graviton defined in \eqref{eq:field_mode}, which is the new ingredient compared to the elastic case in \eqref{eq:amplitude-SWs-momentumSpace}. In eq.\,\eqref{eq:5ptshock}, the equal-rapidity correlator is taken after the operators $e^{-\hat{H}L_i}$ evolved the multi-Reggeons states $|\Psi_{i,j}\rangle$ to the same rapidity of the emitted graviton. 

It is instructive to reproduce the tree-level single-Reggeon (single $t$-channel exchange) contribution to the $2\to3$ amplitude \eqref{eq:5pt+} from the correlator definition \eqref{eq:5ptshock}.  This amounts to determining, at leading order, the insertion rule for the graviton annihilation operator acting on a Wilson line $\Phi$, and hence on the Reggeon field $W$.

In QCD, the corresponding light-cone quantisation on a shock-wave background was worked out originally in~\cite{Balitsky:1995ub} (see also~\cite{Caron-Huot:2013fea,Buccioni:2024gzo,Abreu:2024xoh}), yielding the leading interaction in $g_s$ of a gluon mode with a Wilson-line operator in light-cone gauge. The result for the interaction between $a^{(h)}$ and $W$ is given in terms of the OPE,
\begin{align}
\label{eq:OPEQCD0}
&a^{(h)}(k)W(q)\stackrel{\mathrm{QCD}}{\sim} 2g_s W(q-k)\varepsilon_{\mu}^{(h)}(k) V^{\mu}(q,k)+\mathcal{O}(g^2_s),
\end{align}
with the interaction vertex,
\begin{align}
\label{eq:OPEQCD1}
V^{\mu}(q,k)=\bigg(\frac{q_{\perp}^{\mu}}{q_{\perp}^2}-\frac{k_{\perp}^{\mu}}{k_{\perp}^2}\bigg),
\end{align}
where the higher-order corrections in \eqref{eq:OPEQCD0}, relevant for higher-loop computations, comprise the interaction of the gluon field with more than one Reggeon $W$, as schematically depicted below in Fig.~\ref{fig:reggeons}.

\begin{figure}[t]
\centering
\begin{tikzpicture}[x=1cm,y=1cm,baseline=(base)]

\newcommand{\WLterm}[2]{%
\begin{scope}[shift={(#1,0)}]

  \def\ytop{1.30}
  \def\yc{0.18}
  \def\R{0.33}
  \def\yup{\the\numexpr0\relax} 

  \draw[kernelline,line cap=round] (-1.45,\ytop) -- (1.45,\ytop);

  \fill[gray!55] (0,\yc) circle [radius=\R];
  \draw[kernelline] (0,\yc) circle [radius=\R];

  \pgfmathsetmacro{\yT}{\yc+\R}
  \pgfmathsetmacro{\yB}{\yc-\R}
  \pgfmathsetmacro{\xR}{\R}

  \ifnum#2>0
    \draw[gluon] (0,\ytop) -- (0,\yT);
  \fi
  \ifnum#2>1
    \draw[gluon] (-0.55,\ytop) -- (-0.24,\yT-0.1);
  \fi
  \ifnum#2>2
    \draw[gluon] (0.55,\ytop) -- (0.24,\yT-0.1);
  \fi

  \draw[gluon] (0,\yB) -- (0,-0.90);

  \draw[gluon,draw=blue] (\xR,\yc) -- (1.75,\yc);

\end{scope}
}

\WLterm{0}{1}
\node at (3.15,0.18) {\Large $+$};

\WLterm{4.65}{2}
\node at (7.80,0.18) {\Large $+$};

\WLterm{9.30}{3}
\node at (12.55,0.18) {\Large $+\cdots$};

\coordinate (base) at (0,-1.05);

\end{tikzpicture}
\caption{Graphic representation of the right-hand side of eq. \eqref{eq:OPEQCD0}. The first diagram corresponds to the $\mathcal{O}(g_s)$ vertex in \eqref{eq:OPEQCD0}, while the other two to $\mathcal{O}(g_s^2)$ and $\mathcal{O}(g_s^3)$ diagrams. The thick black line corresponds to the shockwave background, as in Fig. \ref{fig:reggeon-states},  whereas the black and blue wavy lines to Glauber and soft gravitons, respectively. }
\label{fig:reggeons}
\end{figure}
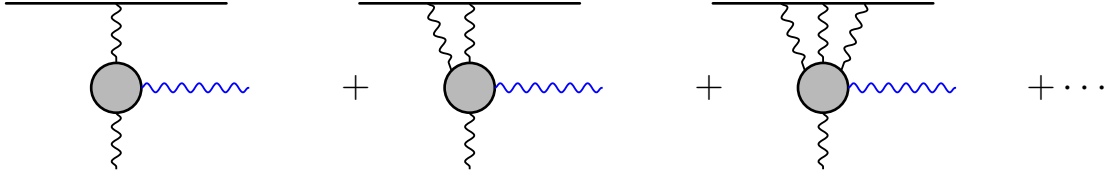

It can be shown~\cite{Buccioni:2024gzo,Abreu:2014cla} that \eqref{eq:OPEQCD1} contains enough information to reproduce the tree-level Lipatov vertex in the light-cone gauge, which is proportional to the Lipatov current $J^{\mu}$ in \eqref{eq:CEV_suba} contracted with the light-cone polarisation vectors in \eqref{eq:polLG}. In fact, it is important to point out that eqs. \eqref{eq:OPEQCD0} and \eqref{eq:OPEQCD1}, as well as  eq.~\eqref{eq:OPEGR} in the gravity case, \textit{only} hold in the light-cone gauge\footnote{The authors thank Federico Buccioni for several clarifying discussions on this point.}.

The gravitational analogue of \eqref{eq:OPEQCD1} would require to perform a light-cone quantisation of Einstein gravity on a shock-wave background. Here, we limit ourselves to conjecture the interaction vertex between a graviton and a Reggeon by requiring that, using the definition of the scattering amplitude in \eqref{eq:5ptshock}, it  yields \eqref{eq:5pt+} upon contraction with $\varepsilon_{\mu\nu}^{(\oplus)}$ and $\varepsilon_{\mu\nu}^{(\ominus)}$. The result for the interaction vertex is\footnote{We find also
\begin{align*}
W(q)a^{(h)}(k)\stackrel{\mathrm{GR}}{\sim}2\frac{\kappa}{\sqrt{\hbar}} W(q+k)\varepsilon_{\mu\nu}^{(h)}(k)\bigg[(q_{\perp}+k_{\perp})^2V^{\mu}(q+k,k)V^{\nu}(q+k,k)-\frac{q_{\perp}^2}{k_{\perp}^4}k^{\mu}_{\perp}k^{\nu}_{\perp}\bigg]\frac{(q_{\perp}+k_{\perp})^2}{q_{\perp}^2}.
\end{align*}
}
\begin{align}
\label{eq:OPEGR}
a^{(h)}(k)W(q)\stackrel{\mathrm{GR}}{\sim} \,& 2 \frac{\kappa}{\sqrt{\hbar}} W(q-k) \varepsilon_{\mu\nu}^{(h)}(k)\bigg[q_{\perp}^2V^{\mu}(q,k)V^{\nu}(q,k)-\frac{(q_{\perp}-k_{\perp})^2}{k_{\perp}^4}k^{\mu}_{\perp}k^{\nu}_{\perp}\bigg],
\end{align}
at leading order. Taking into account \eqref{eq:Reggeon1}, the single-Reggeon contribution for the $\oplus$ polarisation is then
\begin{align}
\nonumber&\frac{i}{2s}\mathcal{M}^{(0)\,\oplus}_{2\rightarrow3 }(q_1,k,q_2)\hat{\delta}^{(d)}(\vec{k}-\vec{q}_1+\vec{q}_2) \simeq \bigg(i\frac{\kappa}{\hbar}\sqrt{\frac{s}{2}}\bigg)^2\langle W(q_2)a^{(\oplus)}(k)W(q_1)\rangle\\\nonumber&=2\frac{\kappa}{\sqrt{\hbar}}\bigg(i\frac{\kappa}{\hbar}\sqrt{\frac{s}{2}}\bigg)^2\langle W(q_2)W(q_1-k)\rangle\bigg[q_{1\perp}^2(\varepsilon^{(\oplus)}\cdot V(q_1,k))^2-\frac{(q_{1\perp}-k_{\perp})^2}{k_{\perp}^4}(\varepsilon^{(\oplus)}\cdot k_{\perp})^2\bigg]\\&=-i\frac{\kappa^3s}{\hbar^{3/2}}\frac{1}{q_{2\perp}^2}\hat{\delta}^{(d)}(\vec{k}-\vec{q}_1+\vec{q}_2)\bigg[q_{1\perp}^2(\varepsilon^{(\oplus)}\cdot V(q_1,k))^2-\frac{(q_{1\perp}-k_{\perp})^2}{k_{\perp}^4}(\varepsilon^{(\oplus)}\cdot k_{\perp})^2\bigg],
\end{align}
which, by explicit computation yields again \eqref{eq:5pt+}, 
\begin{align}
\mathcal{M}^{(0)\,\oplus}_{2\rightarrow3 }(q_1,k,q_2) \simeq \frac{s^2\kappa^3}{\hbar^{3/2}}\frac{k_{\mathrm{T}}^* \, q_{1\mathrm{T}}-q^*_{1\mathrm{T}} \, k_{\mathrm{T}}}{(k_{\mathrm{T}}^*)^2 \, q^*_{1\mathrm{T}} \, q_{2\mathrm{T}}},
\end{align}
confirming \textit{a posteriori} our conjecture in \eqref{eq:OPEGR}.

\section{The ultra-relativistic limit of scattering waveforms}
\label{sec:waveform}

In this section we will compute the tree-level waveform sourced by Kerr black holes scattering in the ultra-relativistic regime using the shock wave formalism.
Starting from the on-shell mode expansion \eqref{eq:field_mode}, we extract the radiative field at future null infinity by evaluating $h_{\mu\nu}(x)$ at $x^\mu=(u+r,r\hat n)$ with retarded time $u=t-r$ fixed and taking $r\to\infty$. In this limit, the leading radiative field can be written in terms of outgoing on-shell momenta $k^\mu=\omega(1,\hat n)$ with $\omega>0$, yielding the standard time-domain representation~\cite{Cristofoli:2021vyo},
\begin{equation}
\label{eq:Wdef-omega}
h_{\mu\nu}(u,\hat n,r)
=
\frac{1}{4\pi i r}\int_{0}^{\infty}\frac{d\omega}{2\pi}\,
\left[e^{-i\omega u/\hbar}\,
\varepsilon^{(h)}_{\mu\nu}(\hat n)\,
\mathcal{W}^{\,h}(\omega,\hat n)+\mathrm{c.c.}\right]\,,
\end{equation}
where $\mathcal{W}^{\,h}(\omega,\hat n)$ is the frequency-domain waveshape. In the shock-wave formalism, $\mathcal{W}^{\,h}$  is obtained by a transverse Fourier transform of the equal-rapidity correlator \eqref{eq:5ptshock},
\begin{equation}
\label{eq:Wmode-def}
i\mathcal{W}^{\,h}(k,b)
\simeq
\,
\langle
\tilde{\Psi}_j(b_2)\,|\,e^{-\hat H L_2}\,\hat a_{(h)}(k)\,e^{-\hat H L_1}\,|\tilde{\Psi}_i(b_1)
\rangle\,
\Big|_{k=\omega(1,\hat n)}\,
\end{equation}
where, using the Fourier transform in \eqref{eq:fourier_WL} one gets the representation
\begin{equation}
\label{eq:Wmode-def}
i\mathcal{W}^{\,h}(k,b)
\simeq
\int_{q_{1\perp},\,q_{2\perp}}
e^{-\,i(\vec q_1\cdot \vec b_1-\vec q_2\cdot \vec b_2)/ \hbar}\,
\langle
\Psi_j(q_2)\,|\,e^{-\hat H L_2}\,\hat a_{(h)}(k)\,e^{-\hat H L_1}\,|\Psi_i(q_1)
\rangle\,
\Big|_{k=\omega(1,\hat n)}\,.
\end{equation}
In \eqref{eq:Wmode-def} the dependence on the transverse locations $b_{1,2}$ of the Wilson lines is implicit; by translation invariance $\mathcal{W}^{\,h}$ depends only on $\vec b\equiv \vec b_1-\vec b_2$ up to an overall phase. At tree level, setting $e^{-\hat H L_{1,2}}\to 1$ and using \eqref{eq:5ptshock}, the definition \eqref{eq:Wmode-def} reduces to the transverse Fourier transform of the five-point amplitude,
\begin{align}
\mathcal{W}^{(0)\,h}(k,b)
&=
\frac{1}{2s}\int_{q_\perp}
\mathcal{M}^{(0)\,h}_{2\to 3}(q,k,q-k)\,
e^{-i \, \vec{b}\cdot \vec{q}/ \hbar}\,,
\qquad k^\mu=\omega(1,\hat n)\,.
\label{eq:wfdef}
\end{align}
Throughout this section, we parametrise transverse vectors in complex form as
\begin{equation}
b_{\mathrm T} \equiv b^x + i b^y\,, \qquad 
k_{\mathrm T} \equiv k^x + i k^y\,,
\end{equation}
so that the standard scalar and antisymmetric products read
\begin{align}
\vec b\cdot\vec k 
= \frac{1}{2}\!\left(b_{\mathrm T} k_{\mathrm T}^* + b_{\mathrm T}^* k_{\mathrm T}\right), \qquad
\epsilon_{ij} b^i k^j 
= \frac{1}{2i}\!\left(b_{\mathrm T} k_{\mathrm T}^* - b_{\mathrm T}^* k_{\mathrm T}\right).
\end{align}
Here and in the following, $\epsilon_{ij}$ denotes the two-dimensional Levi-Civita symbol in the transverse plane with the convention $\epsilon_{12}=+1$.

\subsection{Tree-level waveform}

We now use the expression \eqref{eq:wfdef} to perform an explicit computation of the tree-level waveform. Following \cite{Ciafaloni:2015xsr}, we first perform the integration over $q^x$ by computing the residues, and then the one over $q^y$, as described in Appendix \ref{App:waveform}. We end up with
\begin{align}
\label{eq:Woplus}
\mathcal{W}^{(0)\,\oplus}(b,k)&\simeq -i\frac{s\kappa^3}{4\pi\hbar^{3/2}}\bigg[\frac{b_{\mathrm{T}}k_{\mathrm{T}}^*-b_{\mathrm{T}}^*k_{\mathrm{T}}\,e^{-i\vec{b}\cdot \vec{k}/ \hbar}}{\vec{b}^{2}(k_{\mathrm{T}}^*)^2}\nonumber \\
&\qquad \qquad + \frac{k_{\mathrm{T}}}{k_{\mathrm{T}}^*}\frac{e^{-ib_{\mathrm{T}}^*k_{\mathrm{T}}/ (2\hbar)}}{2 i}\left(\text{E}_i\left(\frac{ib_{\mathrm{T}}^*k_{\mathrm{T}} }{2 \hbar}\right)+\text{E}_i\left(-\frac{ib_{\mathrm{T}}k_{\mathrm{T}}^*}{2 \hbar}\right)\right)\bigg],
\end{align}
where $E_i(z)$ is the \textit{exponential-integral} function,
\begin{align}
\text{E}_i(z)\equiv-\int_{-z}^{\infty}\frac{e^{-t}}{t}dt,
\end{align}
related to the \textit{sine-} and \textit{cosine-integral }functions by 
\begin{align}
\text{C}_i(z)+i\,\text{S}_i(z)-\frac{i\pi}{2}=\text{E}_i(iz).
\end{align}
Similarly, one gets for $\mathcal{W}^{(0)\,\ominus}$
\begin{align}
\mathcal{W}^{(0)\,\ominus}(b,k)=\mathcal{W}^{(0)\,\oplus}\Big|_{(b_{\mathrm{T}},k_{\mathrm{T}})\leftrightarrow(b_{\mathrm{T}}^*,k_{\mathrm{T}}^*)}.
\end{align}
It is instructive to obtain the waveform components also in the standard $(+,\times)$ tensor basis,
\begin{align}
\label{eq:plustimestensors}
\varepsilon_{(\times)}^{\mu\nu}=\frac{1}{2}(\varepsilon^\mu_\phi\varepsilon^\nu_\theta+\varepsilon^\nu_\phi\varepsilon^\mu_\theta),\qquad \varepsilon_{(+)}^{\mu\nu}=\frac{1}{2}(\varepsilon^\mu_\theta\varepsilon^\nu_\theta-\varepsilon^\nu_\phi\varepsilon^\mu_\phi),
\end{align}
where
\begin{align}
\label{eq:polbasis1}
 \varepsilon_{\phi}^{\mu}=(0,0;-\sin\phi,\cos\phi),\qquad \varepsilon_{\theta}^{\mu}=(-\sin\theta,\sin\theta;\cos\theta\cos\phi,\cos\theta\sin\phi).
\end{align}
Using the light-cone parametrisation $k^\mu=\omega \left(2\cos^2(\theta/2),2\sin^2(\theta/2);\sin\theta\cos\phi,\sin\theta\sin\phi \right)\,$, it is easy to check that $\varepsilon_{\theta}\cdot k=\varepsilon_{\phi}\cdot k=0$.
The full waveform can thus be expressed as
\begin{align}
\mathcal{W}^{\mu\nu}=\mathcal{W}^{\ominus}\varepsilon^{\mu\nu}_{(\oplus)}+\mathcal{W}^{\oplus}\varepsilon^{\mu\nu}_{(\ominus)}=\mathcal{W}^{+}\varepsilon^{\mu\nu}_{(+)}+\mathcal{W}^{\times}\varepsilon_{(\times)}^{\mu\nu},
\end{align}
where, using \eqref{eq:polbasis2} $\mathcal{W}^{+,\times}=2\varepsilon^{(+),(\times)}_{\mu\nu}\mathcal{W}^{\mu\nu}$. Thus one gets from \eqref{eq:polrotations}
\begin{align}
&i\mathcal{W}^\times=e^{2i\phi}\mathcal{W}^{\ominus}-e^{-2i\phi}\mathcal{W}^{\oplus},\qquad\mathcal{W}^+=e^{2i\phi}\mathcal{W}^{\ominus}+e^{-2i\phi}\mathcal{W}^{\oplus}.
\end{align}
Explicitly,
\begin{align}
\mathcal{W}^{(0)+}&\simeq \frac{is\kappa^3}{4\pi\hbar^{3/2}}\bigg[2\bigg(\frac{1-e^{-i\vec{b}\cdot\vec{k}/\hbar}}{\vec{b}\cdot\vec{k}}\bigg)\frac{(\vec{b}\cdot\vec{k})^2}{\vec{b}^2\vec{k}^{\,2}} +\frac{e^{-ib_{\mathrm{T}}k_{\mathrm{T}}^*/ (2\hbar)}}{2 i} \left(\mathrm{E}_i\left(-\frac{ib_{\mathrm{T}}^*k_{\mathrm{T}}}{2 \hbar}\right)+\mathrm{E}_i\left(\frac{ib_{\mathrm{T}}k_{\mathrm{T}}^*}{2 \hbar}\right)\right) \nonumber \\
&\qquad\qquad+\frac{e^{-ib_{\mathrm{T}}^*k_{\mathrm{T}}/ (2\hbar)}}{2 i} \left(\mathrm{E}_i\left(\frac{ib_{\mathrm{T}}^*k_{\mathrm{T}}}{2 \hbar}\right)+\mathrm{E}_i\left(-\frac{ib_{\mathrm{T}}k_{\mathrm{T}}^*}{2 \hbar}\right) \right)\bigg], \label{eq:Wplus} \\
\mathcal{W}^{(0)\times}&=\frac{s\kappa^3}{4\pi\hbar^{3/2}}\bigg[2i\bigg(\frac{e^{-i\vec{b}\cdot\vec{k}/\hbar}+1}{\vec{b}^2\vec{k}^{\,2}}\bigg)\epsilon_{ij}k^i b^j -\frac{e^{-ib_{\mathrm{T}}k_{\mathrm{T}}^*/(2\hbar)}}{2 i} \left(\mathrm{E}_i\left(-\frac{ib_{\mathrm{T}}^*k_{\mathrm{T}}}{2 \hbar}\right)+\mathrm{E}_i\left(\frac{ib_{\mathrm{T}}k_{\mathrm{T}}^*}{2 \hbar}\right)\right) \nonumber \\
&\qquad\qquad+\frac{e^{-ib_{\mathrm{T}}^*k_{\mathrm{T}}/ (2\hbar)}}{2 i} \left(\mathrm{E}_i\left(\frac{ib_{\mathrm{T}}^*k_{\mathrm{T}}}{2 \hbar}\right)+\mathrm{E}_i\left(-\frac{ib_{\mathrm{T}}k_{\mathrm{T}}^*}{2 \hbar}\right) \right)\bigg]. \label{eq:Wtimes}
\end{align}
Using the high-energy factorisation in \eqref{eq:npt} and the impact factors for Kerr black holes in \eqref{eq:impactfactorsKerr}, it is easy to extend the above formulae to the spinning case. We start by noticing that on the five-point kinematics and in the classical limit we have \cite{Guevara:2019fsj} 
\begin{align}
a_1\cdot q_1= \frac{2i}{s}\epsilon(p_1,p_2,q_1,a_1),\qquad a_2\cdot q_2= \frac{2i}{s}\epsilon(p_1,p_2,q_2,a_2),
\end{align}
with $\epsilon(a,b,c,d)\equiv\epsilon_{\mu\nu\rho\sigma}a^{\mu}b^{\nu}c^{\rho}d^{\sigma}$ and $\epsilon_{\mu\nu\rho\sigma}$ the four-dimensional Levi-Civita symbol. Therefore the  $2\rightarrow 3$ amplitude for the scattering of Kerr black holes can be written as
\begin{align}
\mathcal{M}^{(0)\,h}_{2\rightarrow3 }(q_1,k,q_2,a_i)=\mathcal{M}^{(0)\,h}_{2\rightarrow3 }(q_1,k,q_2)\mathrm{exp}\bigg\{\frac{2i}{\hbar s}\bigg[\epsilon(p_1,p_2,q_1,a_1)+\epsilon(p_1,p_2,q_2,a_2)\bigg]\bigg\}.
\end{align}
In the CM frame one has
\begin{align}
\epsilon(p_1,p_2,q_1,a_1)=\frac{s}{2}\epsilon_{ij}q_1^ia_1^j,\qquad\epsilon(p_1,p_2,q_2,a_2) =\frac{s}{2}\epsilon_{ij}q_2^ia_2^j.
\end{align}
where we used $\epsilon_{0ij3}=-\epsilon_{ij}$. Hence the spinning tree-level waveform reads
\begin{align}
\mathcal{W}^{(0)\,h}(b,k,a_i)=\frac{1}{2s}e^{i\epsilon_{ij}k^ia_2^j/\hbar}\int_{q_{\perp}}
\mathcal{M}_{2\rightarrow3}^{(0)\,h}(q,k,q-k)e^{-iq^{i}(b_{i}-\epsilon_{ij}(a_1+a_2)^j)/\hbar}.
\end{align}
Therefore, up to an overall phase, the spinning waveform is obtained from the spinless one by a Newman--Janis shift of the impact parameter~\cite{Alessio:2022kwv,Alessio:2023kgf},
\begin{align}
 \mathcal{W}^{(0)\,h}(b,k,a_i)=e^{i\epsilon_{ij}k^ia_2^j/\hbar}\mathcal{W}^{(0)\,h}(\tilde{b},k_{\mathrm{T}}),\qquad\tilde{b}_{i}=b_{i}-\epsilon_{ij}(a_1+a_2)^j.
\end{align}

\subsection{Recovering the soft expansion: Weinberg and logarithmic soft theorems}

As emphasized in \eqref{eq:softlimit}, the Lipatov current reduces, in the soft limit, to the high-energy limit of the tree level Weinberg current. As a consequence, the waveform in \eqref{eq:Wplus} and \eqref{eq:Wtimes}, should match the high-energy limit of the leading memory waveform. Infact, as it is clear from the expressions in \eqref{eq:Wplus} and \eqref{eq:Wtimes},  a $1/\omega$ pole can only come from the small-$\omega$ expansion of \eqref{eq:Wtimes}. This is also in agreement with \cite{DiVecchia:2022nna,Alessio:2024wmz}. The leading $1/\omega$ Weinberg current is 
\begin{align}
\nonumber
\mathcal{W}^{(0)\mu\nu}_{\frac{1}{\omega}}=&i\frac{2\kappa G m_1m_2}{\vec{b}^2 \, \hbar^{3/2}}\frac{2\sigma^2-1}{\sqrt{\sigma^2-1}}\bigg[b\cdot k\bigg(\frac{p_1^{\mu}p_1^{\nu}}{(p_1\cdot k)^2}-\frac{p_2^{\mu}p_2^{\nu}}{(p_2\cdot k)^2}\bigg)\\ &\qquad\qquad +\frac{1}{p_2\cdot k}(p_2^{\mu}b^{\nu} + p_2^{\nu}b^{\mu}) - \frac{1}{p_1\cdot k}(p_1^{\mu}b^{\nu} + p_1^{\nu}b^{\mu})\bigg].
\end{align}
In the $\sigma\to\infty$ limit, it is simple to show that
\begin{align}
\mathcal{W}^{(0)+}_{\frac{1}{\omega}}\simeq 0,\qquad\mathcal{W}^{(0)\times}_{\frac{1}{\omega}}=2\mathcal{W}^{(0)\mu\nu}_{\frac{1}{\omega}}\varepsilon^{(\times)}_{\mu\nu}\simeq-i\frac{s\kappa^3}{\pi \hbar^{3/2}}\frac{\epsilon_{ij}b^i k^j}{\vec{b}^2\vec{k}^{\,2}}.
\end{align}
This agrees exactly with the soft limit of \eqref{eq:Wtimes}, which is entirely controlled by the first term in that expression. A further non-trivial consistency check is provided by eq.~\eqref{eq:Wplus}: its small-$\omega$ expansion contains a logarithmic contribution that must reproduce the ultra-relativistic limit of the subleading classical soft graviton theorem~\cite{Sahoo:2018lxl}, which can be written as
\begin{align}
\mathcal{W}^{(0)\mu\nu}_{\log\omega}=\frac{2\kappa G\sigma(2\sigma^2-3)}{\hbar^{3/2}(\sigma^2-1)^{\frac{3}{2}}}\log\omega\bigg(\frac{p_2\cdot k}{p_1\cdot k}p_1^{\mu}p_1^{\nu}+\frac{p_1\cdot k}{p_2\cdot k}p_2^{\mu}p_2^{\nu}-p_1^{\mu}p_2^{\nu}-p_1^{\nu}p_2^{\mu}\bigg).
\end{align}
In the high-energy limit one has
\begin{align}
\mathcal{W}^{(0)+}_{\log\omega}=2\mathcal{W}^{(0)\mu\nu}_{\log\omega}\varepsilon^{(+)}_{\mu\nu}\simeq \frac{s\kappa^3}{2\pi\hbar^{3/2}}\log\omega,\qquad\mathcal{W}^{(0)\times}_{\log\omega}=2\mathcal{W}^{(0)\mu\nu}_{\log\omega}\varepsilon^{(\times)}_{\mu\nu}\simeq 0,
\end{align}
which agrees with the $\log\omega$ term coming from \eqref{eq:Wplus}. The latter in particular comes from the soft expansion of the $\mathrm{E}_i(x)$, which reads
\begin{align}
\mathrm{E}_i(x)=\gamma_E+\log(x)+\dots.
\end{align}
%


\section{Conclusions and future directions}
\label{sec:conclusions}

The motivation for studying the high-energy limit of the gravitational two-body problem is stronger than ever, given its growing relevance for the theoretical and phenomenological description of compact-binary dynamics. Building on earlier developments in Regge theory, SCET and shock-wave methods, in this paper we have taken some initial steps towards a systematic framework for studying the ultra-relativistic limit of $2 \to 2 + n$ gravitational massive amplitudes at leading order in the multi-Regge expansion.

In sec.~\ref{sec:MRK}, we reviewed Regge theory as applied to gravitational scattering amplitudes of two incoming and two outgoing minimally coupled massive scalar particles, emphasising the factorisation of high-energy dynamics into impact factors and Lipatov currents~\cite{Lipatov:1982vv,Lipatov:1982it}. This structure applies equally to spinless and spinning particles, including those describing Kerr black holes, and underlies the universality of leading logarithmic contributions observed at high PM orders. Focusing on the classical regime, we reformulated the Regge expansion in sec.~\ref{sec:regge} using an exponential representation of the S-matrix, which provides a generating functional for amplitudes at leading logarithmic accuracy. This representation yields a transparent description of the $s$-channel multi-$H$ sequence~\cite{Amati:1990xe,Amati:2007ak,Rothstein:2024nlq} in terms of iterative three-particle unitarity cuts between tree-level amplitudes in multi-Regge kinematics, and simultaneously reproduces the leading BFKL $t$-channel evolution related to the imaginary part of the $2\to2$ amplitude~\cite{Lipatov:1982it}. In particular, our computation of the leading logarithmic contribution at 5PM in the 2SF sector reproduces the recently obtained massless result~\cite{Alessio:2025isu}, showing the universality of these leading logarithms which we also expect to appear at higher odd PM orders, namely at $(2N+1)$PM with $N\geq 3$, at leading order in the Regge limit.

We further observed that higher-loop contributions to the $2\to2$ amplitude in the MRK approximation exhibit a simple factorised structure, naturally organised in terms of the Lipatov kernel obtained by gluing two Lipatov currents. This observation motivated the development, in sec.~\ref{sec:shock-wave}, of a gravitational extension of the shock-wave formalism originally formulated in QCD~\cite{Balitsky:1995ub,Caron-Huot:2013fea}. In this framework, highly boosted Wilson lines generate shock-wave backgrounds that interact through a boost-invariant Hamiltonian. We analysed the resulting structure and symmetries, and demonstrated its equivalence with the recently developed SCET description for forward scattering~\cite{Rothstein:2024nlq}. Finally, in sec.~\ref{sec:waveform} we applied the shock-wave formalism to the computation of the $2\to3$ amplitude, and extracted the corresponding tree-level scattering waveform in the ultra-relativistic regime. We recovered the spinless result~\cite{Ciafaloni:2015xsr} and, for the first time, extended the analysis to include spin effects.

In the high-energy picture, many issues remain to be understood. At next-to-leading logarithmic (NLL) accuracy, one expects the NLL contributions associated with graviton ladder exchange to be universal, and therefore identical for the scattering of massive or massless particles~\cite{Barcaro:2025ifi}. However, additional effects might arise from next-to-leading order impact factors attached to the ladder, which may depend on the mass and spin of the external states.

A further unresolved issue concerns the role of high-energy logarithms at $(2N)$PM with $N\geq 2$, starting already at 4PM order~\cite{Dlapa:2022lmu}. At present, neither the EFT framework~\cite{Rothstein:2024nlq} nor the shock-wave formalism provides guidance in this sector: within these approaches, contributions involving different numbers of Glauber gravitons attached to the ends of the ladder vanish identically, suggesting that high-energy logarithms at even PM orders lie beyond leading logarithmic accuracy and originate from qualitatively different mechanisms.

Finally, the precise impact of high-energy logarithms on gauge-invariant physical observables remains to be clarified~\cite{Alessio:2025flu}. Understanding how these logarithms enter quantities such as the impulse, the waveform and other inclusive radiative observables at higher orders is an important step toward assessing their phenomenological relevance and their possible resummation. We plan to return to these questions in future work.

\section*{Acknowledgements}

We thank Federico Buccioni, Giulio Falcioni, Michael Saavedra and Ira Rothstein for invaluable discussions, as well for collaborating with us on related topics. We also thank Jan Plefka for comments on the draft.
The work of R.G.\ is supported by the Royal Society grant RF\textbackslash ERE\textbackslash 231084.

\appendix


\section{Multi-particle kinematics}
\label{sec:appa}

We consider the amplitude $\mathcal{M}_{2\rightarrow2+n}$ involving two incoming and two outgoing massive particles with momenta $p_1,p_2$ and $k_0,k_{n+1}$, respectively, and $n$ outgoing gravitons with momenta $k_r$ with $r=1,\dots,n$, as depicted in Fig.~\ref{fig:MRK_amplitude} (see \textit{e.g.}~\cite{DelDuca:1995hf,DelDuca:1999iql}). We adopt the all-outgoing convention for amplitudes, so that physical incoming legs correspond to momenta with negative energy (equivalently, $p_1\to -p_1$, $p_2\to -p_2$ in the physical region).
On-shellness and total momentum conservation read
\begin{align}
\label{eq:BFKLkin}
p^2_1=m^2_1=k_0^2,\qquad p_2^2=m_2^2=k_{n+1}^2,\qquad k^2_i=0,\qquad p_1+p_2+\sum_{j=0}^{n+1}k_j=0.
\end{align}
We define $D$-dimensional light-cone coordinates for each momentum,
\begin{align}
p^{\mu}=(p^+,p^-,\vec{p}\,),\qquad p^{\pm}=p^0\pm p^{D-1},
\end{align}
where $\vec{p}$ is $d$-dimensional with $d=D-2$. It is convenient to introduce the rapidity variable $y$ and the transverse mass $m_{\perp}$ as
\begin{align}
y=\frac{1}{2}\log\frac{p^+}{p^-},\qquad m^2_{\,\perp}=m^2+\vec{p}^{\,2},
\end{align}
in terms of which one gets the on-shell parametrisation,
\begin{align}
p^{\mu}=(m_{\perp}e^y,m_{\perp}e^{-y},\vec{p}\,).
\end{align}
 Without loss of generality we take 
\begin{subequations}
\label{eq:rapidityparametrisation}
\begin{align}
\label{eq:appAkin}
&p_1^{\mu}=-(m_1e^{Y_1},m_1e^{-Y_1},\vec{0}), &&p_2^{\mu}=-(m_2\,e^{Y_{2}},m_2\,e^{-Y_{2}},\vec{0}),\\ &k_0^{\mu}=(m_{1\,\perp}e^{y_0},m_{1\,\perp}e^{-y_0},-\vec{q}_{1}), &&k_{n+1}^{\mu}=(m_{2\,\perp}e^{y_{n+1}},m_{2\,\perp}\,e^{-y_{n+1}},\vec{q}_{n+1}),\\&k_r^{\mu}=(|\vec{k}_{r}|e^{y_r},|\vec{k}_{r}|e^{-y_r},\vec{k}_{r}) \,, \label{eq:mtmmassless}
\end{align}
\end{subequations}
with 
\begin{align}
    m_{{1}\,\perp}^2=m_{1}^2+\vec{q}_{{1}}^{\,2},\qquad  m_{{2}\,\perp}^2=m_{2}^2+\vec{q}_{{n+1}}^{\,2}.
\end{align}
For later convenience, it is also useful to define $q_1=-(p_1+k_0)$, $q_{n+1}=k_{n+1}+p_2$ and $q_{r+1}=q_r-k_r$. In general, any vector $v$ can be decomposed into its parallel component, lying on the two-dimensional plane spanned by $p_1$ and $p_2$, and its orthogonal one, perpendicular to $p_1$ and $p_2$,
\begin{subequations}
\label{eq:decomp}
\begin{align}
&v^{\mu}=v_{\parallel}^{\mu}+v_{\perp}^{\mu},\qquad v_{\parallel}^{\mu}=P^{\mu}_{\parallel}{}_{\nu}v^{\nu},\qquad v_{\perp}^{\mu}=P^{\mu}_{\perp}{}_{\nu}v^{\nu}\,,\\
&P^{\mu}_{\parallel}{}_{\nu}=-\frac{1}{\sigma^2-1}\Big(\frac{p_1^{\mu}p_1^{\nu}}{m_1^2}+\frac{p_2^{\mu}p_2^{\nu}}{m_2^2}-2\sigma \frac{p_1^{(\mu}p_2^{\nu)}}{m_1m_2}\Big),\qquad\qquad\eta^{\mu\nu}=P_{\parallel}^{\mu\nu}+P_{\perp}^{\mu\nu}\,.
\end{align}
\end{subequations}
and $\sigma=(p_1\cdot p_2/m_1m_2)=(s-m_1^2-m_2^2)/(2m_1m_2)$. In the frame where the incoming momenta $p_1$ and $p_2$ have vanishing transverse components, $\vec p_{1\perp}=\vec p_{2\perp}=0$, the plane spanned by $p_1$ and $p_2$ coincides with the $(+,-)$ light-cone directions and we have
\begin{align}
v_{\parallel}^{\mu}=(v^+,v^-,\vec{0}),\qquad v^{\mu}_{\perp}=(0,0,\vec{v}),
\end{align}
so that $v^2_{\perp}=-\vec{v}^2$.

\subsection{Multi-Regge kinematics}
\label{sec:appareg}

Next, we consider the amplitude $\mathcal{M}_{2\rightarrow2+n}$ in multi-Regge kinematics (MRK). We start with the simplest case $r=1$ in \eqn{eq:mtmmassless}, corresponding to the five-point amplitude $\mathcal{M}_{2\rightarrow2+1}$ in the double-Regge kinematics discussed in detail in the last part of subsection \ref{sec:reggemrk}. Double-Regge kinematics reads
\begin{align}
k_0^+\gg k_1^+\gg k_2^+,\qquad k_0^-\ll k_1^-\ll k_2^-,\qquad |\vec{k}_{0}|\simeq|\vec{k}_{1}|\simeq|\vec{k}_{2}|.
\end{align}
Using the parametrisation \eqref{eq:rapidityparametrisation}, the first two conditions are equivalent to 
\begin{align}
\label{eq:rapidityordering1}
y_0\gg y_1\gg y_2.
\end{align}
Momentum conservation implies in this limit,
\begin{align}
\label{eq:momcons}
k_0^+\simeq p_1^+,\qquad k_2^-\simeq p_2^-,
\end{align}
and hence, the Mandelstam invariants are 
\begin{subequations}
\begin{align}
\label{eq:svariables}
    &s=(p_1+p_2)^2\simeq k_0^+k_2^-=m_{1\,{\perp}}m_{2{\perp}}\,e^{y_0-y_2},\\&
    s_1=(k_0+k_1)^2\simeq k_0^+k_1^-=m_{1\,{\perp}}|\vec{k}|\,e^{y_0-y_1},\\&
    s_2=(k_1+k_2)^2\simeq k_1^+k_2^- =m_{2\,{\perp}}|\vec{k}|\,e^{y_1-y_2}.
\end{align}
\end{subequations}
Furthermore eq. \eqref{eq:momcons} implies that $q_i^{\mu}\simeq q_{i\,\perp}^{\mu}$ and
\begin{align}
 &t_1= (p_1+k_0)^2=q^2_1\simeq-\vec{q}^{\,2}_{1},\\&
 t_2= (p_2+k_2)^2=q^2_2\simeq-\vec{q}^{\,2}_{2}.
\end{align}
Alternatively, the double-Regge limit can be recast as $s\gg s_1,s_2\rightarrow\infty$ with $t_i,m^2_i$ fixed \cite{DiVecchia:2020ymx}. The ratio $(s_1s_2/s)\simeq |\vec{k}|^2$ is kept fixed as well. 

In the general $2\rightarrow2+n$ case, the MRK is defined as the kinematical region which generalises eq. \eqref{eq:rapidityordering1} by taking the rapidities of the outgoing particles strongly ordered and all the transverse momenta to be comparable as
\begin{align}
\label{eq:StrongRapOrd}
y_0\gg y_1\gg\dots\gg y_n\gg y_{n+1},\qquad |\vec{k}_{r}|\simeq|\vec{k}_{}|, \qquad r=1,2,\dots,n.
\end{align}
In MRK, the Mandelstam invariants are reduced to
\begin{subequations}
\begin{align}
 &s=(p_1+p_2)^2\simeq k_0^+k_{n+1}^-=m_{1\,{\perp}}m_{2\,{\perp}}\,e^{y_0-y_{n+1}},\\&
    s_{1r}=(k_0+k_r)^2\simeq k_0^+k_r^-=m_{1\,{\perp}}|\vec{k}|\,e^{y_0-y_r},\\&
    s_{2r}=(k_{n+1}+k_r)^2\simeq k_r^+k_{n+1}^- =m_{2\,{\perp}}|\vec{k}|\,e^{y_r-y_{n+1}},\\&s_{rs}=(k_r+k_s)^2\simeq k_r^+k_s^-=|\vec{k}|^2\,e^{y_r-y_s},\qquad r<s
\\
\label{eq:trMRK}
& t_r=q_r^2\simeq-\vec{q}_r^{\,2}.
 \end{align}
\end{subequations}

\subsection{Phase-space}\label{app:phase-space}

The on-shell phase space measure for the production of $2+n$ particles is defined as
\begin{subequations}
    \label{eq:phasespace2n}
\begin{align}
&\int\mathrm{d}\mathcal{P}_{2+n}=\prod_{i=0}^{n+1}\int \mathrm{d}\Phi(k_i)\hat{\delta}^{(D)}\Big(p_1+p_2+\sum_{j=0}^{n+1}k_j\Big),\\& \mathrm{d}\Phi(k_0)=\hat{\mathrm{d}}^Dk_0\hat{\delta}^+(k_0^2-m^2_1),\qquad \mathrm{d}\Phi(k_{n+1})=\hat{\mathrm{d}}^Dk_{n+1}\hat{\delta}^+(k_{n+1}^2-m^2_2),\\&\mathrm{d}\Phi(k_{r})=\hat{\mathrm{d}}^Dk_{r}\hat{\delta}^+(k_{r}^2)
\end{align}
\end{subequations}
Solving the on-shell delta functions and employing light-cone coordinates and the definition of rapidity previously introduced, one finds $\mathrm{d}\Phi(k_i)=\frac{1}{2}\hat{\mathrm{d}} y_i\,\hat{\mathrm{d}}^{d}\vec{k}_{i}$  and thus, using the $D$-dimensional delta function decomposition,
\begin{align}
\hat{\delta}^{(D)}(p)=2\hat{\delta}^{(d)}(\vec{p})\hat{\delta}(p^+)\hat{\delta}(p^-),
\end{align}
the phase space becomes
\begin{align}
\label{eq:P2n}
\int\mathrm{d}\mathcal{P}_{2+n}=\frac{1}{2^{n+1}}\bigg(\prod_{i=0}^{n+1}\int\hat{\mathrm{d}}y_i\,\int_{k_{i \perp}}\bigg)\hat{\delta}^{(d)}\Big(\sum_{j=0}^{n+1} \vec{k}_{j}\Big)\prod_{\lambda=\pm}\hat{\delta}\Big(p_1^\lambda+p_2^\lambda+\sum_{j=0}^{n+1}k_{j}^\lambda\Big),
\end{align}
where we explicitly used the parametrisation in \eqref{eq:rapidityparametrisation}. It is useful to momentarily introduce the  variables $(\Delta Y_{12},\bar{Y}_{12})$ and $(x_1,x_2)$,
\begin{align}
Y_1=\frac{\Delta Y_{12}}{2}+\bar{Y}_{12},\quad Y_2=-\frac{\Delta Y_{12}}{2}+\bar{Y}_{12},\qquad x_1=e^{\bar{Y}_{12}}\sqrt{\frac{m_1}{m_2}},\quad x_2=\frac{1}{x_1}.
\end{align}
in terms of which the last two delta function in \eqref{eq:P2n} become, in the high-energy limit,
\begin{align}
\prod_{\lambda=\pm} \hat{\delta}\Big(p_1^\lambda+p_2^\lambda+\sum_{j=0}^{n+1}k_{j}^\lambda\Big) \simeq\hat{\delta}(x_1\sqrt{s}-m_{1\,{\perp}}e^{y_0})\hat{\delta}(x_2\sqrt{s}-m_{2\,{\perp}}e^{y_{n+1}}).
\end{align}
Assuming that the integrand does not depend on the rapidities $y_0$ and $y_{n+1}$, which is the case for MRK amplitudes at leading logarithmic accuracy we consider in this work, the integral over the two external massive particles rapidities can be performed, yielding 
\begin{align}
\int \hat{\mathrm{d}}y_0\int\hat{\mathrm{d}}y_{n+1}\hat{\delta}(x_1\sqrt{s}-m_{1 \perp}e^{y_0})\hat{\delta}(x_2\sqrt{s}-m_{2 \perp}e^{y_{n+1}})=\frac{1}{s},
\end{align}
so that, solving also the last $d$-dimensional delta function one arrives at
\begin{align}
\label{eq:MRKphasespace}
\int\mathrm{d}\mathcal{P}_{2+n}\simeq\frac{1}{2^{n+1}s}\bigg(\prod_{j=1}^n\int\hat{\mathrm{d}}y_j\bigg)\bigg(\prod_{i=1}^{n+1}\int_{q_{i \perp}}\bigg).
\end{align}
Let us focus on the integral over the rapidities $y_j$. If the integrand does not depend on the graviton rapidities $y_r$ as well, as it happens for the MRK amplitudes we consider, also the integration over these variables can be performed directly, yielding\footnote{In order to disentangle the integrals, one needs to perform a Laplace transform.}
\begin{align}
\prod_{j=1}^n\int\hat{\mathrm{d}}y_j
&=\int_{y_{n+1}}^{y_0}\hat{\mathrm{d}}y_1
\int_{y_{n+1}}^{y_1}\hat{\mathrm{d}}y_2
\cdots
\int_{y_{n+1}}^{y_{n-1}}\hat{\mathrm{d}}y_n \nonumber \\
&=\frac{1}{n!}\left(\frac{y_0-y_{n+1}}{2\pi}\right)^{\!n}
\simeq \frac{1}{n!}\left(\frac{\log|s/t|}{2\pi}\right)^{\!n}\,.
\label{eq:rapidity-simplex}
\end{align}
In MRK one has $s\simeq k_0^+k_{n+1}^-=m_{1\perp}m_{2\perp}e^{y_0-y_{n+1}}$ with $m_{i\perp}^2=m_i^2+\vec q_i^{\,2}\sim |t|$, hence $y_0-y_{n+1}\simeq \log(s/|t|)$ at leading power in the Regge expansion. Therefore one gets 
\begin{align}\label{eq:phase-space}
\int\mathrm{d}\mathcal{P}_{2+n}\simeq\frac{1}{2^{n+1}s}\frac{1}{n!}\bigg(\frac{\log(s/|t|)}{2\pi}\bigg)^n
\left( \prod_{i=1}^{n+1}\int_{q_{i \perp}} \right) \,.
\end{align}
It is manifest that the classical leading logarithms are generated in the amplitude through the integrations over the strongly ordered rapidities of the soft gravitons.


\section{Integrals in multi-Regge amplitudes}
\label{app:integrals}

We review here the results in \cite{Alessio:2025isu}, in particular the computation of the integrals $H^1(q^2)$ \eqref{eq:h1} and $H^2(q^2)$ \eqref{eq:double-h}. As outlined in the main text, the LL $2\to2$ amplitudes exhibit a factorisation between light cone and transverse kinematics. This reduces the general form of four point Feynman integrals to the computation of massless two-point integrals in $d=D-2$ Euclidean transverse dimensions, with external momentum $\vec q$ such that $t=q^2\simeq -\vec q^{\,2}\neq0$.
In order to compute \eqref{eq:h1} and \eqref{eq:double-h}, we introduce the following class of {\it generalised bubbles} \cite{Kotikov:2018wxe},
\begin{equation}\label{eq:generalised-bubbles}
\begin{gathered}
    \mathcal{B}_{a_1,a_2}(q) = \int \frac{\hat{\mathrm{d}}^d\vec{l}}{[(\vec{q}-\vec{l})^2]^{a_1}[\vec{l}^{\,2}]^{a_2}} = \frac{(\vec{q}^{\,2})^{\frac{d}{2}-a_1-a_2}}{(4\pi)^{\frac{d}{2}}} \, B_{a_1,a_2}, \\
    B_{a_1,a_2} = \frac{F\left(a_1\right) F\left(a_2\right)}{F\left(a_1+a_2-\frac{d}{2}\right)}, \hspace{20pt} F(a) = \frac{\Gamma\left(\frac{d}{2}-a\right)}{\Gamma\left(a\right)},
\end{gathered}
\end{equation}
functions in $d$ Euclidean dimensions, depending on $q_\perp$. Strictly related are the {\it iterated bubbles} of order $n$. They are defined as
\begin{equation}
    \mathcal{B}_{1,1}^{\,(n)}(\vec{q}) = \int \left(\prod_{i=1}^n \hat{\mathrm{d}}^d\vec{l}_{i} \right) \left[\bigg(\vec{q} - \sum_{j=1}^n \vec{l}_{j}\bigg)^2\prod_{i=1}^n \vec{l}_{i}^{\,2}\right]^{-1}
\end{equation}
and it is easy to see that it is solved applying $n$ times eq.\eqref{eq:generalised-bubbles}. Here the subscripts $(1,1)$ refer to the first integration, that's a simple bubble $\mathcal{B}_{1,1}$. By matching the notation with \eqref{eq:generalised-bubbles} we have $\mathcal{B}_{a_1,a_2} = \mathcal{B}_{a_1,a_2}^{\,(1)}$. Iterations give a recursive structure for the powers of the momenta in the bubble integrands,
\begin{equation}
    \underbrace{\left(1,1\right) \equiv (1,p_1)}_\text{1st iteration} \to \underbrace{\left(1,2-d/2\right) \equiv (1,p_2)}_\text{2nd iteration} \to \cdots \to \underbrace{(1,p_n)}_{n-\text{th iter.}}, \quad p_n = n+\frac{d}{2}(1-n)
\end{equation}
that gives an overall factor $(\vec{q}^{\,2})^{n(d/2-1)-1}/(4\pi)^{nd/2}$. The non trivial part is the computation of the Gamma dependence,
\begin{equation}
    \prod_{i=1}^n B_{1,p_i} = \frac{\Gamma\left(\frac{d}{2}-1\right)^{n+1}}{F(p_{n+1})}\,,
\end{equation}
that gives
\begin{equation}\label{eq:iterated-bubbles}
    \mathcal{B}_{1,1}^{\,(n)}(q) = \frac{(\vec{q}^{\,2})^{n\frac{d-2}{2}-1}}{(4\pi)^{nd/2}} \frac{\Gamma\left(\frac{d}{2}-1\right)^{n+1}}{F(p_{n+1})}\,.
\end{equation}

\subsection{$H^1$ integral}

The integral $H^1(q^2)$ in equation \eqref{eq:h1} is reduced using \texttt{LiteRed2}~\cite{Lee:2012cn} to perform integration by parts identities (IBPs). We adopt the basis,
\begin{equation}\label{eq:basis}
    \mathcal{P}_{H^1} = \left\{ (\vec{q}-\vec{q}_1)^2, \vec{q}_1^{\,2}, (\vec{q}-\vec{q}_2)^2, (\vec{q}_1-\vec{q}_2)^2, \vec{q}_2^{\,2} \right\},
\end{equation}
leading to the following two master integrals,
\begin{align}\label{eq:MIs-def}
    & \mathcal{I}_{01110}^{H^1} (q) = \int_{q_{1\!\perp},q_{2\!\perp}} \frac{1}{\vec{q}_1^{\,2} \, (\vec{q}-\vec{q}_2)^2 \, (\vec{q}_1-\vec{q}_2)^2} = \left(\mu^{2}e^{\gamma_E}\right)^{2\epsilon}\mathcal{B}_{1,1}^{\,(2)} (q)\,, \\
    & \mathcal{I}_{11101}^{H^1} (q) = \int_{q_{1\!\perp},q_{2\!\perp}} \frac{1}{(\vec{q}-\vec{q}_1)^2 \, \vec{q}_1^{\,2} \, (\vec{q}-\vec{q}_2)^2 \, \vec{q}_2^{\,2}} = \left(\mu^{2}e^{\gamma_E}\right)^{2\epsilon}\big(\mathcal{B}_{1,1}(q)\big)^2\,,
\end{align}
solved using bubbles \eqref{eq:generalised-bubbles} and iterated bubbles \eqref{eq:iterated-bubbles}. The result of the IBPs is
\begin{align}
    H^1(q^2) = & -\frac{2}{3} (d+1) \, \vec{q}^{\,2} \, \mathcal{I}_{01110}^{H^1} (q) + \left(\vec{q}^{\,2}\right)^2 \mathcal{I}_{11101}^{H^1} (q) = \\
    = &\, \frac{1}{16\pi^2} \left(\frac{4\pi \mu^2 e^{\gamma_E}}{\vec{q}^{\,2}}\right)^{2\epsilon} \left( \frac{\Gamma (-\epsilon )^4 \Gamma (\epsilon +1)^2}{\Gamma (-2 \epsilon )^2} + \frac{2 (2 \epsilon -3) \Gamma (-\epsilon )^3 \Gamma (2 \epsilon +1)}{3 \Gamma (-3 \epsilon )}\right),
\end{align}
while the $\epsilon$-expanded result is
\begin{align}
    H^1(q^2) = \frac{1}{8\pi^2} \left(\frac{4\pi\mu^2}{\vec{q}^{\,2}}\right)^{2\epsilon} \bigg[ & -\frac{1}{\epsilon ^2}+\frac{2}{\epsilon }+\zeta _2+\frac{1}{3} \left(68 \zeta _3-6 \zeta _2\right) \epsilon +\left(\frac{129\zeta_4}{4}-\frac{64 \zeta _3}{3}\right) \epsilon^2 \\
    & + \left(-\frac{68}{3} \zeta _2 \zeta _3+\frac{692 \zeta _5}{5}-\frac{57 \zeta_4}{2}\right) \epsilon ^3+O\left(\epsilon ^4\right)\bigg] \nonumber\,,
\end{align}
reproducing \eqref{eq:3pcc} and \eqref{eq:H1expanded}. This agrees with the calculation of Amati, Ciafaloni and Veneziano~\cite{Amati:1990xe}.

\subsection{$H^2$ integral}

The definition of the $H^2(q^2)$ in eq.\eqref{eq:definition-H2-perp} shows a factorised dependence on the variables $\vec{q}_{1}$ and $\vec{q}_{2}$ in the kernels, making easy to integrate them. We use
\begin{equation}\label{eq:J-integral}
    I(k_2,l) = \int \hat{\mathrm{d}}^d\vec{k}_{1}\frac{\mathcal{H}_{\mathrm{GR}}(k_1,k_2;l)}{\vec{k}_1^{\,2} \, \vec{k}_2^{\,2} \, (\vec{k}_1-\vec{l})^2 \, (\vec{k}_2-\vec{l})^2} \equiv I_1(k_2,l) + I_2(k_2,l)\,,
\end{equation}
with
\begin{align}
    I_1(k_2,l) = &\, \frac{B_{1,1}}{(4\pi)^{d/2}} \frac{(\vec{l}^{\,2})^\frac{d}{2}}{\vec{k}_2^{\,2} \, (\vec{k}_2-\vec{l})^2}\,,\\
    I_2(k_2,l) = & \left(\frac{d}{2} - \frac{d}{2}\frac{\vec{l}^{\,2}}{\vec{k}_2^{\,2}}-\left(1-\frac{d}{2}\right)\frac{(\vec{l}-\vec{k}_{2})^2}{\vec{k}_2^{\,2}}\right)\mathcal{B}_{1,1}(l-k_{2}) + \\
    & + \left(\frac{d}{2} - \frac{d}{2}\frac{\vec{l}^{\,2}}{(\vec{l}-\vec{k}_{2})^2}-\left(1-\frac{d}{2}\right)\frac{\vec{k}_2^{\,2}}{(\vec{l}-\vec{k}_2)^2}\right)\mathcal{B}_{1,1}(k_{2})\,.
\end{align}
We insert a transverse delta function to enforce momentum conservation and rewrite $H^2$ as
\begin{equation}
    H^2(q^2) = \mu^{8\epsilon} e^{\epsilon\gamma_E}\int_{k_{2\perp},l_{1\perp},l_{2\perp}} \hat{\delta}^{(d)}(\vec{l}_1+\vec{l}_2-\vec{q}-\vec{k}_2)\, \vec{k}_2^{\,2} I(k_{2},l_{1})\, I(k_{2},l_{2})\,.
\end{equation}
The integrands are now entangled only by the Dirac delta. We disentangle them by computing the Fourier transform of $H^2$ in $b$-space,\footnote{We stress that the FT brings an additional factor $e^{\epsilon\gamma_E}$, which simplifies the organisation of the result of the amplitude in $b$-space, see for example~\cite{Alessio:2025isu}.}
\begin{equation}
    \tilde{H}^2(b^2) = \int_{q_\perp} \,e^{i\vec{b}\cdot\vec{q}} \, H^2(q^2) = \mu^{8\epsilon} e^{4\epsilon\gamma_E} \int_{q_{2\perp}} e^{-i\vec{b} \cdot \vec{q}_2}\, \vec{q}_2^{\,2} \left(\tilde{I}(q_{2},b)\right)^2\,.
\end{equation}
We now handle the integrand separating the following contributions,
\begin{equation}\label{eq:H2-b-space}
\begin{aligned}
    \tilde{H}^2(b) = & \mu^{8\epsilon} e^{4\epsilon\gamma_E} \int_{q_{2\perp}} e^{-i\vec{b} \cdot \vec{q}_2}\,\vec{q}_2^{\,2}\left(\tilde{I}_1(q_{2},b)^2 +2 \tilde{I}_1(q_{2},b)\tilde{I}_2(q_{2},b) + \tilde{I}_2(q_{2},b)^2\right) \equiv \\
    & \equiv \tilde{H}_{(a)}^2(b) + \tilde{H}_{(b)}^2(b) + \tilde{H}_{(c)}^2(b)\,.
\end{aligned}
\end{equation}
The second and third terms are computed \eqref{eq:generalised-bubbles} and \eqref{eq:iterated-bubbles} together with the Fourier transform
\begin{equation}
\int\hd^d\vec{q}\,\frac{e^{i\vec{b}\cdot\vec{q}}}{\left(\vec{q}^{\,2}\right)^a} = \frac{(\vec{b}^{\,2}/4)^{a}}{(\pi \vec{b}^{\,2})^{d/2}}\,F(a)\,,
\end{equation}
obtaining, in $q$-space
\begin{equation}
    H_{(b)}^2(q^2) + H_{(c)}^2(q^2) = \frac{\vec{q}^{\,2}}{(4\pi)^4}\left(\frac{4\pi\mu^2 e^{\gamma_E}}{\vec{q}^{\,2}}\right)^{4\epsilon} g(d)  \Big[f_1(d)+f_2(d)\big(h_a(d)+h_b(d)\big)\Big]\,,
\end{equation}
with
\begin{align}
    & g(d) = \frac{\pi  2^{4(1-d)} \Gamma \left(\frac{d}{2}\right)^2}{\Gamma \left(\frac{d-1}{2}\right)^2 \Gamma \left(\frac{5 d}{2}-3\right)}\,, \nonumber \\
    & f_1(d) = 4^{d+1} \left(d \left(5 d^2+d-9\right)-6\right) \Gamma (3-2 d) \Gamma (d-2)^2 \Gamma \left(\frac{d}{2}\right)\,, \nonumber \\
    & f_2(d) = \Gamma \left(1-\frac{d}{2}\right) \Gamma \left(\frac{d}{2}-1\right)\,, \nonumber \\
    & h_a(d) = -2^{2 d+1} (d (2 d-1)-4) \frac{\Gamma (1-2 d) \Gamma (1-d) \Gamma \left(\frac{d}{2}-1\right) \Gamma (d) \Gamma (2 d)}{\Gamma \left(1-\frac{3 d}{2}\right)\Gamma\left(\frac{3 d}{2}\right)}\,, \nonumber \\
    & h_b(d) = -\frac{1}{\sqrt{\pi}}3 ((d-3) d+4) (d (3 d-2)-4) \Gamma \left(\frac{3}{2}-d\right) \Gamma \left(1-\frac{d}{2}\right) \Gamma \left(\frac{d}{2}\right) \Gamma \left(\frac{3 d}{2}-3\right)\,. \nonumber
\end{align}
The term containing $\tilde{I}_1^{\,2}$ in \eqref{eq:H2-b-space}, once expressed in $q$-space, is related to a known integral. It contains a massless kite topology in which two powers of the propagators are non integer, as already announced in~\cite{Alessio:2025isu}. The family of scalar massless kites is defined as
\begin{equation}\label{eq:general-kite}
    \hspace{-10pt}\text{K}(\nu_1,\nu_2,\nu_3,\nu_4,\nu_5)
    = \int_{l_{1\perp},l_{2\perp}} \frac{1}{[(\vec l_1)^{\,2}]^{\nu_1}[(\vec l_1-\vec q)^2]^{\nu_4} [(\vec l_2)^{\,2}]^{\nu_3}[(\vec l_2-\vec q)^2]^{\nu_2} [(\vec l_1+\vec l_2-\vec q)^2]^{\nu_5}}\,,
\end{equation}
and the relation with our term is
\begin{equation}
    H^{2}_{(a)}(q^2) = \frac{\mu^{8\epsilon}\,e^{2\epsilon\gamma_E}}{(4\pi)^d}\,(B_{1,1})^2\,\text{K}\left(-d/2,1,-d/2,1,1\right)\,.
\end{equation}
This integral has known solution~\cite{Bierenbaum:2003ud,Kotikov:2013eha}, achieved through both a double Mellin-Barnes representation and Gegenbauer polynomial technique~\cite{Kotikov:1995cw,Kotikov:2018wxe}. Using the result in appendix B of~\cite{Kotikov:2016rgs}, we obtain
\begin{equation}
  H^{2}_{(a)}(q^2) = \frac{\vec{q}^{\,2}}{(4\pi)^4}\left(\frac{4\pi\mu^2e^{\gamma_E}}{\vec{q}^{\,2}}\right)^{4\epsilon} \Big[\bar f_1(d)+\bar f_2(d)\big(\bar h_a(d)+\bar h_b(d)\big)\Big]\,,
\end{equation}
with
\begin{align}
    & \bar f_1(d) = \frac{8 \Gamma \left(1-\frac{3 d}{2}\right) \Gamma (1-d) \Gamma \left(1-\frac{d}{2}\right) \Gamma \left(\frac{d}{2}\right)^4 \Gamma (d) \Gamma \left(\frac{3 d}{2}\right)}{\Gamma
   (d-1) \Gamma \left(\frac{5 d}{2}-3\right)}\,, \quad \bar f_2(d) = \frac{(d-2) \Gamma \left(1-\frac{d}{2}\right)^5 \Gamma \left(\frac{d}{2}\right)^5}{\Gamma \left(2-\frac{d}{2}\right)^4 \Gamma \left(-\frac{d}{2}\right)}\,, \nonumber\\
    & \bar h_a(d) = -\frac{(d-2)^2 \Gamma (1-2 d) \Gamma (2 d) \, _3F_2\left(1,d-2,\frac{3 d}{2}-2;1-\frac{d}{2},\frac{3 d}{2}-1;1\right)}{(3 d-4) \Gamma (2 d-2) \Gamma \left(\frac{5 d}{2}-3\right)}\,, \nonumber\\
    & \bar h_b(d) = -\frac{\Gamma (1-d)^2 \Gamma \left(1-\frac{d}{2}\right) \Gamma \left(\frac{d}{2}\right) \Gamma (d)^2 \, _3\tilde{F}_2\left(d-2,\frac{3 d}{2}-2,\frac{5 d}{2}-3;\frac{3 d}{2}-1,2
   d-3;1\right)}{\Gamma \left(1-\frac{3 d}{2}\right) \Gamma (d-2) \Gamma \left(\frac{3 d}{2}\right)}\,,
\end{align}
which is the result that we obtained in appendix B of~\cite{Alessio:2025isu}.


\section{Absence of conformal invariance in the gravitational Regge theory }
\label{app:conformal}

The JIMWLK-Balitsky evolution equation in QCD shows non-trivial conformal invariance properties, as outlined in~\cite{Caron-Huot:2013fea}.
Yang--Mills theory is classically conformal in four dimensions and has a dimensionless coupling. In the high-energy limit (at fixed coupling) this implies an $SL(2,\mathbb{C})$ invariance acting on the transverse plane, which becomes manifest upon identifying $\vec x\in\mathbb{R}^2$ with $z=x^1+i x^2$.
In fact, the introduction of infinite Wilson lines $\Phi_{i/j}$ along a fixed direction breaks the $SO(4,2)$ conformal symmetry down to $SO(3,1)\simeq SL(2,\mathbb{C})$, which acts as the global conformal group on the transverse plane. This symmetry becomes manifest when the transverse coordinates $\vec x$ are identified with complex variables $z=x^1+i x^2$, so that $\vec x\cdot\vec p=(z p_{\mathrm{T}}^*+z^*p_{\mathrm{T}})/2$. Furthermore, the elements composing the Hamiltonian in QCD at leading logarithmic accuracy, {\it i.e.} the gluon Regge trajectory and Lipatov kernel, show explicit conformal invariance. That invariance has been used to solve the BFKL equation in QCD~\cite{Lipatov:1985uk}, see also~\cite{Barone:2002cv,Ioffe:2010zz}. Thus we may ask whether this property holds in our gravitational framework.
To understand it, we analyse the two main components of the theory, i.e. the Wilson lines \eqref{eq:grav-WL} and the evolution Hamiltonian \eqref{eq:Hamiltonian} at the LL accuracy.

As regards the Wilson lines, from \eqref{eq:W-field} and \eqref{eq:W-field-def} we see that the high-energy limit is characterized by a dimensionless effective coupling $\kappa\sqrt{s}$. This is an accident of the Planckian regime of gravity, which combines a dimensionful coupling with a double copy numerator containing $p_{1,2} \sim \sqrt{s}$. 
Moreover, the Wilson line depends on the transverse coordinates only through the insertion point of the eikonal field $W(x_\perp)$ (equivalently $h_{++}$ or $h_{--}$ along the line), while Möbius transformations act solely on the transverse plane. Accordingly, the gravitational Wilson lines transform as scalar primaries of weight zero under $SL(2,\mathbb{C})$
\begin{equation}
    z \mapsto z(\rho) = \frac{a\rho + b}{c\rho + d}\bigg|_{ad-bc=1}\,,\qquad a,b,c,d \in \mathbb{C}\,.
\end{equation}

We now turn to the Regge trajectory operator $\hat{\mathcal{R}}_1$ and the boost Hamiltonian $\hat{\mathcal{R}}_2$. They both appear exponentiated in the evolution operator, then must be dimensionless. However, they do not contain the dimensionless coupling $\kappa\sqrt{s}$ but the standard $\kappa$, from which we argue that the evolution operator breaks the conformal invariance. It is very easy to see this breakdown for the one-loop Regge-trajectory insertion.
We start studying the application of $\hat{\mathcal{R}}_1$ to a Wilson line,
\begin{equation}
    \frac{\delta \Phi_i(q)}{\delta W(k)}|0\rangle = \frac{\delta}{\delta W(k)} \sum_{n=0}^{\infty} \frac{(i\lambda)^n}{n!} \big|W^n(q)\big\rangle = i\lambda \sum_{n=0}^\infty \big|W^n(q-k)\big\rangle = i\lambda \, \Phi_i(q-k)|0\rangle\,,
\end{equation}
where $\lambda = \sqrt{{s}/{2}}\cdot\kappa/\hbar$.
Then the Fourier transform of $\hat{\mathcal{R}}_1\Phi_i(q)$ is
\begin{align}
    \text{F.T.}\big[\hat{\mathcal{R}}_1\Phi_i\big](\vec{x}) &= \int_{q_\perp} 
    e^{i \vec{x} \cdot \vec{q}_\perp}
    \,\hat{\mathcal{R}}_1 \Phi_i(q) \nonumber \\
    &= -i\lambda \int d^d \vec{x}'\, \tilde{\alpha}^{(1)}(\vec{x}'-\vec{x}) W(\vec{x}') \Phi_i(\vec{x})
\end{align}
where $\tilde{\alpha}^{(1)}(\vec{x})$ is the Fourier transform of the one-loop Regge trajectory of eq.~\eqref{eq:gravitonRegge}. In $d=2$ and using the complex notation, it behaves as $\kappa^2/|\vec{x}|^4$, thus the integral above is proportional to
\begin{equation}
    \kappa^2 \int \frac{d^2\vec{x}'}{|\vec{x}'|^4} W(\vec{x}+\vec{x}')\,.
\end{equation}
The Reggeon field $W$ has no dimensions, and the integral is not scale invariant. For example, under inversion $z'=1/\rho$ it scales as $ {d^2\vec{x}'}/{|\vec{x}'|^4} = d^2\rho$, showing that $\hat{\mathcal{R}}_1$ breaks conformal invariance. A more detailed analysis would be required to establish the breaking mechanism for $\hat{\mathcal{R}}_2$, but this is not necessary for our purposes. Using scaling arguments, we consider
\begin{equation}
    \text{F.T.}\big[\hat{\mathcal{R}}_2\Phi_i\big](\vec{x}) \propto \int d^2\vec{x}_1\,d^2\vec{x}_2\, \tilde{H}_{22}(\vec{x}_2-\vec{x}_1,\vec{x}_1,\vec{x}_2)W(\vec{x}-\vec{x}_1)W(\vec{x}-\vec{x}_2)\Phi_i(\vec{x})\,,
\end{equation}
where
\begin{align}
    \tilde{H}_{22}(\vec{x},\vec{x}_1,\vec{x}_2) &= 
    \int_{l_\perp, q_{1 \perp}, q_{2 \perp}}  e^{i(\vec{x} \cdot \vec{l}_\perp+\vec{x}_1 \cdot \vec{q}_{1\perp}+\vec{x}_2 \cdot \vec{q}_{2\perp})}
   H_{22}(l;q_1,q_2)\,.
\end{align}
Under a global rescaling $(\vec{x},\vec{x}_1,\vec{x}_2)\mapsto a(\vec{x},\vec{x}_1,\vec{x}_2)$, we rescale the momenta $(\vec{l}_\perp,\vec{q}_{1\perp},\vec{q}_{2\perp})\mapsto (\vec{l}_\perp,\vec{q}_{1\perp},\vec{q}_{2\perp})/a$ so that the Fourier phases remain invariant. The measure then transforms as $\hat{\mathrm{d}}^2 l_{\perp}\,\hat{\mathrm{d}}^2 q_{1\perp}\,\hat{\mathrm{d}}^2 q_{2\perp}\mapsto a^{-6}\hat{\mathrm{d}}^2 l_{\perp}\,\hat{\mathrm{d}}^2 q_{1\perp}\,\hat{\mathrm{d}}^2 q_{2\perp}$, while $H_{22}$ is unchanged. Hence
\begin{equation}
\tilde{H}_{22}(a\vec{x},a\vec{x}_1,a\vec{x}_2)=a^{-6}\,\tilde{H}_{22}(\vec{x},\vec{x}_1,\vec{x}_2)\,.
\end{equation}
We conclude that the dimensionality of the gravitational coupling prevents the Regge evolution from being conformally invariant, even at leading logarithmic accuracy.


\section{Details on the scattering waveform computation}
\label{App:waveform}
\subsection{The method of residues}

The integrand in equation \eqref{eq:wfdef} is, for the polarization $\oplus$,
\begin{align}
\label{eq:D11}
\nonumber\mathcal{M}_{2\to 3}^{(0)\,\oplus}(q,k,q-k)e^{-i\vec{q}\cdot \vec{b}/\hbar}&=\frac{-2is^2\kappa^3(k^yq^x-k^xq^y)e^{-i(b^xq^x+b^yq^y)/\hbar}}{\hbar^{3/2}(k^x-ik^y)^2(q^x-k^x+iq^y-ik^y)(q^x-iq^y)}\\&\equiv f^{\oplus}(q^x,q^y),
\end{align}
that has to be integrated over $(q^x,q^y)$. Following \cite{Ciafaloni:2015xsr}, we first choose choose $\vec b \parallel \hat x$ with $b^x>0$ and integrate over $q^x$ using the residue theorem. The function in \eqref{eq:D11} has two simple poles in $q^{x}_1=iq^y$ and $q^{x}_2= k^x - i(q^y-k^y)$. Therefore, by closing the contour in the lower-half $q^x$ complex plane, we get
\begin{align}
\int_{-\infty}^{\infty}\hat{\mathrm{d}}q^x f^{\oplus}(q^x,q^y)=
-i\Big[\Theta(-q^y)\,\mathrm{Res}_{q^x=i q^y}f^\oplus
+\Theta(q^y-k^y) \,\mathrm{Res}_{q^x=k^x-i(q^y-k^y)}f^\oplus\Big].
\end{align}
The pole at $q^x=i q^y$ lies inside the contour for $q^y<0$, while the pole at $q^x = k^x - i\,(q^y-k^y)$ is enclosed only when $q^y>k^y$. As a result, the $q^x$ integral evaluates to
\begin{align}
\nonumber&\int_{-\infty}^{\infty}\hat{\mathrm{d}}q^x f^{\oplus}(q^x,q^y)=-\frac{2s^2\kappa^3}{\hbar^{3/2}}e^{-ib^yq^y/\hbar}\Bigg[\frac{q^y\,e^{b^xq^y /\hbar}}{(k^x-ik^y)(k^x+ik^y-2iq^y)}\Theta(-q^y)\\
&\qquad \qquad\qquad \qquad \qquad \qquad +\frac{e^{-b^x(ik^x-k^y+q^y)/\hbar}(k^x+ik^y)(k^y-q^y)}{(k^x-ik^y)^2(k^x+ik^y-2iq^y)}\Theta(q^y-k^y)\Bigg].
\end{align}
Further integrating the above expression over $q^y$ yields the result in \eqref{eq:Woplus}.

\subsection{$(\oplus,\ominus)$ and $(+,\times)$ basis}
\label{appDvectors}

The vectors in \eqref{eq:polLG} satisfy the relations
\begin{align}
\label{eq:relationsplusminus}
\varepsilon^{(\oplus)} \cdot \varepsilon^{(\oplus)}=0,\qquad\varepsilon^{(\ominus)} \cdot \varepsilon^{(\ominus)}=0,\qquad \varepsilon^{(\oplus)}\cdot \varepsilon^{(\ominus)}=-1.\end{align}
Consequently, the tensors $\varepsilon_{\mu\nu}^{(\oplus)}=\varepsilon_{\mu}^{(\oplus)}\varepsilon_{\nu}^{(\oplus)}$ and $\varepsilon_{\mu\nu}^{(\ominus)}=\varepsilon_{\mu}^{(\ominus)}\varepsilon_{\nu}^{(\ominus)}$ satisfy
\begin{align}
\label{eq:relationsplusminus2}
\varepsilon_{\mu\nu}^{(\oplus)}\varepsilon^{\mu\nu}_{(\oplus)}=0,\qquad \varepsilon_{\mu\nu}^{(\ominus)}\varepsilon^{\mu\nu}_{(\ominus)}=0,\qquad \varepsilon_{\mu\nu}^{(\oplus)}\varepsilon^{\mu\nu}_{(\ominus)}=1.
\end{align}
The vectors in \eqref{eq:polbasis1} satisfy
\begin{align}
\varepsilon_{\theta}^2=-1,\qquad \varepsilon_{\phi}^2=-1,\qquad \varepsilon_{\theta}\cdot \varepsilon_{\phi}=0,
\end{align}
and the tensors in \eqref{eq:plustimestensors} 
\begin{align}
\label{eq:polbasis2}
\varepsilon_{(\times)}^{\mu\nu}\varepsilon_{(\times)\mu\nu}=\frac{1}{2},\qquad\varepsilon_{(+)}^{\mu\nu}\varepsilon_{(+)\mu\nu}=\frac{1}{2},\qquad\varepsilon_{(\times)}^{\mu\nu}\varepsilon_{(+)\mu\nu}=0.
\end{align}
The $\{\oplus,\ominus\}$ and $\{+,\times\}$ tensors satisfy 
\begin{subequations}
\label{eq:polrotations}
\begin{align}
&\varepsilon_{\mu\nu}^{(\times)}\varepsilon^{\mu\nu}_{(\oplus)}=-\frac{i}{2}e^{2i\phi},\qquad\varepsilon_{\mu\nu}^{(\times)}\varepsilon^{\mu\nu}_{(\ominus)}=\frac{i}{2}e^{-2i\phi},\\&\varepsilon_{\mu\nu}^{(+)}\varepsilon^{\mu\nu}_{(\oplus)}=\frac{1}{2}e^{2i\phi},\qquad\varepsilon_{\mu\nu}^{(+)}\varepsilon^{\mu\nu}_{(\ominus)}=\frac{1}{2}e^{-2i\phi}.
\end{align}
\end{subequations}

\bibliographystyle{JHEP}
\bibliography{references}
\end{document}